\documentclass[12pt]{article}
\pdfoutput=1
\usepackage{jheppub}
\usepackage[utf8]{inputenc}
\usepackage{verbatim}
\usepackage{amsmath}
\usepackage{amssymb}
\usepackage{amsthm}
\usepackage{slashed}
\usepackage{amsfonts}
\usepackage{subfigure}
\usepackage{float}
\usepackage{physics}

\newtheorem{theorem}{Theorem}
\newtheorem{definition}[theorem]{Definition}

\newcommand{\eq} {equation}
\newcommand{\eqa} {eqnarray}
\newcommand{\NN} {\nonumber}

\preprint{CALT-TH-2020-013, DMUS-MP-20/10, YITP-20-146}

\begin{document}

\title{Toward simulating Superstring/M-theory \\
on a quantum computer}

\author[a]{Hrant Gharibyan,}
\author[b,c]{Masanori Hanada,}
\author[c]{Masazumi Honda,}
\author[a]{Junyu Liu}

\affiliation[a]{Walter Burke Institute for Theoretical Physics and Institute for Quantum Information and Matter, California Institute of Technology, Pasadena, CA 91125, USA}
\affiliation[b]{Department of Mathematics, University of Surrey, Guildford, Surrey, GU2 7XH, UK}
\affiliation[c]{Yukawa Institute for Theoretical Physics,  Kyoto University,\\ Kitashirakawa Oiwakecho, Sakyo-ku, Kyoto 606-8502, Japan}


\abstract{
We present a novel framework for simulating matrix models on a quantum computer. 
Supersymmetric matrix models have natural applications to superstring/M-theory and gravitational physics, in an appropriate limit of parameters. Furthermore, for certain states in the Berenstein-Maldacena-Nastase (BMN) matrix model, 
several supersymmetric quantum field theories dual to superstring/M-theory can be realized on a quantum device.
Our prescription consists of four steps: 
regularization of the Hilbert space, adiabatic state preparation, simulation of real-time dynamics, and measurements. 
Regularization is performed for the BMN matrix model with the introduction of energy cut-off 
via the truncation in the Fock space. 
We use the Wan-Kim algorithm for fast digital adiabatic state preparation to prepare the low-energy eigenstates of this model as well as thermofield double state. 
Then, we provide an explicit construction for simulating real-time dynamics utilizing techniques of block-encoding, qubitization, and quantum signal processing. 
Lastly, we present a set of measurements and experiments that can be carried out on a quantum computer to further our understanding of superstring/M-theory beyond analytic results. 
}
\maketitle

\section{Introduction}
{}
Quantum Field Theory (QFT) is the language of nature. In order to understand nature, we have to define QFT and solve it.   
In high energy physics, the lattice approach to QFT is a powerful conceptual and computational tool \cite{Wilson:1974sk,Creutz:1980zw}. Supersymmetry (SUSY) is another important piece in modern theoretical physics. 
It may exist at a low-energy scale within reach by the LHC or next-generation particle accelerators, and at very least, 
it plays an important role in the holographic approach to quantum gravity. 
More specifically, via gauge/gravity duality \cite{Maldacena:1997re}, 
certain supersymmetric QFTs can give a nonperturbative formulation of superstring/M-theory.
This gives us a strong motivation to define supersymmetric QFTs and solve them. 

Typically, the traditional lattice QFT approach considers the Euclidean spacetime and uses the Markov Chain Monte Carlo simulation method.\footnote{
See, e.g.,~Refs.~\cite{Hanada:2018fnp,Joseph:2019zer} for reviews for dummies and beginners.
} 
Such approach is effective for various important problems, such as the derivation of spectrum of hadrons from QCD \cite{Aoki:2008sm,Durr:2008zz}, 
determination of nuclear potential \cite{Ishii:2006ec}, and black hole thermodynamics in the holographic setup \cite{Anagnostopoulos:2007fw,Catterall:2008yz}. 
Still, there are many problems that cannot be accessed in this manner, most notably the real-time dynamics. 
Quantum simulation is a promising approach to such problems. 

Despite a lot of effort and impressive progress, for many QFTs, the realizations on quantum computers remain challenging. 
The primary reason is that the lattice Hamiltonian is technically very complicated. 
Therefore, in this paper, we give an alternative approach, avoiding the use of a lattice. 
In order to explain the basic idea, let us recall a famous quote from Feynman ---
{\it if you want to make a simulation of nature, you'd better make it quantum mechanical.}
An important idea implicit in this quote is that nature itself is a gigantic quantum computer, and systems that have natural physical realization can be simulated more easily. 
The lattice regularization is rather artificial.
It may look natural to us humans, but it is safe to assume that nature is much smarter than us. 
Therefore, we should look for {\it physical} realization of QFTs in simpler quantum mechanical systems. 
In string theory, the system of D0-branes and open strings can be described by quantum mechanics. 
D0-branes and open strings can have rich dynamics, and certain bound states in this system are equivalent to supersymmetric QFTs. 
By using such property, we can give {\it physical realizations} of those supersymmetric QFTs in a world described by certain quantum mechanics. 
As we can easily imagine, such physical realizations can be put on the quantum computer much more easily; 
essentially, we only have to put the quantum mechanics of D0-branes and open strings on a quantum computer.  
After that point, we only have to mimic what string/M-theory does. 
In some sense, it is similar to the Hamiltonian engineering for the analog quantum simulation; 
we realize a particular bound state of D0-branes and open strings, which is equivalent to the QFT we want to simulate, 
in a world described by the matrix model. 

This paper is organized as follows. In the rest of this section, we give several important remarks regarding the lattice regularization, 
in order to motivate the use of an alternative method introduced in this paper. 
In Section~\ref{sec:BMN} and Section~\ref{sec:QFT-from-BMN}, 
we show how the matrix model and QFT can be realized in the Hamiltonian formulation. 
We introduce an explicit regularization scheme, such that it can be realized on a digital quantum computer with a large but finite number of qubits. 

In Section~\ref{sec:Quantum-algorithms}, we present an explicit formalism for simulating regularized matrix model Hamiltonian on a quantum computer. We use the Wan-Kim fast adiabatic algorithm~\cite{wan2020fast} to prepare ground state as well as thermofield double states on qubits. We then use block-encoding, qubitization, and quantum signal processing technique~\cite{low2016hamiltonian, low2017optimal} to approximate the unitary real-time evolution. Lastly, we discuss several experiments and measurements that can be performed on this simulation of the matrix model to develop a deeper understanding of superstring/M-theory and holography beyond current analytic results.
 
\subsection{Euclidean lattice: how it (sometimes) works on classical computers}
{}
When a quantum field theory in the Euclidean space is regularized on a lattice,
it is important to keep the symmetries of the theory exactly at the regularized level. 
For example, Wilson's plaquette action \cite{Wilson:1974sk} for the Yang-Mills theory preserves gauge symmetry, 
discrete translation (shift of one lattice unit), discrete rotation (90-degree rotation), parity, and charge conjugation. 
These exact symmetries control the radiative corrections such that the correct continuum limit is realized. 
Unless sufficiently many symmetries are preserved at the regularized level, the right continuum limit will not be obtained, 
because radiative corrections can break necessary symmetries.\footnote{
In principle, one can add various additional terms (counter-terms) to the action in order to compensate for symmetry-breaking radiative corrections. 
However, it is usually a very complicated task and does not work practically. 
} 

One of the well-known cases is the chiral symmetry \cite{Nambu:1961tp}. 
The Nielsen-Ninomiya no-go theorem \cite{Nielsen:1980rz} claims that the chiral symmetry cannot be preserved on the lattice with a few natural assumptions. 
For the vector-like models (i.e.~left-handed and right-handed fermions appear in a pair, e.g.,~QCD), the overlap fermion~\cite{Neuberger:1997fp} and domain-wall fermion~\cite{Kaplan:1992bt} provide the ways to circumvent the Nielsen-Ninomiya theorem. 
However, a generic solution applicable to chiral gauge theories --- including the standard model of particle physics, whose electroweak sector is chiral --- is not known. 
Another well-known case is supersymmetry on the lattice; because supersymmetry algebra contains the infinitesimal translation, 
which is broken on the lattice by definition, it is impossible to keep the entire supersymmetry algebra on the lattice. 
For $(1+1)$- and $(2+1)$-dimensional theories, by keeping a part of supersymmetry, the right continuum limit can be taken \cite{Kaplan:2002wv,Cohen:2003xe,Cohen:2003qw,Kaplan:2005ta,Sugino:2003yb,Sugino:2004qd,Sugino:2004uv,Catterall:2003wd,Catterall:2004np}.
But for $(3+1)$ dimensions, no fine-tuning-free formulation is known.\footnote{
Strictly speaking, 4d ${\cal N}=1$ pure super Yang-Mills theory is an exception: 
by forbidding the gaugino mass by using the chiral symmetry, the supersymmetric continuum limit is guaranteed \cite{Kaplan:1983sk,Curci:1986sm}. 
This approach does not work for supersymmetric QCD or theories with extended supersymmetry, because the chiral symmetry cannot forbid scalar mass terms. 
} 

The situation changes drastically for quantum mechanics, due to the lack of the ultraviolet divergence. 
Very often, naive regularizations that do not respect the symmetries lead to the correct continuum limit. 
Refs.~\cite{Hanada:2007ti,Catterall:2007fp} pointed out that this property is useful for the study of the supersymmetric matrix models. 
Such matrix models are important for quantum gravity via holography, 
and provided nontrivial tests of holographic duality at finite temperature and stringy level, 
which are out of reach with other approaches; see, e.g.~Refs.~\cite{Anagnostopoulos:2007fw,Catterall:2008yz,Hanada:2013rga,Berkowitz:2016jlq,Asano:2016kxo} for original references and Ref.~\cite{Hanada:2016jok} for a review.

\subsection{Minkowski time and Hamiltonian formulation: why it is hard even on a quantum computer}
{}
The situation is similar in the Hamiltonian formulation. Let us consider the lattice Hamiltonian of pure Yang-Mills proposed by Kogut and Susskind \cite{Kogut:1974ag}.  
It preserves various symmetries and hence leads to the desired continuum limit. Here, by continuum limit, we mean the continuum limit along the spatial dimensions; 
by definition, the time direction is continuous in the Hamiltonian formulation. 
On the other hand, for quantum mechanics, there is no need for the continuum limit in this sense, because there is no space by definition. 

When the Hamiltonian formulation is used on the digital quantum computer, yet another limit is needed: 
because we are considering the theories with bosonic degrees of freedom, whose Hilbert space is infinite-dimensional, 
the Hilbert space has to be truncated and expressed by using a finite number of qubits. 
There are two regularization parameters for QFT: the lattice spacing $a$ and the dimension of the Hilbert space $D=2^{n_q}$, where $n_q$ is the number of qubits. 
The two-step limiting procedure is required: 
\begin{itemize}
\item
For fixed lattice spacing $a$, we send $n_q\to\infty$, such that 
the correct lattice Hamiltonian acting on the infinite-dimensional Hilbert space is obtained;  

\item
Then we take the continuum limit, $a\to 0$. 

\end{itemize}
The first step is already nontrivial; for example, how can we express the unitary link variables by using qubits? 
We have to regularize the group manifold, which is doable but rather complicated; see, e.g., Ref.~\cite{Zohar:2014qma}. 
The second step is also highly nontrivial; actually, the situation can be worse than in the case of the Euclidean lattice, 
because now space and time are treated separately, and hence it is harder to keep large enough symmetry. 

For quantum mechanics, the second step is absent. The first step is also simplified because there is no dynamical gauge field. 
Gauge-singlet constraint is imposed on the states; the Hamiltonian only has to have the `global' symmetry. 
In the class of theories we will consider, the dynamical variables are Hermitian matrices, which can be expressed 
as a collection of multiple harmonic oscillators interacting with each other in a certain manner. 
It allows us to use a simple truncation scheme of the Hilbert space. 
\subsection{QFT from Matrix Model: why it can work on quantum computer}\label{sec:QFT-from-MM}
{}
The key idea we use in this paper is to embed the space to matrices, 
{\it following concrete physical processes} in string/M-theory.  
When we use lattice, the technical difficulty was that it is difficult to preserve sufficiently large symmetry at the discretized level. 
Hence we will borrow the idea from `nature,' which is in this case string/M-theory.
Regularizations utilizing actual physical setups in the matrix model, which have the origins in string/M-theory, respect necessary symmetries. 
Roughly speaking, the counterpart of the lattice volume is the matrix size $N$, and the detail of the lattice such as dimension corresponds to the choice of supersymmetric background in the matrix model. 
Hence the counterpart of the lattice-regularized QFT is the finite-$N$ matrix model about a supersymmetric background. 

Historically the first example of this kind of phenomenon is the Eguchi-Kawai equivalence \cite{Eguchi:1982nm}: 
at large $N$, gauge theory living at a point (essentially matrix model) is equivalent to infinite volume theory, if certain conditions are satisfied. 
The `twisted' version of the Eguchi-Kawai equivalence \cite{GonzalezArroyo:1982ub,GonzalezArroyo:1982hz} provides us with a natural way to embed 
the noncommutative space to matrices, which has a counterpart in string theory via the Myers effect \cite{Myers:1999ps}. 
Essentially three classes of quantum field theories can naturally be realized in matrix models:\footnote{
By combining the matrix model approach and lattice regularization, other theories can be realized as well. 
For example, by constructing 2d Euclidean theory by using lattice and then generating two more dimensions by the Myers effect, 
4d ${\cal N}=4$ super Yang-Mills can be obtained without performing the parameter-fine-tuning \cite{Hanada:2010kt,Hanada:2010gs}. 
} 
supersymmetric gauge theories that naturally arise from string theory, gauge theories on noncommutative space, and large-$N$ gauge theory.
The counterpart of the continuum limit on lattice $a\to 0$ is the appropriate large-$N$ limit; the parameter-fine-tuning is not needed. 
The details of the construction of QFT from the matrix model will be explained in Sec.~\ref{sec:QFT-from-BMN}. 
Note that the limit of $n_q\to\infty$ is needed at each fixed $N$, but as we will see, this limit can be taken in a straightforward manner. See Sec.~\ref{sec:BMN} for details.

\subsection{Quantum Gravity in the Lab: matrix model on a quantum computer}
\label{sec:QG_lab}
{}
A substantially large fraction of our motivation for the quantum simulation of the matrix model and supersymmetric QFT lies in quantum gravity. 
{\it Quantum gravity in the lab} \cite{Danshita:2016xbo,Brown:2019hmk} is a line of thinking which uses the holographic duality and experiments in the boundary QFT side to investigate hard problems in the bulk quantum gravity. For example, the quantum teleportation experiment can be used to test the existence of a wormhole \cite{Brown:2019hmk}. Several other experiments have been carried out to test scrambling of quantum information and saturation of quantum chaos in physical models \cite{Landsman2019, PhysRevLett.124.240505}. In addition to physically realizable models, quantum simulation on a digital quantum computer is also a possible avenue for progress. 
Recently, a quantum simulation proposal was made 
for the Sachdev-Ye-Kitaev (SYK) model \cite{Babbush:2018mlj,Garcia-Alvarez:2016wem} and sparse SYK model \cite{Xu:2020shn} that are good toy models for the holographic description of gravity \cite{Maldacena:2016upp}. 

In this paper, we take a similar approach and propose a framework for simulating a matrix quantum mechanics on a universal quantum computer. This theory is dual to superstring/M-theory and in the appropriate limit of parameters describes gravitational physics. Considering this duality, one can study various quantum gravitational phenomena in superstring/M-theory, for example, the presence of traversable wormholes, a saturation of quantum chaos, as well as the sub-AdS locality on this simulation models.

If the QFT side is sufficiently simple, it might be interesting also to address analog quantum simulation in the future. 
One could consider simulate matrix models in cold-atomic physics, for instance, the Rydberg systems. 
Recently, there are some proposal of experimental implementations of some typical chaotic models, for instance Refs.~\cite{Danshita:2016xbo,Garcia-Alvarez:2016wem,Landsman:2018jpm,Brown:2019hmk,Kruchkov:2019idx,Liu:2020sqb}.

\section{Regularizing the BMN matrix model}
\label{sec:BMN}
{}
As we have briefly mentioned, we can realize various supersymmetric quantum field theories in terms of the matrix model.  
For example, we can realize 3d super Yang-Mills theory (SYM), 6d superconformal field theory (SCFT), and 4d SYM by taking suitable vacua in the BMN matrix model \cite{Berenstein:2002jq}.
Hence giving an appropriate regularization scheme of the matrix model is a good starting point toward the realization of such QFTs on the quantum computer.
In this section, we explain how to regularize the BMN matrix model
in a suitable way to realize it on a quantum computer.
In Sec.~\ref{sec:QFT-from-BMN}, we explain how QFT is realized by using the BMN matrix model. 
\subsection{Lagrangian formulation}
\label{sec:BMN-Lagrangian}
{}
Let us start with the Lagrangian formulation of the BMN matrix model \cite{Berenstein:2002jq}. 
All the dynamical variables of the BMN matrix model are $N\times N$ Hermitian matrices.
Although there are various ways\footnote{
1. Mass deformation of the Banks-Fishler-Shenker-Susskind (BFSS) matrix model \cite{Banks:1996vh}
which was introduced as a matrix regularization of supermembrane in 11d \cite{deWit:1988wri}. 
Therefore the massless limit $\mu\rightarrow 0$ in \eqref{BMN-Lagrangian-SO(9)}
becomes the Lagrangian of the BFSS model. 
2. Matrix regularization of supermembrane action in 11d pp-wave background~\cite{Dasgupta:2002hx}.
3. Dimensional reduction of the 4d $\mathcal{N}=4$ super Yang-Mills theory on $\mathbb{R}\times S^3$
along $S^3$~\cite{Kim:2003rza}.
} to construct the action of the BMN matrix model \cite{Berenstein:2002jq,Kim:2003rza},
here we skip their details and simply write the resulting Lagrangian.
The theory is the U$(N)$ gauged supersymmetric matrix quantum mechanics
whose field content can be interpreted as the dimensional reduction of $(9+1)$-dimensional super Yang-Mills theory:\footnote{
The scalars can be interpreted as the dimensional reduction of the spatial component of the gauge field
while the fermions are the dimensional reduction of the Majorana-Weyl spinor (gaugino)
in $(9+1)$-dimensions, and hence, satisfy the Majorana condition.
}
\begin{itemize}
\item The gauge field $A_t$.

\item  9 adjoint scalars $X_I$ with $I=1,\cdots ,9$ .
It is often convenient to decompose them
into 3 scalars $X_i$ with $i=1,2,3$ and 6 scalars $X_a$ with $a=4,5,\cdots ,9$. 

\item 16 adjoint fermions $\Psi_u$ with $u =1,\cdots ,16$.
They are essentially the dimensional reduction of the Majorana-Weyl spinor 
in $(9+1)$-dimensions, which has 16 real degrees of freedom.
We describe it as 16 complex fermions with a kind of reality condition 
called Majorana condition as explained later.  
Below we often write it as $\Psi =(\Psi_1 ,\cdots ,\Psi_{16} )^T$ and do not explicitly write the indices.
\end{itemize}
Following the notation of \cite{Dasgupta:2002hx},
the Lagrangian is given by\footnote{
A parameter $R$ in Ref.~\cite{Dasgupta:2002hx}
is related to our coupling constant by $g^2=R^{-3}$. 
} 
\begin{eqnarray}
L
&=&
\frac{1}{g^2}
{\rm Tr}\Biggl\{
\frac{1}{2}(D_tX_I)^2  + \frac{1}{4}[X_I,X_J]^2  -\frac{\mu^2}{18}X_i^2 
-\frac{\mu^2}{72}X_a^2 
-\frac{i\mu}{6}\epsilon^{ijk}X_iX_jX_k \nonumber\\
& & \quad\qquad
+\frac{i}{2}\Psi^\dagger D_t\Psi
-\frac{1}{2}\Psi^\dagger \gamma_I[X_I,\Psi ]
-\frac{i\mu}{8}\Psi^\dagger \gamma_{123}\Psi
\Biggl\},   
\label{BMN-Lagrangian-SO(9)}
\end{eqnarray}
where $D_t$ is a covariant derivative for the adjoint representation defined by
\begin{\eq}
D_t (\cdot )=\partial_t (\cdot ) -ig [A_t, (\cdot )] .
\end{\eq}
The symbol $\epsilon_{ijk}$ is the structure constant of $\text{SU}(2)$, namely 
$\epsilon_{123}=\epsilon_{231}=\epsilon_{312}=+1$, 
$\epsilon_{132}=\epsilon_{321}=\epsilon_{213}=-1$. 
The matrices $\gamma^I$ are $16\times 16$ 
real symmetric and traceless matrices satisfying
\begin{\eq}
\{\gamma^I,\gamma^J\}=2\delta^{IJ} ,
\label{eq:gamma}
\end{\eq}
which originally come from gamma matrices in higher dimensions\footnote{
In the $(9+1)$-dimensional perspective,
the fermion originally had 32 complex components and we had ten $32\times 32$ gamma matrices.
The fermion is subject to Majorana and Weyl conditions,
and finally has 16 real degrees of freedom.
The Majorana condition for 32-component spinors is 
$\bar{\Psi}\equiv\Psi^\dagger\Gamma^0=\Psi^T C $,
where $C$ is charge conjugation matrix.
The Weyl condition is the projection of the left-handed spinor.
In the Lagrangian~\eqref{BMN-Lagrangian-SO(9)},
the Weyl condition is already used
while imposing the Majorana condition will be discussed later.
}.

There are various representations of the gamma matrices satisfying \eqref{eq:gamma}
but physical observables are independent of how to represent it.
Therefore we can choose convenient representations depending on the problems under consideration.

In this paper, we take the following representation:
\begin{eqnarray}
\gamma^i
= \left( \begin{array}{cc}
-\sigma^i\otimes\textbf{1}_4 & \mathbf{0}_8 \\ 
\mathbf{0}_8 & \sigma^i\otimes\textbf{1}_4
\end{array} \right),  \qquad
\gamma^a
=\left( \begin{array}{cc}
\mathbf{0}_8 & \textbf{1}_2\otimes\textsl{g}^{a}\\ 
\textbf{1}_2\otimes\textsl{g}^{a\dagger} & \mathbf{0}_8
\end{array} \right),   
\end{eqnarray}
where $\sigma^i$ 
are usual Pauli matrices, 
and $\textsl{g}^{a}$ 
are $4\times 4$ matrices satisfying 
$\textsl{g}^{a}\textsl{g}^{b\dagger}+\textsl{g}^{b}\textsl{g}^{a\dagger}=2\delta^{ab}$.
As in the gamma matrices,
there are various representations of $\textsl{g}^{a}$
but below, we do not use explicit forms\footnote{
An explicit example  is
\begin{eqnarray*}
& &
\textsl{g}^{4}
=\sigma_1\otimes\sigma_2,  \qquad
\textsl{g}^{5}
=\sigma_2\otimes\textbf{1}_2, \qquad
\textsl{g}^{6}
=\sigma_3\otimes\sigma_2, \nonumber\\
&&\textsl{g}^{7}
= i\sigma_2\otimes\sigma_3, \qquad
\textsl{g}^{8}
=i\textbf{1}_2\otimes\sigma_2,  \qquad
\textsl{g}^{9}
=i\sigma_2\otimes\sigma_1,
\end{eqnarray*}
} of $\textsl{g}^{a}$. 
An advantage of this choice, particular for our purpose, is that
the matrix $\gamma_{123}$ becomes diagonal
and so does the fermion mass term $\Psi^\dagger \gamma_{123}\Psi$. 
In our choice of the gamma matrices, the Majorana condition is written as
$\Psi^\dagger = \Psi^{\rm T}K $,
where
\begin{eqnarray}
K=
\left(\begin{array}{cc}
\mathbf{0}_8 & -i\sigma_2\otimes\textbf{1}_4\\
 i\sigma_2\otimes\textbf{1}_4 & \mathbf{0}_8
\end{array}\right).  
\end{eqnarray}
The Majorana condition is solved as
\begin{eqnarray}
\Psi
=
\left(
\begin{array}{c}
\psi_{Ip}\\
\epsilon_{pq}\psi^{\dagger Iq}
\end{array}
\right), 
\qquad
\epsilon_{pq}=( i\sigma_2 )_{pq} ,
\end{eqnarray}
where $I=1,2,3,4$ and $p=1,2$.
Thus,
the Lagrangian in our choice of the gamma matrices is~\cite{Dasgupta:2002hx}
\begin{align}
&L=
{\rm Tr}\Biggl\{
\frac{1}{2}(D_tX_I)^2 
+
\frac{g^2}{4}[X_I,X_J]^2 
-
\frac{\mu^2}{18}X_i^2 
-
\frac{\mu^2}{72}X_a^2 
-
\frac{i\mu g}{3}\epsilon^{ijk}X_iX_jX_k
\nonumber\\
&\quad\qquad
+
i\psi^{\dagger Ip} D_t\psi_{Ip}
-
g\psi^{\dagger Ip}\sigma^i_p{}^q[X_i,\psi_{Iq}]
\nonumber\\
&\quad\qquad
+
\frac{g}{2}\epsilon_{pq}\psi^{\dagger Ip}\textsl{g}^{a}_{IJ}[X_a,\psi^{\dagger Jq}]
-
\frac{g}{2}\epsilon^{pq}\psi_{Ip}(\textsl{g}^{a\dagger})^{IJ}[X_a,\psi_{Jq}]
-
\frac{\mu}{4}\psi^{\dagger Ip}\psi_{Ip}
\Biggl\}. 
\end{align}

The BMN matrix model has various symmetries:
\begin{itemize}
\item Translation symmetry along the time $t$.

\item U$(N)$ gauge symmetry (redundancy): 
\begin{eqnarray}
X_I\to\Omega X_I\Omega^{-1}, 
\qquad
\Psi\to\Omega \Psi\Omega^{-1}, 
\qquad
A_t
\to i\Omega\partial_t\Omega^{-1}  +\Omega A_t\Omega^{-1} ,
\end{eqnarray}
where $\Omega(t)$ is a $t$-dependent $N\times N$ unitary matrix\footnote{
Note that $D_t X_I$ also transforms as $\Omega(D_t X_I)\Omega^{-1}$, 
and this fact makes the Lagrangian $L$ invariant under the gauge transformation. 
}. 

\item Supersymmetry:
\begin{eqnarray}
\delta X_I 
&=&  \Psi^\dag  \gamma_I \epsilon  , \NN\\
\delta A_t
&=& \Psi^\dag \epsilon , \nonumber\\
\delta\Psi
&=&
\left[
(D_t X_I)\gamma_I 
+\frac{\mu}{3}X_i \gamma^i\gamma_{123}  -\frac{\mu}{6}X_a \gamma^a\gamma_{123}
+\frac{ig}{2}[X_I,X_J] \gamma^{IJ}   \right] \epsilon .\,
\label{eq:SUSYtrans}
\end{eqnarray}
where $\epsilon$ is the 16 component Killing spinor satisfying
\begin{eqnarray}
\epsilon(t)=e^{-\frac{\mu }{12}\gamma_{123} t}\epsilon_0 ,\quad
\partial_t \epsilon_0 =0.
\end{eqnarray}

\item $\text{SO}(3)$ global symmetry rotating the 3 scalars $X_i$.

\item $\text{SO}(6)$ global symmetry rotating the 6 scalars $X_a$.

\end{itemize}
Note also that the dimension of the parameters and fields are
\begin{\eq}
[\mu ]=[{\rm mass}]^1 ,\quad [g] =[{\rm mass}]^{\frac{3}{2}} ,\quad  
[X_I ] =[{\rm mass}]^{-\frac{1}{2}} ,\quad  [A_t ] =[{\rm mass}]^{-\frac{1}{2}} ,\quad
[\Psi  ] =[{\rm mass}]^0 .
\end{\eq}

\subsection{Hamiltonian formulation
}
\label{sec:BMN-Hamiltonian}
{}
Let us switch to the Hamiltonian formalism.
We expand the matrices as
\begin{equation}
X_I = \sum_{\alpha =1}^{N^2} X_I^\alpha \tau_\alpha , \quad
\psi_{Ip} = \sum_{\alpha =1}^{N^2} \psi_{Ip}^\alpha \tau_\alpha ,
\end{equation}
where  $\tau_\alpha$ is the generator of the U($N$) gauge group satisfying
\begin{equation}
[\tau_\alpha ,\tau_\beta ] = if_{\alpha\beta\gamma} \tau_\gamma ,\quad
{\rm Tr}(\tau_\alpha\tau_\beta)=\delta_{\alpha\beta}.  
\label{eq:generator}
\end{equation}
We take the temporal gauge $A_t =0$. 
The canonical conjugate momentum of $X_I^\alpha$ in this gauge is simply
$P_I^\alpha = \partial_t X_I^\alpha $,
while the conjugate of $\psi$ is $i\psi^\dag$. Note that the conjugate momentum of the gauge field $A_t$ is zero
since the Lagrangian does not contain $\partial_tA_t$.

In the operator formalism, $P_I$, $X_I$, and $\psi$ are promoted to the operators  
with the canonical (anti-)commutation relations:
\begin{eqnarray}
[\hat{X}_{I\alpha},\hat{P}_{J\beta}] =i\delta_{IJ}\delta_{\alpha\beta} ,\quad
\{  \hat{\psi}^{\dag Ip \alpha}  ,\hat{\psi}_{Jq}^\beta \} 
= \delta_{IJ} \delta^{pq} \delta^{\alpha\beta} .
\end{eqnarray}
The Hamiltonian is given by
\begin{eqnarray}
\hat{H}
&=&{\rm Tr}\Biggl\{
\frac{1}{2}(\hat{P}_I)^2 
-\frac{g^2}{4}[\hat{X}_I,\hat{X}_J]^2 
+\frac{\mu^2}{18}\hat{X}_i^2 
+\frac{\mu^2}{72}\hat{X}_a^2 
+\frac{i\mu g}{3}\epsilon^{ijk}\hat{X}_i\hat{X}_j\hat{X}_k \nonumber\\
&&\qquad
+g\hat{\psi}^{\dagger Ip}\sigma^i_p{}^q[\hat{X}_i,\hat{\psi}_{Iq}]
-\frac{g}{2}\epsilon_{pq}\hat{\psi}^{\dagger Ip}\textsl{g}^{a}_{IJ}[\hat{X}_a,\hat{\psi}^{\dagger Jq}]
+\frac{g}{2}\epsilon^{pq}\hat{\psi}_{Ip}(\textsl{g}^{a\dagger})^{IJ}[\hat{X}_a,\hat{\psi}_{Jq}]+\frac{\mu}{4}\hat{\psi}^{\dagger Ip}\hat{\psi}_{Ip} \Biggl\}.   \NN\\
\label{BMN-Hamiltonian-canonical-normalization}
\end{eqnarray}
Our gauge choice $A_t=0$ leads to the Gauss-law constraint upon acting on physical states
\begin{equation}
\hat{G}_\alpha  |{\rm phys} \rangle =0 \quad
{\rm with}\quad 
\hat{G}_\alpha 
\equiv  \sum_{\beta ,\gamma =1}^{N^2} f_{\alpha\beta\gamma}
\left(  \sum_{I=1}^9\hat{X}_I^\beta \hat{P}_I^\gamma 
-i\sum_{I,p} \hat{\psi}^{\dagger Ip \beta} \hat{\psi}_{Ip}^\gamma 
\right) , 
\label{Gauss-law-Lagrangian-BMN}  
\end{equation}
which is equivalent to the equation of motion of $A_t$.
The operator $\hat{G}_\alpha$ is the conserved charge of the U$(N)$ gauge transformation. 
Hence the Gauss-law constraint \eqref{Gauss-law-Lagrangian-BMN} means that the physical states are gauge singlets. 

The supercharge $\hat{Q}$ is given by~\cite{Dasgupta:2002hx}
\begin{eqnarray}
\hat{Q}_{Ip}
&=&{\rm Tr}\Biggl\{\left(
\hat{P}_a-\frac{i\mu}{6}\hat{X}_a
\right)
\textsl{g}^{a}_{IJ}\epsilon_{pq}\hat{\psi}^{\dagger Jq}
-\left( \hat{P}_i+\frac{i\mu}{3}\hat{X}_i \right)
\sigma^i_p{}^q\hat{\psi}_{Iq} \nonumber\\
&&\qquad
+\frac{g}{2}[\hat{X}_i,\hat{X}_j]\epsilon^{ijk}\sigma^k_p{}^q\hat{\psi}_{Iq}
-\frac{ig}{2}[\hat{X}_a,\hat{X}_b] (\textsl{g}^{ab})_I{}^J\hat{\psi}_{Jp}
+ig[\hat{X}_i,\hat{X}_a](\sigma^i\epsilon)_{pq}\textsl{g}^a_{IJ} \hat{\psi}^{\dagger Jq}
\biggl\}, \NN\\
\end{eqnarray}
which satisfy the algebra
\begin{\eq}
\{ \hat{Q}^{\dag Ip} ,\hat{Q}_{Jq} \}
= 2\delta^I_J \delta^p_q \hat{H} 
-\frac{\mu}{3}\epsilon^{ijk}\sigma_q^{kp} \delta^I_J \hat{M}^{ij}
-\frac{i\mu}{6} \delta^p_q (\textsl{g}^{ab})_J^{\ I} \hat{M}^{ab} ,
\label{eq:supercharge_anticommutator}
\end{\eq}
where  $\hat{M}^{ij}$ and $\hat{M}^{ab}$ 
are the generators of $\text{SO}(3)$ and $\text{SO}(6)$ global symmetries defined by
\begin{\eqa}
\hat{M}^{ij}
&=& {\rm Tr} \left[ \hat{X}^i \hat{P}^j -\hat{X}^j \hat{P}^j 
 +i\epsilon^{ijk} \hat{\psi}^\dag \sigma^k \hat{\psi} \right] ,\NN\\
\hat{M}^{ab}
&=& {\rm Tr} \left[ \hat{X}^a \hat{P}^b -\hat{X}^b \hat{P}^a 
 +\frac{1}{2} \hat{\psi}^\dag \textsl{g}^{ab} \hat{\psi} \right] ,  
\end{\eqa}
respectively.
It is important to note that the supercharge is gauge-invariant
\begin{\eq}
[\hat{Q}_{Ip} ,\hat{G}_\alpha ] =0 ,
\end{\eq}
and therefore supersymmetric condition is gauge-invariant.

In the anticommutation relation \eqref{eq:supercharge_anticommutator}, we omitted a term proportional to the gauge generator $\hat{G}$. Such a term vanishes when acting on the gauge-singlet sector, but it can have a nontrivial consequence for the non-singlet sector. 
The Hamiltonian can be expressed without using the gauge generator as $\hat{H}=\frac{1}{32}\{\hat{Q}^{\dag Ip} ,\hat{Q}_{Ip}\}$, and hence, the positive semi-definiteness holds in the non-singlet sector as well~\cite{Maldacena:2018vsr}. 

\subsection{Regularization of the Hilbert space}
\label{sec:regularizatuon-Hilbert}
{}
The total Hilbert space of the BMN matrix model can be decomposed as
\begin{\eq}
\mathcal{H}_{\rm BMN}
= \mathcal{H}_X \otimes \mathcal{H}_\Psi ,
\end{\eq}
where $\mathcal{H}_X$ and $\mathcal{H}_\Psi$ are
subspaces associated with the scalars $X_I$ and fermions $\Psi_u$, respectively.
The dimension of $\mathcal{H}_X$ is infinite because $X_I$ are bosonic,  
while $\mathcal{H}_\Psi$ is finite-dimensional 
since it is associated with a finite number of fermions.
Therefore we need to regularize $\mathcal{H}_X$ in a certain way
while we do not need a regularization for $\mathcal{H}_\Psi$.
In this subsection
we first regularize the subspace $\mathcal{H}_X$
and then discuss the properties of the total Hilbert space after the regularization.
\subsubsection{Fock basis of the bosonic part}
{}
First let us decompose the full Hamiltonian 
into the purely bosonic part and the other part:
\begin{eqnarray}
\hat{H} &=& \hat{H}_X +\hat{H}_\Psi , \NN\\
\hat{H}_X
&=& {\rm Tr}\Bigl\{
\frac{1}{2} \hat{P}_I^2 +\frac{\mu^2}{18}\hat{X}_i^2  +\frac{\mu^2}{72}\hat{X}_a^2 
 - \frac{g^2}{4}[\hat{X}_I,\hat{X}_J]^2  +\frac{i\mu g}{3}\epsilon^{ijk}\hat{X}_i\hat{X}_j\hat{X}_k \Bigl\},    \NN\\
\hat{H}_\Psi
&=&{\rm Tr}\Biggl\{
\frac{\mu}{4}\hat{\psi}^{\dagger Ip}\hat{\psi}_{Ip}
+g\hat{\psi}^{\dagger Ip}\sigma^i_p{}^q[\hat{X}_i,\hat{\psi}_{Iq}]
-\frac{g}{2}\epsilon_{pq}\hat{\psi}^{\dagger Ip}\textsl{g}^{a}_{IJ}[\hat{X}_a,\hat{\psi}^{\dagger Jq}]
+\frac{g}{2}\epsilon^{pq}\hat{\psi}_{Ip}(\textsl{g}^{a\dagger})^{IJ}[\hat{X}_a,\hat{\psi}_{Jq}] \Biggl\}.   \NN\\
\end{eqnarray}
Here we focus on the purely bosonic part $\hat{H}_X$ and further decompose it into free and interacting parts:
\begin{eqnarray}
\hat{H}_X &=& \hat{H}_X^{\rm free} +\hat{H}_X^{\rm int} , \NN\\
\hat{H}_X^{\rm free}
&=& \sum_{I=1}^9 \sum_{\alpha =1}^{N^2}
\left(\frac{1}{2}\hat{P}_{I\alpha}^2 +\frac{\omega_I^2}{2} \hat{X}_{I\alpha}^2 \right) ,  \nonumber \\
\hat{H}_X^{\rm int}
&=&{\rm Tr}\Bigl\{
 - \frac{g^2}{4}[\hat{X}_I,\hat{X}_J]^2  +\frac{i\mu g}{3}\epsilon^{ijk}\hat{X}_i\hat{X}_j\hat{X}_k \Bigl\} .
 \label{eq:BMN-Hamiltonian-free-and-int}
\end{eqnarray}
where
\begin{align}
\omega_I =
\begin{cases}
\frac{\mu}{3} & {\rm for}\  I=1,2,3\cr
\frac{\mu}{6} & {\rm for}\  I=4,5,\cdots,9
\end{cases} 
\end{align}

The free part is essentially $9N^2$ harmonic oscillators described in the Fock basis, 
corresponding to $\alpha=1,2,\cdots,N^2$ and $I=1,2,\cdots,9$. 
The annihilation and creation operators are defined as
\begin{eqnarray}
\hat{A}_{I\alpha}
=
\sqrt{\frac{\omega_I}{2}}
\hat{X}_{I\alpha}
+
\frac{i\hat{P}_{I\alpha}}{\sqrt{2\omega_I}}~,
\qquad
\hat{A}_{I\alpha}^\dagger
=
\sqrt{\frac{\omega_I}{2}}
\hat{X}_{I\alpha}
-
\frac{i\hat{P}_{I\alpha}}{\sqrt{2\omega_I}}, 
\end{eqnarray}
satisfying
\begin{equation}
[\hat{A}_{I\alpha} , \hat{A}_{J\beta}^\dag ] = \delta_{IJ} \delta_{\alpha\beta}.
\label{eq:com_a}
\end{equation}
We can use $n_{I\alpha}$ to specify the excitation level.
Namely, 
\begin{eqnarray}
|\{n_{I\alpha}\}\rangle
\equiv
\otimes_{I,\alpha}
|n_{I\alpha}\rangle_{I\alpha}
=
\left(
\prod_{I,\alpha}
\frac{\hat{A}_{I\alpha}^{\dagger n_{I\alpha}}}{\sqrt{n_{I\alpha}!}}
\right)
|{\rm VAC_{free}}\rangle, 
\end{eqnarray} 
where $|{\rm VAC_{free}}\rangle=\otimes_{I,\alpha}|0\rangle_{{I\alpha}}$ is the Fock vacuum which is annihilated by any annihilation operator $\hat{A}_{I\alpha}$:
\begin{eqnarray}
\hat{A}_{I\alpha}|{\rm VAC_{free} }\rangle=0. 
\end{eqnarray}
Note that 
\begin{eqnarray}
\hat{A}_{I\alpha}
=\sum_{j=0}^\infty \sqrt{j+1}\ket{j}_{I\alpha}\bra{j+1}_{I\alpha}, 
\qquad
\hat{A}^\dagger_{I\alpha}
=\sum_{j=0}^\infty \sqrt{j+1}\ket{j+1}_{I\alpha}\bra{j}_{I\alpha}. 
\label{eq:A-Adag-before-cutoff}
\end{eqnarray}
The free part of the Hamiltonian is essentially the number operator:
\begin{eqnarray}
\hat{H}_X^{\rm free}
=
\sum_{I=1}^9  \sum_{\alpha =1}^{N^2}
\left(
\hat{n}_{I\alpha}
+
\frac{1}{2}
\right)
\omega_I,
\label{eq:freeH}
\end{eqnarray}
where 
\begin{equation}
\hat{n}_{I\alpha}=\hat{A}^\dagger_{I\alpha}\hat{A}_{I\alpha} ,\quad
{\rm and}\quad 
\hat{n}_{I\alpha}|n_{I\alpha}\rangle_{I\alpha}=n_{I\alpha}|n_{I\alpha}\rangle_{I\alpha} .
\end{equation} 
\subsubsection{Truncating the bosonic Fock space}
{}
As a regularization, we simply truncate the Fock space. 
Namely, for all $I$ and $\alpha$, we restrict $n_{I\alpha}$ to be in\footnote{
In principle, we could introduce different cutoffs for different oscillators. 
It would be useful to introduce different cutoffs for $I=1,2,3$ and $I=4,\cdots,9$ keeping the SO($3$)$\times$SO(6) symmetry
(more precisely, discrete rotations which can survive after the truncation). 
} 
\begin{eqnarray}
0\le n_{I\alpha} \leq \Lambda -1,
\label{eq:truncation}
\end{eqnarray}
so that we have only $\Lambda$ states for the Fock space of each harmonic oscillator.
The explicit truncated forms of $\hat{A}$ and $\hat{A}^\dagger$ are
\begin{eqnarray}
\left.\hat{A}_{I\alpha}\right|_{\rm regularized}
&=&
\sum_{j=0}^{\Lambda-2}
\sqrt{j+1}\ket{j}_{I\alpha}\bra{j+1}_{I\alpha}, 
\nonumber\\
\left.\hat{A}^\dagger_{I\alpha}\right|_{\rm regularized}
&=&
\sum_{j=0}^{\Lambda-2}
\sqrt{j+1}\ket{j+1}_{I\alpha}\bra{j}_{I\alpha}. 
\end{eqnarray}
Compared to \eqref{eq:A-Adag-before-cutoff}, our regularization is not appropriate when we study problems where the highest excited state is important.
In other words, the problem is irrelevant
when we are interested in problems where only low energy states are important.

Correspondingly, when restricted to the truncated Hilbert space, 
the commutation relation \eqref{eq:com_a} is modified near the cutoff. 
The position and momentum operators after the truncation are defined as
\begin{\eqa}
\left.\hat{X}_{I\alpha}\right|_{\rm regularized}
&=&
\left. \frac{1}{\sqrt{2\omega_I}} \left(  \hat{A}_{I\alpha} +\hat{A}_{I\alpha}^\dag \right) \right|_{\rm regularized} ,
\nonumber\\
\left.\hat{P}_{I\alpha}\right|_{\rm regularized}
&=&
\left. \frac{1}{i} \sqrt{\frac{\omega_I}{2} } \left(  \hat{A}_{I\alpha} -\hat{A}_{I\alpha}^\dag \right) \right|_{\rm regularized} .
\end{\eqa}
\subsubsection{The full Hilbert space after the truncation}
{}
After the regularization,
the Hilbert subspace coming from $X_I$ has the dimension
\begin{\eq}
\left. {\rm dim }\left( \mathcal{H}_X \right) \right|_{\rm regularized}  =\Lambda^{9N^2} .
\end{\eq}
Regarding the other part $\mathcal{H}_\Psi$,
the fact that the fermion $\Psi$ has $16N^2$ real degrees of freedom,
leads us to
\begin{\eq}
{\rm dim }\left( \mathcal{H}_\Psi \right)   = 2^{8N^2} .
\end{\eq}
Thus, the dimension of the full Hilbert space after the regularization is
\begin{eqnarray}
\left. {\rm dim} \left( {\cal H}_{\rm BMN} \right) \right|_{\rm regularized}
= \Lambda^{9N^2}\cdot 2^{8N^2}.
\end{eqnarray} 

It is convenient to use the Fock basis regarding the fermion as well. We use the same notation $|{\rm VAC}_{\rm free}\rangle$ as before to denote the Fock vacuum both for the bosons and fermions, i.e., 
\begin{eqnarray}
\hat{A}_I^\alpha|{\rm VAC}_{\rm free}\rangle
=
\hat{\psi}_{Ip}^\alpha|{\rm VAC}_{\rm free}\rangle
=
0. 
\end{eqnarray}

The minimum number of qubits needed for the regularization is $9N^2\log_2\Lambda+8N^2$. 
(Note that this is the number of logical qubits with a proper error correction.)
In sec.~\ref{sec:realization},
we will discuss how to realize the regularized theory in terms of qubits.
\subsection{Sparseness of the Hamiltonian}
\label{sec:sparseness}
{}
In this section, we show that the matrix model Hamiltonian is very sparse for $N,\Lambda \gg 1$
as long as potential is polynomial. 
In Sec.~\ref{sec:Quantum-algorithms}, we will see that such sparse Hamiltonian can be simulated efficiently. 
\subsubsection*{Purely bosonic part}
After the truncation \eqref{eq:truncation},
our Hamiltonian is 
$\Lambda^{9N^2} \times \Lambda^{9N^2}$ matrix and therefore has $\Lambda^{18N^2} $ elements.
We are interested in how many elements are nonzero among them.
We first discuss a rough estimate.
For polynomial potential,\footnote{
More precisely, we take the degree of the potential finite as $N\rightarrow \infty$.
}
the Hamiltonian consists of a finite number of products of $X_{I\alpha}$ and $P_{I\alpha}$
whose number is polynomial in $N$.
Noting that 
the operators $\hat{X}_{I\alpha}=\frac{\hat{A}_{I\alpha}+\hat{A}^\dagger_{I\alpha}}{\sqrt{2\omega_I}}$ and $\hat{P}_{I\alpha}=-i\sqrt{\frac{\omega_I}{2}}(\hat{A}_{I\alpha}-\hat{A}^\dagger_{I\alpha})$
change $n_{I\alpha}$ by $\pm 1$, 
a finite product of $X_{I\alpha}$ and $P_{I\alpha}$ has only $O(1)$ nonzero matrix elements in each row and column.
Therefore, as long as we consider the polynomial potential, 
the number of nonzero matrix elements of the Hamiltonian
is a polynomial of $N$ among $\Lambda^{18N^2} $ elements.
So the Hamiltonian in the Fock basis
is very sparse.

Now we make the estimate more precise.
Let us start with the free part $\hat{H}_{\rm free}$.
From the expression \eqref{eq:freeH},
it is obvious that the free Hamiltonian has only $9N^2$ nonzero elements.
Therefore the free Hamiltonian is very sparse.
Next, let us consider the interacting part.
For our purpose, only the terms with the highest degree in the potential are relevant.
Both for the BFSS and BMN matrix models,
it is proportional to
\begin{eqnarray}
\sum_{I\neq J} {\rm Tr}[\hat{X}_I,\hat{X}_J]^2
= -\sum_{I\neq J } \sum_{\alpha ,\beta ,\gamma ,\rho, \sigma =1}^{N^2}
f_{\alpha\beta\sigma}f_{\gamma\rho\sigma}\hat{X}_I^\alpha\hat{X}_J^\beta\hat{X}_I^\gamma\hat{X}_J^\rho. 
\end{eqnarray}
We can easily see that
each term has only 16 nonzero matrix elements at most in each row and column.
So the problem is reduced to the counting of the number of the terms.
To do this,
a little bit of group theory is needed. 
We normalize the generators of U($N$) as ${\rm Tr}\left(\tau_\alpha\tau_\beta\right) =\delta_{\alpha\beta}$,
Here $\alpha$ and $\beta$ runs from 1 to $N^2$. 

Instead of $\alpha$, we can use $p,q=1,2,\cdots,N$ to label the generators, as 
$\tau_{p,q}^{ij}\equiv \tau_\alpha^{ij}$ for $\alpha=(p-1)N+q$.
A convenient choice is\footnote{
The results of this subsection are independent of the choice of the generators
since it comes only from properties of the structure constants. 
} 
\begin{eqnarray}
\tau_{p,q}^{ij}
=
\frac{1}{2}\left( \delta_{pi}\delta_{qj} +\delta_{pj}\delta_{qi} \right)
+\frac{i}{2}\left( \delta_{pi}\delta_{qj} -\delta_{pj}\delta_{qi} \right). 
\end{eqnarray} 
They satisfy ${\rm Tr}(\tau_\alpha\tau_\beta)=\delta_{\alpha\beta}$ and $\sum_\alpha\tau_\alpha^{ij}\tau_\alpha^{kl}=\delta_{il}\delta_{jk}$. 
The structure constant $f_{\alpha\beta\gamma}$ is defined by 
$[\tau_\alpha,\tau_\beta]=if_{\alpha\beta\gamma}\tau_\gamma$.
Written explicitly,\footnote{
To derive the values of $f_{\alpha\beta\gamma}$, it is convenient to use $N\times N$ matrices $M_{p,q}$ whose $(i,j)$-component is 
$(M_{p,q})^{ij}=\delta_{pi}\delta_{qj}$. 
They satisfy $M_{p,q}M_{r,s}=\delta_{qr}M_{p,s}$, and the U($N$) generators are written as 
$\tau_{p,q}
=\frac{1+i}{2}M_{p,q}+\frac{1-i}{2}M_{q,p} $.
The commutator of two generators can be expressed as 
\[
[\tau_{p,q},\tau_{r,s}]
= \frac{i}{2} \Bigl[
 -\delta_{qs}(\tau_{p,r}-\tau_{r,p}) -\delta_{pr}(\tau_{q,s}-\tau_{s,q})
+\delta_{qr}(\tau_{p,s}+\tau_{s,p}) -\delta_{ps}(\tau_{q,r}+\tau_{r,q}) \Bigr] . 
\]
}
\begin{eqnarray}
f_{pq,rs,tu}
&=& \frac{1}{2} \Bigl[
-\delta_{qs} \left( \delta_{pt}\delta_{ru} - \delta_{pu}\delta_{rt}\right)
-\delta_{pr}\left( \delta_{qt}\delta_{su} -\delta_{qu}\delta_{st} \right) \nonumber\\
&& \quad
+\delta_{qr} \left( \delta_{pt}\delta_{su} +\delta_{pu}\delta_{st}\right)
-\delta_{ps} \left( \delta_{qt}\delta_{ru} +\delta_{qu}\delta_{rt} \right) \Bigr]. 
\label{eq:structure-constant-U(N)}
\end{eqnarray}
From this, we can see that, among $\sim N^6$ possible combinations of $\alpha,\beta$ and $\gamma$, 
only $\sim N^3$ leads to nonzero $f_{\alpha\beta\gamma}$. 

The quartic interaction term is 
\begin{eqnarray}
{\rm Tr}[\hat{X}_I,\hat{X}_J]^2
=
-
f_{\alpha\beta\sigma}f_{\gamma\rho\sigma}\hat{X}_I^\alpha\hat{X}_J^\beta\hat{X}_I^\gamma\hat{X}_J^\rho. 
\end{eqnarray}
To see how many terms exist, we have to see how many combinations of $\alpha,\beta,\gamma$ and $\rho$ give nonzero 
$\sum_\sigma f_{\alpha\beta\sigma}f_{\gamma\rho\sigma}$. It is $\sim N^4$, as one can check by using \eqref{eq:structure-constant-U(N)}.
Another way to understand that there are $\sim N^4$ terms is to look at the definition of the trace, 
${\rm Tr}(\hat{X}\hat{Y}\hat{Z}\hat{W})=\sum_{i,j,k,l=1}^N\hat{X}_{ij}\hat{Y}_{jk}\hat{Z}_{kl}\hat{W}_{li}$;  
obviously, there are $N^4$ terms in the sum. 
Therefore, there are $\sim N^4$ nonzero elements in each row and column. 

\subsubsection*{Interaction between bosons and fermions}
The interaction between fermion and boson leads to only $\sim N^3$ terms at each row and column. 
Hence the leading contribution in the full Hamiltonian comes from ${\rm Tr}[X_I,X_J]^2$, which gives $\sim N^4$ terms at each row and column.

\subsection{Remarks on cutoff dependence}
\label{sec:remove-cutoff}
{}
We have truncated the Hilbert space by introducing the cutoff $\Lambda$ in the harmonic oscillator basis.
Needless to say, we have to take $\Lambda$ sufficiently large in order to achieve a good approximation.
Although the details of the cutoff effect depend on the details of problems under consideration, for reasonable physical setups of interest, our truncation prescription is valid, though it may or may not be optimal. For example, as a small perturbation about the ground states, we can imagine several incoming or outgoing objects (e.q.~graviton or D-brane) described by the eigenvalues of matrices $X_I$. If the eigenvalues take large values, then the free part of the Hamiltonian dominates the energies of those objects. As long as we consider the Hamiltonian time evolution, the energy is conserved, and hence those eigenvalues do not become arbitrarily large, which in turn means the arbitrary high-frequency modes in the Fock basis are not needed. 

As we explain briefly in Appendix~\ref{sec:coordinate-basis}, we could also use the coordinate basis for a regularization. The truncation errors in such a truncation scheme are discussed, e.g., in Refs.~\cite{Jordan:2011ne,Klco:2018zqz}.  
\subsection{Remarks regarding gauge invariance}
\label{sec:gauge-invariance-BMN}
{}
We are using the extended Hilbert space containing non-gauge-invariant states. 
The gauge transformation is generated by $\hat{G}_\alpha$ defined in eq.~\eqref{Gauss-law-Lagrangian-BMN}.   
Without the cutoff $\Lambda$,
these generators commute with the Hamiltonian:
\begin{eqnarray}
[\hat{G}_\alpha ,\hat{H} ] = 0.
\end{eqnarray}
Therefore, if an initial state $|\phi \rangle$ is gauge-invariant, 
the state remains gauge-invariant during the Hamiltonian time evolution
\begin{eqnarray}
\hat{G}_\alpha|\phi \rangle=0
\quad \longrightarrow \quad
\hat{G}_\alpha e^{-i\hat{H}t} |\phi \rangle
=0. 
\end{eqnarray}
One can construct gauge-invariant states as follows.
From \eqref{Gauss-law-Lagrangian-BMN} and $f_{\alpha\beta\gamma}\hat{X}_I^\beta \hat{P}_I^\gamma=-if_{\alpha\beta\gamma}\hat{A}_I^{\dagger\beta} \hat{A}_I^\gamma$, 
we can see that 
the vacuum of the free Hamiltonian is gauge-invariant:
\begin{eqnarray}
\hat{G}_\alpha |{\rm VAC}_{\rm free}\rangle =0 ,
\end{eqnarray}
where
\begin{\eq}
\hat{A}_{I\alpha} |{\rm VAC}_{\rm free}\rangle =\hat{\psi}_{Ip} |{\rm VAC}_{\rm free}\rangle =0.
\end{\eq}
Then any state obtained by acting gauge-invariant operators such as 
${\rm Tr}\left(\hat{A}^\dagger_{I_1}\hat{A}^\dagger_{I_2}\cdots\hat{A}^\dagger_{I_l}\right)$ on the Fock vacuum is gauge-invariant as well. 

Note that the gauge invariance can be broken at a finite cutoff due to the regularization effect. 
Technically this comes from the modification of the canonical commutation relation.
For instance, neither the free Hamiltonian nor the interaction part commute with the gauge charge.
Therefore the amount of breaking depends 
on how the states around the cutoff affect problems under consideration.

The gauge-invariant states span only a small fraction of the Hilbert space, and 
the unphysical, gauge-non-singlet states occupy the majority of the Hilbert space. 
Therefore, one might worry that accumulated simulation errors can spoil the gauge-singlet condition. 
One possible way to protect the gauge-singlet condition is to give a penalty to the non-singlet terms
by adding a term proportional to $\sum_\alpha\hat{G}_\alpha^2$ to the Hamiltonian. 
In the case of the BMN matrix model, essentially the same thing might be happening automatically \cite{Maldacena:2018vsr}, 
as we will explain in Sec.~\ref{sec:BMN-Maldacena-Milekhin}. See also Ref.~\cite{Milekhin:2020zpg} which suggests that the non-singlet errors can be corrected rather easily, at least in the confining phase. 

\subsubsection*{Maldacena-Milekhin proposal}\label{sec:BMN-Maldacena-Milekhin}
{}
Maldacena and Milekhin \cite{Maldacena:2018vsr} 
argued that, at sufficiently large $N$ and strong coupling where classical gravity is a good dual description, the non-singlet modes should be heavy and negligible. 
If it is true, once the initial condition is taken to be a singlet 
and the simulation is precise enough to keep the energy approximately constant, 
we do not have to do anything else to protect the singlet constraint. 

This proposal can be tested on classical computers, by the lattice Monte Carlo simulation. 
Ref.~\cite{Berkowitz:2018qhn} performed lattice simulation of the BFSS matrix model 
($\mu\to 0$ limit of the BMN matrix model) and obtained the numerical results supporting the proposal. 
\section{QFT from the BMN matrix model}\label{sec:QFT-from-BMN}
{}
It is known that the BMN matrix model has
a class of supersymmetric configurations called fuzzy spheres
which preserves all the supersymmetries.
By expanding the theory around fuzzy sphere backgrounds appropriately,
we can obtain various supersymmetric QFTs 
which have origins from superstring/M-theory
and are expected to be dual to some semiclassical gravities 
when the theories have large enough degrees of freedom.
In this section,
we give a brief review of the fuzzy sphere and
discuss how it is realized on quantum computers.
\subsection{Fuzzy spheres in the BMN matrix model}\label{sec:BMN-fuzzy-sphere}
{}
\subsubsection{Fuzzy sphere as BPS solution in the Lagrangian formalism}
{}
Let us use the Lagrangian formulation of the BMN matrix model introduced in Sec.~\ref{sec:BMN-Lagrangian}. 
For trivial configurations for the fields except for $X_i$,
we can write the Lagrangian as
\begin{eqnarray}
\left. L \right|_{A_t =X_a =\Psi =0}
&=& {\rm Tr}\Biggl\{
\frac{1}{2}(\partial_t X_i )^2  +\frac{g^2}{4} \left( [X_i,X_j]  -\frac{i\mu}{3g} \epsilon_{ij k} X_k \right)^2 
\Biggl\} .
\end{eqnarray}
This implies that we have the following class of classical solutions
\begin{eqnarray}
X_i =\frac{\mu}{3g}J_i , \qquad X_a =0, \qquad A_t=0, \qquad \Psi=0,
\label{eq:fuzzy_Lag}
\end{eqnarray}
where $J_i$ is the generator of $\text{SU}(2)$ algebra in $N$-dimensional representation  (not necessarily irreducible),
satisfying
\begin{eqnarray}
[J_i,J_j]= i \epsilon^{ijk}J_k. 
\end{eqnarray}
If we regard $X_i$ as a position,
then the solution describes a space whose coordinates do not commute each other 
but obey the constraint $X_i^2 =( \mu /3g )^2 {\rm Tr}(J_i^2 )$ as in sphere.
Therefore this solution describes a non-commutative version of the sphere
and it is often called the fuzzy sphere solution.
In addition, substituting the soliton \eqref{eq:fuzzy_Lag}
into the supersymmetric transformation \eqref{eq:SUSYtrans},
one can show that the fuzzy sphere solution preserves all the supersymmetries.

Note that the fuzzy sphere solution \eqref{eq:fuzzy_Lag} is not invariant under generic gauge transformation,
unless the representation of $J_i$ is $N$ copies of one dimensional representation of $\text{SU}(2)$ i.e., $J_i =0$
(we call this case ``trivial'').
For the generic representation of $J_i$, the fuzzy sphere solution is invariant
under a subset of $U(N)$ gauge transformation such that $[\zeta ,J_i ] =0$. 

\subsubsection{Fuzzy sphere in the Hilbert space}\label{sec:fuzzy_sphere_in_Hilbert_space}
{}
Let us see how the quantum states corresponding to the fuzzy sphere solutions \eqref{eq:fuzzy_Lag}
of the classical equations of motion, 
which preserves all supersymmetries, can be constructed in the Hamiltonian formalism. 
(A few ways to explicitly construct such states on quantum devices will be explained in  Sec.~\ref{sec:Quantum-algorithms}.)
Denoting such a state by $| J_i \rangle $, 
the state has the following properties:
\begin{itemize}
\item Invariance under supersymmetry
\begin{equation}
\hat{Q}_{ Ip} |J_i \rangle =\hat{Q}^{\dag Ip} |J_i \rangle = 0 .
\label{eq:SUSY_state}
\end{equation}

\item Gauge transformation property
\begin{\eq}
i\zeta^\alpha \hat{G}_\alpha |J_i  \rangle  = |J_i +i[\zeta , J_i ] \rangle  -|J_i  \rangle
\begin{cases}
=0         & {\rm for}\ [\zeta , J_i ] =0 \cr
\neq 0 & {\rm for}\ [\zeta , J_i ]  \neq 0
\end{cases} \quad \left( {\rm up\ to}\ \mathcal{O}(\zeta^2 ) \right) .
\label{eq:gauge_state}
\end{\eq}

\end{itemize}

Although it is hard to construct such states analytically for strong coupling,
we can explicitly construct it for weak coupling as follows. 
First, for the trivial case $J_i =0$, it is simply the free vacuum:
\begin{\eq}
\left. |J_i =0 \rangle \right|_{g\rightarrow 0} = |{\rm VAC}_{\rm free}\rangle.
\end{\eq}
Indeed one can show that 
this state satisfies the SUSY condition \eqref{eq:SUSY_state} up to $\mathcal{O}(g)$.
In addition, this state is gauge-invariant and obviously satisfies \eqref{eq:gauge_state}.

In order to construct generic fuzzy-sphere states, 
it is convenient to 
redefine the operator $\hat{X}_i$ as
as \cite{Dasgupta:2002ru}
\begin{eqnarray}
\hat{X}_i
=\frac{\mu}{3g}J_i + \hat{Y}_i. 
\end{eqnarray}
Equivalently, we can use the translation operator $e^{\frac{i\mu}{3g}\sum_{i,\alpha}\hat{P}_i^\alpha J_i^\alpha}$ as 
\begin{eqnarray}
e^{-\frac{i\mu}{3g}\sum_{i,\alpha}\hat{P}_i^\alpha J_i^\alpha}
\hat{X}_i
e^{\frac{i\mu}{3g}\sum_{i,\alpha}\hat{P}_i^\alpha J_i^\alpha}
=\hat{X}_i -\frac{\mu}{3g}J_i
=\hat{Y}_i. 
\end{eqnarray}
Then the Hamiltonian can be written as\footnote{
 Note that $\hat{H}_{\rm free}^{(J_i )}$ and $\hat{H}_{\rm int}^{(J_i )}$ depends explicitly on the representation of SU(2), and they are different from the `free' and `interaction' parts defined in \eqref{eq:BMN-Hamiltonian-free-and-int}. 
}
\begin{eqnarray}
\hat{H} &=& \hat{H}_{\rm free}^{(J_i )} +\hat{H}_{\rm int}^{(J_i )} ,\NN\\
\hat{H}_{\rm free}^{(J_i )}
&=&{\rm Tr}\Bigl\{
\frac{1}{2} \hat{P}_I^2 +\frac{\mu^2}{36}\left( 2 [J_i,\hat{Y}_j] -i\epsilon_{ijk}\hat{Y}_k \right)^2
+\frac{\mu^2}{18}\hat{X}_a^2 +\frac{\mu^2}{18} [J_i,\hat{X}_a]^2 \nonumber\\
&& \qquad
+\frac{\mu}{4}\hat{\psi}^{\dagger Ip}\hat{\psi}_{Ip}
+\frac{\mu}{3}\hat{\psi}^{\dagger Ip}\sigma_p^{iq} [J_i,\hat{\psi}_{Iq}]
\Bigl\} , \NN\\
\hat{H}_{\rm int}^{(J_i )}
&=&  {\rm Tr}\Bigl\{
-\frac{g\mu}{6} [\hat{Y}_i ,\hat{Y}_j ] \left( 2[J_i ,\hat{Y}_j ] -i\epsilon_{ijk}\hat{Y}_k \right)
-\frac{g\mu}{6} [J_i ,\hat{X}_a ] [\hat{Y}_i ,\hat{X}_a ] 
+g\hat{\psi}^{\dagger Ip}\sigma^i_p{}^q[\hat{Y}_i,\hat{\psi}_{Iq}] \NN\\
&& \quad 
 - \frac{g^2}{4}[\hat{Y}_i, \hat{Y}_j ]^2   - \frac{g^2}{2}[\hat{Y}_i ,\hat{X}_a ]^2 
\Biggl\} .
\end{eqnarray}
Note that the total Hamiltonian $\hat{H}$ is unchanged
and we have just decomposed the Hamiltonian in a different way from the trivial fuzzy sphere case.
The ``free part'' $\hat{H}_{\rm free}^{(J_i )}$ is quadratic in all the operators
and therefore, we can rewrite it as a collection of simple harmonic oscillators
after appropriate diagonalization of the mass terms.\footnote{
To do this, it is most appropriate to expand the operators in terms of fuzzy sphere harmonics
which is a non-commutative version of spherical harmonics (see, e.g., ~\cite{Ishii:2008ib} and references therein).
Here we do not explicitly write their details.
}
Then we can construct the Fock space associated with the harmonic oscillators and
and its Fock vacuum $|J_i ; {\rm VAC}_{\rm free}  \rangle$
is the fuzzy-sphere state in the weak-coupling limit:
\begin{\eq}
\left.  |J_i  \rangle \right|_{g\rightarrow 0} = |J_i ; {\rm VAC}_{\rm free}  \rangle \quad
{\rm with}\quad \hat{H}_{\rm free}^{(J_i )} |J_i ;{\rm VAC}_{\rm free}  \rangle =0 .
\end{\eq}
In  Sec.~\ref{sec:Quantum-algorithms}, we will explain how the ground state of the Hamiltonian including the interaction term can be obtained on quantum device. 

Strictly speaking, the quadratic Hamiltonian $\hat{H}_{\rm free}^{(J_i )}$ has zero modes, which corresponds to the gauge transformation of the fuzzy sphere, or in other words, the Nambu-Goldstone (NG) modes associated with the spontaneous breaking of the U($N$) symmetry due to the choice of a particular fuzzy-sphere configuration. Let us look into this point.\footnote{
When we discuss the realization of $\ket{J_i}$ on the quantum device in  Sec.~\ref{sec:Quantum-algorithms}, we will give a prescription to control this flat direction. 
}

From the state $| J_i  \rangle$, we can obtain a gauge-invariant state simply by a `symmetrization':
\begin{equation}
|J_i  \rangle 
\quad \rightarrow \quad
\mathcal{S} ( |J_i  \rangle )
\equiv \frac{1}{ {\cal N}^{1/2} } \int d\Omega \ | \Omega  J_i \Omega^{-1} \rangle ,
\end{equation}
where
\begin{eqnarray}
| \Omega  J_i \Omega^{-1} \rangle 
= e^{i\zeta^\alpha \hat{G}_\alpha } |J_i \rangle 
\quad {\rm with}\ \Omega =e^{i\zeta} .
\end{eqnarray}
Here the integral is over Haar measure of the $U(N)$ gauge group
and ${\cal N}^{1/2}$ is a normalization constant.
The symmetrization commutes with the Hamiltonian since the Hamiltonian commutes with $\hat{G}_\alpha$. Therefore, $\mathcal{S} ( |J_i  \rangle )$ is also the ground state. 
By construction, this state is gauge-invariant and hence satisfies the Gauss-law constraint. 
This can be seen explicitly by using  the invariance of the Haar measure:
\begin{eqnarray}
 e^{i\tilde{\zeta}^\alpha \hat{G}_\alpha }  \cdot \mathcal{S} ( |J_i  \rangle )
&=& \frac{1}{ {\cal N}^{1/2} } \int d\Omega \ 
| e^{i\tilde{\zeta}}  \Omega  J_i (e^{i\tilde{\zeta}} \Omega )^{-1} \rangle \NN\\
&=& \frac{1}{ {\cal N}^{1/2} } \int d(e^{i\tilde{\zeta}} \Omega ) \ 
| e^{i\tilde{\zeta}}  \Omega  J_i (e^{i\tilde{\zeta}} \Omega )^{-1} \rangle 
=\mathcal{S} ( |J_i  \rangle ) .
\end{eqnarray}
Hence $\mathcal{S} ( |J_i  \rangle )$ is the gauge-invariant ground state.

Now we have two states, $|J_i\rangle$ and ${\cal S}|J_i\rangle$. Apparently, $|J_i\rangle$ is much simpler when the realization on the quantum device is concerned. Hence, if possible, we would want to use $|J_i\rangle$ for the simulation; but does it make sense to use $|J_i\rangle$?
A conjecture by Maldacena and Milekhin~\cite{Maldacena:2018vsr} has an interesting consequence regarding this point. 
According to their conjecture, at sufficiently large $N$ and strong coupling, 
the energy of the non-singlet modes can be bounded from below. 
Below this lower bound, only the singlet modes can contribute to the dynamics, and hence, it does not matter whether we use $|J_i\rangle$ or ${\cal S}|J_i\rangle$. In some interesting cases (including the strongly-coupled QFT limits we consider below in Sec.~\ref{sec:mapping-rule}), this lower bound is rather high, and we can study interesting quantum gravity problems by using $|J_i\rangle$. 

Even if the Maldacena-Milekhin proposal is correct, 
$|J_i\rangle$ and ${\cal S}|J_i\rangle$ can give different results in certain parameter regions of interest. 
Therefore, preferably we should consider the singlet state ${\cal S}(|J_i\rangle)$. 
We will provide possible protocol for preparing the state ${\cal S}(|J_i\rangle)$ in Sec.~\ref{sec:adiabatic_state_preparation}.
We will comment on the Maldacena-Milekhin proposal further in Sec.~\ref{sec:gauge-invariance-QFT}. 

\subsection{Mapping rule}\label{sec:mapping-rule}
{}
We review the correspondence between the fuzzy-sphere states and QFTs. 
\subsubsection{3d maximally supersymmetric Yang-Mills theory (M2-branes and D2-branes)}
\label{sec:mapping-rule-M2}
{}
The 3d maximally supersymmetric Yang-Mills theory (maximal SYM)
describes the worldvolume theory of D2-branes in type IIA superstring \cite{Witten:1995im,Itzhaki:1998dd}.
The fuzzy sphere in the BMN matrix model can be regarded as a D2-brane 
made of D0-branes via the Myers effect \cite{Myers:1999ps}.  
Hence the 3d maximal SYM can be realized by taking (an appropriate) fuzzy-sphere configuration. 
At low energy, the same theory 
is expected to describe the M2-branes. 

The precise construction is as follows \cite{Maldacena:2002rb}. 
Let us consider $N_2$ fuzzy spheres
with spin $s$.
The size of each fuzzy sphere is $N_5\equiv 2s+1$,
and the size of the gauge group of the matrix model is $N=N_2N_5$. 
Hence the SU(2)-generators in the fuzzy sphere state take the following form:
\begin{eqnarray}
J_i
=
J_i^{(s)}\otimes\textbf{1}_{N_2}, 
\qquad
s=\frac{N_5-1}{2}. 
\label{eq:fuzzy3d}
\end{eqnarray}
Here $J_i^{(s)}$ are the generators in the spin-$s$ representation. 
This configuration can be interpreted as the $N_2$-coincident spherical D2-branes. 
If we focus on the fluctuation about this configuration, 
then we can obtain the 3d $U(N_2 )$ maximal SYM on fuzzy sphere\footnote{
One can also realize the maximal 3d SYM in monopole background by choosing the representation of $J_i$ 
in a different way but here we have turned off the monopole background for simplicity of explanations.
See e.g.~\cite{Ishii:2008ib} for details.
}.
The ultraviolet momentum cutoff is proportional to $\mu N_5$. 
The coupling constant of the 3d theory $g_{\rm 3d}$ and the noncommutativity parameter\footnote{
By using the standard embedding of the sphere to $\mathbb{R}^3$ and zooming in the neighborhood of the north pole, we get two-dimensional noncommutative plane parametrized by $x$ and $y$, with the noncommutative product $x\ast y - y\ast x = i\theta$.
The coordinate and matrices are related as follows. 
Firstly, $gX_i$ is identified with the gauge-covariant derivative.  
By writing $gX_i=\frac{\mu}{3}J_i+ga_i$, 
the first term $\frac{\mu}{3}J_i$ is regarded as the momentum $p_i$, or equivalently the derivative $-\partial_i$, 
and the second term is regarded as the gauge field. At the north pole ($J_3=s\sim N_5$), we have 
$[p_1,p_2]\sim i\mu^2N_5 \equiv i\theta^{-1}$. The coordinate $x$ and $y$ are defined as $x=-\theta p_2$ and $x=\theta p_1$, such that $[x,p_1]=[y,p_2]=i$. 
} 
$\theta$ are related to $g_{\rm 1d}$ and $\mu$ by\footnote{
Here $g_{\rm 1d} =g$. 
We have added the subscript ``1d'' to emphasize that
it is the gauge coupling in one dimension.
} 
\begin{eqnarray}
\frac{g_{\rm 1d}^2}{\mu^2 N_5} = g_{\rm 3d}^2, 
\qquad
\theta=\frac{1}{\mu^2 N_5}. 
\label{g3d-vs-g1d}
\end{eqnarray}
This is the standard relation between the matrix model and noncommutative space, 
which was first pointed out in Refs.~\cite{GonzalezArroyo:1982ub,GonzalezArroyo:1982hz}
for the case of the fuzzy torus, in the context of the Eguchi-Kawai reduction \cite{Eguchi:1982nm}.
In order to obtain the usual 3d SYM on noncompact and commutative space, we take three limits, namely 
\begin{itemize}
\item
The continuum limit $\mu N_5\to\infty$,  

\item
Decompactification limit (flat limit) $r_{{\rm S}^2} = \frac{3}{\mu} \to\infty$, 

\item
Commutative limit $\theta\to 0$, 

\end{itemize}
while fixing the coupling constant and gauge group of the 3d theory, $g_{\rm 3d}$ and $N_2$. 
Note that one does not have to take the decompactification limit and the commutative limit, 
if one is interested in the theory at finite volume or with finite noncommutativity. 

In this construction, the existence of maximal supersymmetry is crucial: 
in general, the commutative limit of QFT on noncommutative space is not the same as the corresponding theory on the commutative space, due to the UV/IR mixing \cite{Minwalla:1999px},  
but this problem is absent for the theories with maximal supersymmetry \cite{Matusis:2000jf,Hanada:2014ima}. 

The coupling constant $g_{\rm 3d}^2$ has the dimension of mass. 
At a given energy scale $\varepsilon$, the effective, dimensionless coupling is $g_{\rm 3d}^2/\varepsilon$. 
M-theory and IIA string description is valid at $g_{\rm 3d}^2/\varepsilon\gtrsim N_2^{-1/5}$
and $N_2^{-1}\lesssim g_{\rm 3d}^2/\varepsilon\lesssim N_2^{-1/5}$, respectively \cite{Itzhaki:1998dd}. 
If we are interested in the low-energy spectrum, the characteristic energy scale is the inverse of the radius of S$^2$
: $\varepsilon\sim\mu$. 
In the M-theory regime, 
the dual gravity description is the $N_2$-coincident spherical M2-branes in the pp-wave geometry with the flux parameter $\mu$, with tension $T_{\rm M2}=\frac{1}{(2\pi)^2l_{\rm P}^3}$
and radius $r_{{\rm M2}}=\frac{\mu g_{\rm 1d}^{2/3}N_5}{12\pi T_{\rm M2}}$ \cite{Maldacena:2002rb,Asano:2017nxw}.

\subsubsection{6d ${\cal N}=(2,0)$ superconformal field theory (M5-branes)}
\label{sec:6dSCFT}
The 6d ${\cal N}=(2,0)$ SCFT is the worldvolume theory of M5-branes in M-theory.
It is also expected to be dual to M-theory on $AdS_7 \times S^4$.
Ref.~\cite{Maldacena:2002rb} proposed that various nontrivial brane configurations can be described by the fuzzy-sphere vacua of the BMN matrix model. 
Among them, the M5-brane configuration is obtained 
in the following scaling limit in \eqref{eq:fuzzy3d} \cite{Maldacena:2002rb,Asano:2017nxw,Asano:2017xiy}:
\begin{eqnarray}
N_2\to\infty, 
\qquad
N_5\ :\ {\rm fixed}, 
\qquad
g_{\rm 1d}^2N_2\to\infty, 
\qquad
g_{\rm 1d}^2\to 0. 
\end{eqnarray}
The dual gravity description is the $N_5$-coincident spherical M5-branes in the pp-wave geometry with the flux parameter $\mu$, 
with tension $T_{\rm M5}=\frac{1}{(2\pi)^5l_{\rm P}^6}=\frac{\mu^3g_{\rm 1d}^2}{(12\pi N_2)^3}$ and 
radius $r_{{\rm M5}}=\left(\frac{\mu g^{2/3}N_5}{6\pi^3T_{\rm M5}}\right)^{1/4}
=
\left(
\frac{288N_5N_2^3}{\mu^{2}g_{\rm 1d}^{4/3}}
\right)^{1/4}$. 
The scalars $X_I$ in the matrix model with the canonical normalization \eqref{BMN-Hamiltonian-canonical-normalization}, whose dimension is $({\rm mass})^{-1/2}$, 
is obtained by multiplying $\sqrt{\frac{\mu}{6N_2}}$ to the coordinate in the gravity side: 
\begin{eqnarray}
r^{(X)}_{{\rm S}^5}
=
\sqrt{\frac{\mu}{6N_2}}
r_{{\rm M5}}
=
\left(
\frac{8N_5N_2}{g_{\rm 1d}^{4/3}}
\right)^{1/4}. 
\end{eqnarray}
This parametrically large S$^5$ should be realized as the spherical distribution of the elements of six scalars $X_4,X_5,\cdots,X_9$. 

\subsubsection{4d $\mathcal{N}=4$ SYM in the large-$N$ limit}\label{sec:4dSYM-large-N}
{}
The 4d $\mathcal{N}=4$ SYM describes the worldvolume theory of D3-branes in type IIB superstring.
It also provides the canonical example of the AdS/CFT correspondence with type II superstring on $AdS_5 \times S^5$.
Around a certain concentric-fuzzy-sphere background, 
the BMN matrix model can describe the 4d ${\cal N}=4$ SYM on S$^3$ in the planar limit \cite{Ishii:2008ib}. 
(Note that, unlike the M2/D2 and M5 theories, this construction is valid only in the planar sector.)
The SU(2)-generators are chosen to be the following form\footnote{
See \cite{Kawai:2009vb} for another choice.
The large-$N$ reduction on $S^3$ has been tested in various ways \cite{Ishiki:2008te,Ishiki:2009sg,Honda:2010nx,Honda:2011qk,Honda:2013nfa,Honda:2012ni,Asano:2012gt}.
}:
\begin{eqnarray}
J_i
=
\oplus_{s=n-T}^{n+T}
\left(J_i^{(s)}\otimes\textbf{1}_{k}\right), 
\end{eqnarray}
and the large-$N$ limit is taken as
\begin{eqnarray}
k,n,T,n-T\to\infty, 
\qquad
\lambda_{4d}=\frac{8\pi^2g_{\rm 1d}^2k}{\mu n}: {\rm fixed}. 
\end{eqnarray}
Note that the sum with respect to $s$ is taken over the integer and half-integer. 
The radius of S$^3$ is given by $6/\mu$.
An intuitive way to understand this construction is to regard S$^3$ as the S$^1$-fibration of S$^2$;  
for each $s$, 
the 3d maximal SYM on S$^2$ is obtained as in Sec.~\ref{sec:mapping-rule-M2}, 
and the additional S$^1$ direction is generated 
via the summation over $s$ interpreted as summing up KK momenta, 
in a way similar to the quenched Eguchi-Kawai reduction \cite{Bhanot:1982sh,Parisi:1982gp,Gross:1982at}. 
In this case, note that
any decompactication limit is not required even if we are interested in flat noncompact space
since $\mathbb{R}\times S^3$ is conformally equivalent to the flat space.

\subsubsection{Comments on QFTs without string theory dual}
\label{sec:theories-without-string-dual}
{}
It is straightforward to construct other matrix models, which admit the fuzzy sphere background; see, e.g.,~Ref.~\cite{Kim:2006wg}. By using them, 3d QFTs on the noncommutative space can be regularized. 
Without supersymmetry, the fuzzy sphere background is usually unstable in the QFT limit, 
and hence, only supersymmetric theories admit simple constructions \cite{VanRaamsdonk:2001jd,Azeyanagi:2007su,Azeyanagi:2008bk,Hanada:2016bnb}.\footnote{ 
Non-supersymmetric theories are not impossible, but it requires more complicated setup such as the unitary type matrix model,  which are not easy on quantum computer \cite{VanRaamsdonk:2001jd,Hanada:2016bnb}. 
}
Note that the commutative limit ($\theta\to 0$) of such noncommutative theory is different from the theory on the commutative space in general.
In the opposite limit ($\theta\to\infty$), only the planar diagrams survive, and the large-$N$ limit of the commutative theory is realized \cite{GonzalezArroyo:1982ub,GonzalezArroyo:1982hz}. 
The construction of the large-$N$ limit of the 4d ${\cal N}=4$ SYM \cite{Ishii:2008ib,Ishiki:2008te}
explained in Sec.~\ref{sec:4dSYM-large-N} can be generalized to other supersymmetric theories including supersymmetric QCD~\cite{Hanada:2009hd}. 

\subsection{Comments on the cutoff effect}
\label{sec:cutoff-QFT-from-BMN}
{}
In practical simulations, 
we shall approximate the fuzzy sphere state by imposing the truncation on the Hilbert space
and the approximations depend on the cutoff $\Lambda$.
Although it is difficult to estimate the cutoff effect precisely,
one can estimate lower bound on $\Lambda$ to get reasonable approximations by a simple argument below.
Note that the bounds estimated below are very minimal requirement, and some amount of error can remain as long as $\Lambda$ is large but finite.
In this sense, our estimation is crude.

Let us give a crude estimate of the cutoff effect in the fuzzy-sphere states. 
Here we only discuss the necessary conditions for the true ground state to be approximated 
by a certain linear combination of the Fock states in the truncated Hilbert space. 
In order for the ground state of the truncated Hamiltonian to be close to the true ground state, additional conditions should be required. 

Let us first consider the non-gauge-invariant state $|J_i\rangle$ and take the gauge in which $J_3$ is diagonal. 
Then the diagonal elements of $J_3$ runs from $-s$ to $+s$. 
Hence, depending on the value of the coupling $g_{\rm 1d}$, the flux parameter $\mu$ and the spin $s$, the fuzzy sphere can be rather large in the coordinate basis. 
Similar properties hold for other gauge choices as well. 
Therefore, in order to express the wave function properly, we have to be able to describe the functions which spreads from $X_i^\alpha\sim -\frac{s\mu}{g_{\rm 1d}}$ to $X_i^\alpha\sim +\frac{s\mu}{g_{\rm 1d}}$, for all $\alpha=1,2,\cdots,N^2-1$. 
Similar consideration is needed for $X_a^\alpha$ and $P_I^\alpha$ as well. 
Then higher modes in the Fock basis, whose wave function in the coordinate basis is larger, are needed; 
the $n$-th excited mode of the harmonic basis
is localized at $X\lesssim \sqrt{\frac{n}{\mu}}$, and linear combination of the modes at $n<\Lambda$ can describe only the states localized at $X\lesssim \sqrt{\frac{\Lambda}{\mu}}$. Hence the following conditions are needed:  
\begin{itemize}
\item
The width of the wave function of the $n$-th excited state is 
$\sqrt{\langle n|\hat{X}^2|n\rangle}\sim\sqrt{\frac{n}{\mu}}$. (Note that $\langle n|\hat{X}|n\rangle =0$.)
The size of the wave function of the $\Lambda$-th excited state $\sim\sqrt{\frac{\Lambda}{\mu}}$ has to be larger than the typical value of $X_I^\alpha$ in the true ground state
defined through the expectation value of $\hat{X}^2$. 

\item
Typical value of the momentum of the $n$-th excited state is 
$\sqrt{ \langle n|\hat{P}^2|n\rangle} \sim\sqrt{n\mu}$. (Note that $\langle n|\hat{P}|n\rangle=0$.)
Hence $\sqrt{\Lambda\mu}$ has to be larger than the typical value of $P_I^\alpha$.  

\end{itemize}
In addition to these, let us require the following: 
\begin{itemize}
\item
If we want some resolution to distinguish the possible values of $X_I^\alpha$, 
then $\frac{1}{\sqrt{ \langle \Lambda|\hat{P}^2|\Lambda\rangle}}$ $\sim$ $\frac{1}{\sqrt{\Lambda\mu}}$ has to be smaller than that resolution scale.
In order for the fuzzy sphere background to make sense, 
the resolution has to be finer than the difference between the values of 
the matrix entries of the fuzzy sphere, $\frac{\mu}{3g_{\rm 1d}}$. 
\end{itemize}

Below, we consider the $N_2$-coincident fuzzy sphere background \eqref{eq:fuzzy3d}. 
The spin is $s=\frac{1}{2}\left(N_5-1\right)$, where $N=N_2N_5$. 
The largest value among the matrix entries is $\frac{\mu s}{3g_{\rm 1d}}=\frac{\mu\left(N_5-1\right)}{6g_{\rm 1d}}$. 
The M2/D2 theory and M5 theory (Sec.~\ref{sec:mapping-rule-M2} and Sec.~\ref{sec:6dSCFT}) are described by such configuration. 
Similar estimates can be made for slightly different backgrounds discussed in Sec.~\ref{sec:4dSYM-large-N} and Sec.~\ref{sec:theories-without-string-dual}.  

We emphasize again that the three conditions listed above are very crude, 
and the bound obtained here is weaker than the requirement for a stronger condition that the ground state of the truncated Hamiltonian is close to the true ground state.
Different realization of the fuzzy sphere background, which are related by the gauge transformation, can lead to different estimate. However it is not necessarily a bad news, in the following sense. Let us consider all possible realizations of the fuzzy sphere background and take the strongest bound; we use $\Lambda_{\rm crude,max}$ to denote this. 
The requirement that the ground state of the truncated Hilbert space is close to the true ground state is a gauge-invariant notion, so there must be the value of cutoff appropriate for any realization of the fuzzy sphere; we call it $\Lambda_{\rm strict}$. Then $\Lambda_{\rm crude,max}\le \Lambda_{\rm strict}$.  
\subsubsection*{Weak coupling}
{}
Eventually we want to consider the gauge-invariant state ${\cal S}(\ket{J_i})$, but we start with a simpler case, 
the non-gauge-invariant state $|J_i\rangle$. Furthermore take the specific realization of SU(2) in which $J_3$ is diagonal. 
The first condition for $X_{1,2,3}$ becomes $\sqrt{\frac{\Lambda}{\mu}}\gtrsim\frac{\mu N_5}{g_{\rm 1d}}$, and hence
\begin{eqnarray}
\Lambda
\gtrsim
\frac{\mu^3N_5^2}{g_{\rm 1d}^2}
= \frac{\mu N_5}{g_{\rm 3d}^2}. 
\label{eq:cutoff-fuzzy-sphere-weak-coupling}
\end{eqnarray}
Here we have used the relation \eqref{g3d-vs-g1d}, which relates the couplings of the 1d and 3d theories.
The other two conditions are much weaker than \eqref{eq:cutoff-fuzzy-sphere-weak-coupling}. 
Note that the off-diagonal modes are heavy about this background. 
Therefore, it is likely that \eqref{eq:cutoff-fuzzy-sphere-weak-coupling} can be a good estimate for the stronger requirement: the ground state of the truncated Hilbert space is close to the true ground state. 
As mentioned before, this stronger requirement is a gauge-invariant notion, and hence, we expect \eqref{eq:cutoff-fuzzy-sphere-weak-coupling} gives a good estimate for the gauge-invariant state ${\cal S}(\ket{J_i})$ as well (i.e., $\Lambda_{\rm crude}$ and $\Lambda_{\rm strict}$ can be roughly the same.)

The wave function looks different depending on the embeddings of SU$(2)$ to SU($N$).
For example, we can perform a generic gauge transformation and make all the matrix entries to of order $\frac{\mu}{g_{\rm 1d}}\times N^0$. 
If we use such embedding, much finer resolution ($\sim\frac{\mu}{g_{\rm 1d}N_5}$) is needed in order to distinguish 
the fuzzy sphere and other backgrounds. Hence the third condition matters and $\frac{1}{\sqrt{\Lambda\mu}}\lesssim \frac{\mu}{g_{\rm 1d}N_5}$ is needed, which in turn becomes $\Lambda\gtrsim\frac{g_{\rm 1d}^2N_5^2}{\mu^3}$. This is weaker than \eqref{eq:cutoff-fuzzy-sphere-weak-coupling}; however note that, with such gauge choice, off-diagonal modes are not heavy and the expansion about this specific background cannot be truncated at low order; the condition \eqref{eq:cutoff-fuzzy-sphere-weak-coupling} is required in order for the ground state of the truncated Hamiltonian to be sufficiently close to the true ground state.  
\subsubsection*{M2-brane limit}
{}
At strong coupling, the M2-brane limit ($N_2, \mu$ fixed, $N_5\to\infty$):  
By using \eqref{g3d-vs-g1d} which relates the couplings of the 1d and 3d theories, 
we obtain $\frac{\mu s}{3g_{\rm 1d}}\sim\frac{\sqrt{N_5}}{6g_{\rm 3d}}$. 
It is reasonable to assume that the fluctuation is smaller than this classical value. 
(Otherwise, we cannot take the M2-brane limit.)
Then the requirement is formally the same as 
the weak-coupling result \eqref{eq:cutoff-fuzzy-sphere-weak-coupling}, 
namely
$\sqrt{\frac{\Lambda}{\mu}}\gtrsim \frac{\mu N_5}{g_{\rm 1d}}\sim\frac{\sqrt{N_5}}{g_{\rm 3d}}$ is required. 
Other conditions are much weaker.  
\subsubsection*{M5-brane limit}
{}
According to Ref.~\cite{Asano:2017xiy}, the eigenvalues of $X_{4,\cdots,9}$ form S$^5$, 
whose radius scales as $\left(\frac{8N_5N_2}{g_{\rm 1d}^{4/3}}\right)^{1/4}$ (see Sec.~\ref{sec:6dSCFT}). 
Therefore, 
$\sqrt{\langle\Lambda|\hat{X}^2|\Lambda\rangle}\sim\sqrt{\frac{\Lambda}{\mu}}$ has to be larger than this radius, 
and hence, 
\begin{eqnarray}
\Lambda
\gtrsim
\sqrt{\frac{\mu^2N_2N_5}{g_{\rm 1d}^{4/3}}}
=
\sqrt{\frac{\mu^2N}{g_{\rm 1d}^{4/3}}}.
\end{eqnarray}

\subsection{Maldacena-Milekhin proposal applied to QFT limit}\label{sec:gauge-invariance-QFT}
{}
As mentioned in Sec.~\ref{sec:BMN-Maldacena-Milekhin}, 
Maldacena and Milekhin \cite{Maldacena:2018vsr} made a surprising proposal: 
in the BFSS and the BMN matrix models, at sufficiently strong coupling (sufficiently low energy), gauge non-singlets are too heavy and negligible in the low-energy dynamics. 
(See Ref.~\cite{Berkowitz:2018qhn} for a numerical test via Markov Chain Monte Carlo.)
Then we can simply forget about the non-singlet modes.
Let us consider the consequence of this proposal for the fuzzy-sphere ground states.

Suppose we took the fuzzy sphere state $\ket{J_i}$, which is not gauge-invariant. 
This choice appears to break the SU($N$) symmetry, and hence, we would expect the existence of the Nambu-Goldstone modes, which connect different realization of the fuzzy sphere. 
According to the Maldacena-Milekhin conjecture~\cite{Maldacena:2018vsr}, it is not necessarily the case. 
In the matrix model, the spontaneous symmetry breaking can take place only in the strict large-$N$ limit. 
In the M2/D2 or M5/NS5 limit, the fuzzy sphere configuration and the coupling constant are varied nontrivially with $N$, 
and hence, whether the Nambu-Goldstone mode can actually appear is a highly nontrivial issue.\footnote{We thank A.~Milekhin for useful comments regarding this point.} 
According to the Maldacena-Milekhin conjecture, the energy of the would-be Nambu-Goldstone modes are bounded from below, as 
\begin{eqnarray}
E_{\rm adj}\gtrsim\frac{g_{\rm 1d}^2N_2}{\mu^2N_5}=g_{\rm 3d}^2N_2 = \lambda_{\rm 3d}. 
\end{eqnarray}
If we are interested in the parameter region dual to weakly-coupled type IIA string or M-theory, 
we need to consider the energy scale much lower than $\lambda_{\rm 3d}$. 
There, such modes are negligible, and a sort of super-selection takes place. 
In the same manner, in the scaling region where the M5/NS5 brane theory appears (fixed $N_5$, $N_2\to\infty$ with $g_{\rm 1d}^2N_2\to\infty$), 
the would-be Nambu-Goldstone modes are negligible. 

If the Maldacena-Milekhin conjecture is correct, then we do not have to prepare the gauge-invariant fuzzy sphere state for quantum simulation at strong coupling. This can simplify the state preparation significantly. 
\section{Realizations on quantum computer}\label{sec:Quantum-algorithms}
\label{sec:realization}
{}
So far, we have explained how certain supersymmetric quantum field theories can be formulated by using the matrix models, 
in such a way that they can be put on a universal quantum computer.
We have not yet specified the detail of the implementation. 
In this section, we will give concrete protocols and show the potential benefit of using quantum simulation. 
\subsection{Encoding the bosonic part into qubits (compact mapping)}\label{sec:compact-mapping}
{}
Now we wish to address the generic strategies we usually use to encode the Hamiltonian (see \cite{somma2005quantum,mcardle2019digital} for references). Firstly we focus on the bosonic part. 

As we have seen, we regularize the Fock space by introducing the cutoff $\Lambda$ to the excited modes of harmonic oscillators.
For each harmonic oscillator, we assign $K=\log_2\Lambda$ qubits. 
We use the compact mapping
\begin{align}
|j\rangle  = \left| {{b_{K - 1}}} \right\rangle \left| {{b_{K - 2}}} \right\rangle  \ldots \left| {{b_0}} \right\rangle ~,
\end{align}
for the energy level $j=0,1,\cdots,\Lambda-1$, where we use the binary decomposition
\begin{align}
j = {b_{K - 1}}{2^{K - 1}} + {b_{K - 2}}{2^{K - 2}} +  \ldots  + {b_0}{2^0}. 
\end{align}
One could map the creation operator $A^\dagger$ as 
\begin{align}
\hat{A}^\dag = \sum\limits_{j = 0}^{\Lambda - 2} {\sqrt {j + 1} } |j + 1\rangle \langle j|. 
\end{align}
By using $|j\rangle  = \left| {{b_{K - 1}}} \right\rangle \left| {{b_{K - 2}}} \right\rangle  \ldots \left| {{b_0}} \right\rangle$
and $|j+1\rangle  = \left| {{b'_{K - 1}}} \right\rangle \left| {{b'_{K - 2}}} \right\rangle  \ldots \left| {{b'_0}} \right\rangle$, 
we can write $|j + 1\rangle \langle j|$ as
\begin{align}
 |j + 1\rangle \langle j|
 =
 \otimes_{l=0}^{K-1}
 \left(|b'_l\rangle\langle b_l|\right). 
\end{align}
Note that each $|b'_l\rangle\langle b_l|$ is just a Pauli spin operator: 
\begin{eqnarray}
& &
|0\rangle\langle 0|=\frac{\textbf{1}_2-\sigma_z}{2}, 
\qquad
|1\rangle\langle 1|=\frac{\textbf{1}_2+\sigma_z}{2}, 
\nonumber\\
& &
|0\rangle\langle 1|=\frac{\sigma_x+i\sigma_y}{2}, 
\qquad
|1\rangle\langle 0|=\frac{\sigma_x-i\sigma_y}{2}.  
\end{eqnarray}
Therefore, $\hat{A}^\dagger$ can be written as a linear combination of 
less than $\Lambda^2$ Pauli strings of at most length $K$ (i.e., tensor products of $K$ Pauli spin operators).\footnote{
Each $\ket{b'_l}\bra{b_l}$ contains at most two Pauli matrices, and $l$ runs through $K$ different values
in each $\ket{j+1}\bra{j}$, so there are $2^K=\Lambda$ or less Pauli strings in each $\ket{j+1}\bra{j}$, and the maximum length is $K$. 
There are $\Lambda-1$ different values of $j$, so the number of Pauli strings can be bounded by $\Lambda(\Lambda-1)$. 
More or less, the same bound can be obtained by noticing that the number of possible Pauli strings of length $K$ or less, including the identity, is $4^K=\Lambda^2$. 
} 
The same holds for $\hat{A}$. 
In $\hat{X}_I^\alpha=\frac{\hat{A}+\hat{A}^\dagger}{\sqrt{2\omega_I}}$, the non-Hermitian part in $\hat{A}$ and $\hat{A}^\dagger$ (odd number of $i\sigma_y$)
cancel out and the number of Pauli strings becomes smaller compared to $\hat{A}$ and $\hat{A}^\dagger$.  

In our simulation problem, we prefer to use the compact mapping approach, since it has much fewer costs about Hilbert space, although it does not have a good locality.
Still, however, $\hat{X}_I^\alpha$ and $\hat{P}_I^\alpha$ are $K$-local, 
where $K=\log_2\Lambda$ and $\Lambda$ is at most some powers of $N$ in the situations under consideration; 
see Sec.~\ref{sec:cutoff-QFT-from-BMN}. Therefore, the growth of $K$ is slow ($K\sim\log N$) 
and hence the lack of the locality may not be too problematic. 
\subsection{Encoding the fermionic part into qubits}
The Jordan-Wigner transformation is a standard way to relate fermions and qubits. 
In order to express $\hat{\Psi}$ explicitly by using the Jordan-Wigner transformation, let us use a standard representation with the complex fermions shown in Sec.~\ref{sec:BMN-Hamiltonian}.
The 16-component Majorana fermion $\Psi$ can be expressed by using four ($I=1,2,3,4$) 2-component ($p=1,2$) complex fermions $\psi_{Ip}$ which satisfy the anticommutation relation $\{\hat{\psi}^{\dagger Ip\alpha},\hat{\psi}_{Jq}^\beta\}=\delta^I{}_J\delta^p{}_q\delta^{\alpha\beta}$.
To simplify the notation, we combine three indices to one index which runs from $1$ to $8N^2$: $\{\hat{\psi}^\dagger_{m},\hat{\psi}_n\}=\delta_{mn}$ ($m,n=1,2,\cdots,8N^2$).  Then we can express $\hat{\psi}_{m}$ acting on ${\cal H}_{\Psi}$ as
\begin{eqnarray}
\hat{\psi}_m
=
\sigma_z^{\otimes (m-1)}
\otimes
\frac{\sigma_x-i\sigma_y}{2}
\otimes
\textbf{1}^{\otimes (8N^2-m)}. 
\end{eqnarray}

This encoding is simple, but has a disadvantage: the operators $\hat{\psi}$ is highly non-local. 
In the BMN matrix model, 
the maximum length of the Pauli strings appearing in the realization of  $\hat{\psi}$ is $8N^2$. Such long Pauli strings make the quantum simulation inefficient. 

Alternatively, we can use the Bravyi-Kitaev transformation \cite{bravyi2002fermionic}, which provides us with a good basis, which makes the length of the Pauli strings to scale $\log N$. (For explicit form in terms of Pauli matrices, see, e.g.,~Ref.~\cite{seeley2012bravyi}.)
Whether the Bravyi-Kitaev basis is better than the Jordan-Wigner can depend on the value of $N$. At very large $N$,  the Bravyi-Kitaev basis performs better. 

\subsection{Real-time dynamics: qubitization, quantum signal processing}\label{sec:QSP-real-time}
\label{sec:real-time}
{}
For the real-time evolution, we consider the time-independent Hamiltonian. 
While we could apply various algorithms listed in Appendix~\ref{overview}, 
here we consider another efficient algorithm, taking advantage of the encoding of matrix models. This algorithm, used in the context of quantum signal processing, has an explicit construction of oracles. Here, we will review an algorithm that is related to Refs.~\cite{low2017optimal,Babbush:2018mlj}.

The Hamiltonians of the matrix model is written in terms of Pauli strings,
\begin{align}
\hat{H}=\sum_{i=1}^{L} \alpha_{i} \hat{\Pi}_{i}, \quad \alpha_{i}>0, 
\label{Hamiltonian-as-sum-of-Pauli-chains}
\end{align}
with $L\lesssim \Lambda^8 N^4$.  The Pauli strings $\hat{\Pi}_i$'s are unitary and Hermitian. 
The longest Pauli strings appear from the quartic interaction term ${\rm Tr}[\hat{X}_I,\hat{X}_J]^2$. 
As we have seen in Sec.~\ref{sec:compact-mapping}, with the compact mapping 
each $\hat{X}_I^\alpha$ is a linear combination of less than $\Lambda^2$ Pauli strings of length (at most) $K=\log_2\Lambda$.  
Hence $\hat{X}_I^\alpha\hat{X}_I^\beta\hat{X}_I^\gamma\hat{X}_I^\rho$ is a linear combination of at most $\lesssim\Lambda^8$ Pauli strings of length (at most) $4K$. We have $\sim N^4$ terms of this kind, and hence, 
the number of Pauli strings in eq.\eqref{Hamiltonian-as-sum-of-Pauli-chains} is $L\lesssim \Lambda^8 N^4$. 

The first step in the quantum simulation algorithm is the block-encoding of our Hamiltonian into a unitary matrix. 
We introduce ancilla states $|i\rangle$ ($i=1,2,\cdots,L$) and prepare a state $|G\rangle$ defined by  
\begin{eqnarray}
|G\rangle  =  \sum_{i=1}^L g_i \ket{i},
\qquad
|g_i|^2
=
\frac{\alpha_i}{\lambda}, 
\qquad
\lambda=\sum_{i=1}^L\alpha_i. 
\end{eqnarray} 
Then we can prepare a unitary operator $\hat{U}$ acting on the Hilbert space times $\mathbb{C}^L$, 
which satisfies 
\begin{equation}
\frac{\hat{H}}{\lambda} = \Big( \bra{G} \otimes \hat{I} \Big)  \hat{U}  \Big( \ket{G} \otimes\hat{I} \Big).
\end{equation}
Here $\hat{I}$ is the identity operator acting on the Hilbert space.
Such unitary operator $\hat{U}$ can be constructed as a control-$\hat{\Pi}_i$ operation for $K$-local Pauli string $\hat{\Pi}_i$, 
\begin{equation}
\hat{U} \ket{i} \ket{\psi}=\ket{i} \left(\hat{\Pi}_i \ket{\psi}\right). 
\end{equation}
Equivalently, 
\begin{equation}
\hat{U} 
=
\sum_{i=1}^L
\left(
\ket{i}\bra{i}
\otimes
\hat{\Pi}_i
\right). 
\end{equation}
This unitary is the so-called block encoding of the Hamiltonian of the BMN model. The complexity of building this is at most $C_U = O(KL)$, because there are $L$ terms and each of them involves at most $O(K)$ multiplications of Pauli matrices. Our construction requires $\log L = \log (\Lambda^8 N^4)$ additional qubits for block-encoding procedure, in addition initially needed $9N^2\log_2\Lambda+8N^2$ qubits. 

As a next step, we define a relation operator (which is unitary {\it and} Hermitian)
\begin{equation}
\hat{R} = 2  \ket{G} \bra{G} - \hat{I},  
\end{equation} 
and by using it, we define
\begin{equation}
\hat{W} = \hat{R}\hat{U}. 
\end{equation} 
Then one can show 
\begin{equation}
\bra{G} \hat{W}^n  \ket{G}  = T_n\Big(\frac{\hat{H}}{\lambda}\Big)
\label{eq:chebyshev}
\end{equation}
where $T_n(\cdot)$ stands for the $n$-th order Chebyshev polynomial of the first kind. 
(See the appendix~\ref{sec:chebyshev-proof} for a proof of \eqref{eq:chebyshev}.)

We are interested in approximating the time evolution operator $e^{-i\hat{H}t}$. In fact, the approximation can be made with the help of the following expansion (Jacobi-Anger expansion) of the time evolution operator   
\begin{equation}
e^{-i\hat{H}t} = J_0(-\lambda t) + 2 \sum_{n=1}^\infty i^n  J_{n}(-\lambda t) \times T_n\Big(\frac{\hat{H}}{\lambda}\Big), 
\label{Jacobi-Anger}
\end{equation}
where $J_n$ is the Bessel function of the first kind.  
The right hand side can be handled by the quantum signal processing algorithm \cite{low2017optimal},
namely quantum signal processing can be used as a black box to compute the Chebyshev polynomials efficiently when approximating $e^{-i\hat{H}t}$ 
up to an error $\epsilon$ for repetition $n \sim \lambda t + \log (1/\epsilon)$ \cite{low2017optimal}. 
(See the appendix~\ref{sec:QSP} for a review of quantum signal processing.)
Thus, the total complexity of simulation $e^{-i\hat{H}t}$ is 
\begin{equation}
\label{eq:unitary-scaling}
O\Big((C_U +C_G) \Big( ||\alpha|| \cdot t + \log{\frac{1}{\epsilon}} \Big)\Big)~,
\end{equation}
where\footnote{
The value of $||\alpha||$ is dominated by the quartic interaction, which has $O(N^4)$ combinations of color degrees of freedom. There is an overall factor $g^2$, and the sum over the excitation levels gives a factor of the order  $\sum_{j_1,j_2,j_3,j_4}\sqrt{j_1j_2j_3j_4}\sim\left(\Lambda^{3/2}\right)^4=\Lambda^6$. Therefore,  $||\alpha||\sim g^2N^4\Lambda^6$. The cubic interaction gives a sub-leading correction of order $\mu gN^3\Lambda^{9/2}$. 
} $||\alpha||=\sum_{i=1}^L|\alpha_i|$, $C_U = O(K L)$ and $C_G = O(L)$.

For more detailed discussions, we recommend the readers to read the appendix~\ref{sec:QSP} and the original references \cite{low2017optimal,Babbush:2018mlj}. We could regard this as an alternative algorithm for the real-time evolution of a fixed Hamiltonian, especially when we need to know the precise construction of the oracles. It will be interesting if one could also construct a time-dependent version of the algorithm. 
\subsection{Adiabatic state preparation}
\label{sec:adiabatic_state_preparation}
The simulation algorithms for the real-time dynamics have to be accompanied with the preparation methods for appropriate initial states. Furthermore the states themselves have rich information about the system. 
Hence, we will explain how the ground state can be constructed 
by using the adiabatic state preparation method. 
In this section, we present a novel application of the Wan-Kim algorithm~\cite{wan2020fast} for fast digital state preparation that utilizes block encoding and the theoretical concept of quasi-adiabatic continuation. The interested reader can also look into alternative algorithms in Appendices~\ref{sec:alternative_adiabatic_state_preparation} and \ref{sec:Berry-et-al-algorithm} that are based on time-dependent Hamiltonian simulation methods in Ref.~\cite{berry2019time}. 
 \subsubsection*{Trivial vacuum }
First, let us consider the trivial vacuum.
In the weak-coupling limit, this vacuum is literally `trivial': 
it is just the Fock vacuum, which is `$N$ fuzzy spheres with spin zero.'  In the M5-brane limit (Sec.~\ref{sec:6dSCFT}), it is expected that one M5-brane is described~\cite{Maldacena:2002rb}. 
The nature of the wave function at strong coupling is expected to be not trivial at all, and we could see some crucial properties of the trivial vacuum by quantum simulation. 

The full Hamiltonian of the BMN matrix model consists of the free part and the interaction part,  
$\hat{H}_{\rm BMN} = \hat{H}_{\rm{free}} + \hat{H}_{\rm{int}}(g)$,
where $\hat{H}_{\rm{int}}$ disappears when the coupling constant $g$ is set to zero. 
The free part $\hat{H}_{\rm{free}}$ does not contain $g$. 
In the weak-coupling limit $g\to 0$, the ground state is the Fock vacuum.
Using the notation of Ref.~\cite{wan2020fast} we introduce a notation 
\begin{align}
\hat{H}_0& = \hat{H}_{\rm{free}},  \\
\hat{H}_1 &= \hat{H}_{\rm{free}} +\hat{H}_{\rm{int}}(g)=\hat{H}_{\rm BMN}, 
\end{align}
and use a parameter $s \in [0,1]$ to interpolate $\hat{H}_0$ and $\hat{H}_1$ as
\begin{align}
\hat{H}(s) = (1-s) \hat{H}_0  + s \hat{H}_1.
\end{align}
To use the Wan-Kim algorithm, first we need to construct the block encoding of $\hat{H}_0$, $\hat{H}_1$ and $\hat{H}' = \hat{H}_1 - \hat{H}_0=\hat{H}_{\rm int}$,  
\begin{align}
\frac{\hat{H}_0}{\beta'} = \Big( \bra{G} \otimes \hat{I} \Big) U_{\hat{H}_0}  \Big( \ket{G} \otimes \hat{I} \Big) \nonumber\\
\frac{\hat{H}_1}{\beta'} = \Big( \bra{G} \otimes \hat{I} \Big) U_{\hat{H}_1}  \Big( \ket{G} \otimes \hat{I} \Big)  \nonumber\\
\frac{\hat{H}'}{\beta} = \Big( \bra{G} \otimes \hat{I} \Big) U_{\hat{H}'}  \Big( \ket{G} \otimes \hat{I} \Big) 
\end{align}
where $\beta' \approx \max \{||\hat{H_0} ||, ||\hat{H_1} || \}$ and $\beta \approx ||\hat{H}' ||$.  The block encoding can be done as described in Sec. \ref{sec:real-time} on real-time dynamics and complexity of the construction of these unitaries is  $O(K L)$. 
The norm $||\ \cdot\  ||$ used here is the 1-norm, i.e.~$||\hat{H} ||=\sum_{i,j}|\hat{H}_{ij}|$. 
We denote the scale of the mass gap of the Hamiltonian during the entire evolution from $s=0$ to $s=1$ by $\Delta_{\text{gap}}$. Then, we can use the Algorithm~1 of Ref.~\cite{wan2020fast} with inputs $U_{H_0}, U_{H_1}, U_{H'} $ and $\Delta_{\text{gap}}$ to construct a digital adiabatic unitary $\widetilde U$. According to Theorem 1 of Ref.~\cite{wan2020fast}, if we denote by $\ket{\Omega(s)}$ ground state of $\hat{H}(s) $, an operator $\widetilde U$ can be implemented such that
\begin{align}
\Big | \Big |  \ket{\Omega(1)} - \widetilde U \ket{\Omega(0)} \Big | \Big | \leq \delta 
\end{align}
with probability $1 - O(\delta)$ using
\begin{align}
O\left( \frac{\beta}{\Delta_{\text{gap}}}  \Big[ \frac{\beta'}{\Delta_{\text{gap}}} +\log\Big( \frac{1}{\delta}\Big) \Big] \frac{\log^{2.5} \frac{\beta}{ \Delta_{\text{gap}}} \frac{1}{\delta}}{\log \log \frac{\beta}{ \Delta_{\text{gap}} }\frac{1}{\delta} } \right) 
\end{align}
queries to $U_{\hat{H}_0}$ and $U_{\hat{H}_1}$ and 
\begin{align}
O\left( \frac{\beta}{\Delta_{\text{gap}}}  \frac{\log^{1.5} \frac{\beta}{ \Delta_{\text{gap}}} \frac{1}{\delta}}{\log \log \frac{\beta}{ \Delta_{\text{gap}} }\frac{1}{\delta} } \right) 
\end{align}
queries to $U_{\hat{H}'}$. The gate complexity for all three block-encoded unitaries is $O(K L)$. In addition, to prepare the ground state of the $\hat H_{\text{free}}$ we will need only $O(L)$ number of gates. Thus, the total gate complexity for preparing the state adiabatically is 
\begin{equation}
O\left(KL\frac{\beta' \beta}{\Delta^2_{\text{gap}}} \text{polylog}  \Big( \frac{\beta}{\Delta_{\text{gap}}}  \frac{1}{\delta} \Big)\right).
\end{equation}
For the BMN matrix model, $\beta$ and $\beta'$ are also dominated by quartic interaction term (analogous to $|\alpha|$ in Eq.~$\eqref{eq:unitary-scaling}$) and thus, $\beta, \beta' \sim g^2N^4\Lambda^6$. The cubic interaction gives a sub-leading correction of order $\mu gN^3\Lambda^{9/2}$.

In Appendix~\ref{sec:alternative_adiabatic_state_preparation}, we discuss an alternative, more straightforward algorithm based on quantum simulation technique for time-dependent Hamiltonian evolution and the adiabatic theorem. The advantage of the Wan-Kim's fast digital algorithm is that it gives a poly-log scaling in $1/\delta$ error, as opposed to polynomial scaling for the algorithm described in Appendix~\ref{sec:alternative_adiabatic_state_preparation}. 
 
 \subsubsection*{Fuzzy-sphere vacuum }
 Next we consider the fuzzy-sphere states discussed in Sec.~\ref{sec:fuzzy_sphere_in_Hilbert_space}. The starting point is $|J_i ; {\rm VAC}_{\rm free}  \rangle$, which is the ground state of $\hat{H}_{\rm free}^{(J_i )}$. 
 Because $\hat{H}_{\rm free}^{(J_i )}$ is quadratic, it is easy to determine the ground state in terms of $\hat{P}_I, \hat{Y}_i, \hat{X}_a$ and the representation $J_i$. 
 But there is one issue, which is problematic in our specific regularization with the Fock basis:  $\hat{H}_{\rm free}^{(J_i )}$ has the flat direction, along which the ground state is zero momentum state which is not well described by using the truncated Fock space. Practically, we can lift the flat direction by adding a mass term proportional to $\sum_{i,\alpha,\beta}[\hat{G}^\alpha,\hat{Y}_{i}^\beta]^2$. Let this modification be `gauge fixing' term, $\hat{H}_{\rm g.f.}$. Then $\hat{H}_{\rm free}^{(J_i )}+\hat{H}_{\rm g.f.}$ does not have flat directions and the ground state can be expressed as the Fock vacuum of this quadratic Hamiltonian. 
We will use this state, which we denote by $\ket{J_i;{\rm VAC}}_{\rm g.f.}$, for the state preparation.  
In Appendix~\ref{sec:how-to-construct-|J>} we will show how $\ket{J_i;{\rm VAC}}_{\rm g.f.}$ can be expressed in the original basis used for the regularization. 
To obtain $\ket{J_i}$,  we perform the adiabatic state preparation by taking $\ket{J_i;{\rm VAC}}_{\rm g.f.}$  to be the initial state and 
 \begin{align}
\hat{H}_0& = \hat{H}_{\rm free}^{(J_i )}+\hat{H}_{\rm g.f.}, \\
\hat{H}_1 &= \hat{H}_{\rm free}^{(J_i )} +\hat{H}_{\rm int}^{(J_i )}(g)=\hat{H}_{\rm BMN}.  
\end{align}

Now we consider the gauge-invariant state ${\cal S}(|J_i\rangle)$. 
In order to eliminate potential  NG modes (which can be lifted if the Maldacena-Milekhin conjecture is true), we can force the gauge-singlet constraint by adding a term proportional to $\sum_\alpha\hat{G}_\alpha^2$ to the BMN Hamiltonian $\hat{H}$. Then only the gauge-invariant fuzzy-sphere states can be the ground states. 
Hence it looks reasonable to use $\hat{H}_0=\hat{H}_{\rm free}^{(J_i )}+\hat{H}_{\rm g.f.}$ and $\hat{H}_1=\hat{H}_{\rm BMN} + c\sum_\alpha\hat{G}_\alpha^2$ with a positive coefficient $c$. 

Another possible option would be as follows. 
Once we prepare a gauge-invariant state close to ${\cal S}\left(|J_i \rangle\right)$ at weak coupling, the standard adiabatic state preparation can be used to go to strong coupling.
It might be possible to obtain such state by starting with the Fock vacuum, which is gauge-invariant, 
and perform certain adiabatic state preparation compatible with the gauge invariance to obtain a gauge-invariant state close to ${\cal S}\left(|J_i \rangle\right)$, 
for example, by interpolating $\hat{H}_{\rm free}$ and\footnote{
In order to check whether this particular Hamiltonian, or some other choices, are suitable for the adiabatic state preparation, we need to see whether the vacua can be smoothly interpolated, and whether the low-energy spectrum is gapped during the interpolation; see Sec.~\ref{sec:adiabatic_state_preparation}. Such a check can be done via the lattice Monte Carlo simulation. 
}
\begin{eqnarray}
\frac{1}{2}{\rm Tr}P_I^2
+
\frac{1}{2}{\rm Tr}X_a^2
+
{\rm Tr}\left(\sum_i\hat{X}_i^2-\frac{\mu^2s(s+1)}{9g^2}\hat{I}\right)^2
+
{\rm Tr}\left(
[\hat{X}_i,\hat{X}_j]-\frac{i\mu}{3g}\epsilon^{ijk}\hat{X}_k
\right)^2
+
\sum_\alpha\hat{G}_\alpha^2
\nonumber\\
\end{eqnarray}
for coincident fuzzy spheres with spin $s$. We can repeat the adiabatic state preparation once more to get the fuzzy-sphere state in the matrix model, by interpolating this Hamiltonian and the BMN Hamiltonian. 
In order to make sure that the gauge invariance is preserved,  we can add a term like $\sum_\alpha\hat{G}_\alpha^2$ to the Hamiltonian, or we could keep doing measurement using $\hat{G}_\alpha$ such that the charge will thus approximately be preserved during the time evolution. 

The adiabatic state preparation can work when there ground states at $s=0$ and $s=1$ are smoothly connected, without a level-crossing. Further investigation is needed in order to check this property. 
\subsubsection*{Thermofield double state (TFD)}
{}
Another important state is the thermofield double state,
\begin{equation}
\ket{\text{TFD}} = \frac{1}{Z_\beta} \sum_i  e^{-\frac{\beta E_i}{2}} \ket{E_i}_L \ket{E_i}_R, 
\end{equation}
where we have two copies (left and right) of the same system with total Hamiltonian $\hat{H} = \hat{H}_L \otimes \hat{I}_R + \hat{I}_L \otimes \hat{H}_R$, with $\hat{H}_L = \hat{H}_R $ that have eigenvalues $E_i$. There are numerous recent studies for CFT, SYK model, and generic chaotic systems \cite{Cottrell:2018ash, Wu:2018nrn, Maldacena:2018lmt, Alet:2020ehp} that show that, with good accuracy, TFD is realized as the ground state of the two independent copies of the system that are coupled as 
\begin{equation}\label{eq:target}
\hat{H} 
=  \hat{H}_{\rm L} \otimes \hat{I}_{\rm R} + \hat{I}_{\rm L} \otimes \hat{H}_{\rm R} 
+ \frac{g_{LR}}{k}\sum_{i=1}^k \hat{O}^{(i)}_{\rm L}  \hat{O}^{(i)}_{\rm R}. 
\end{equation}
The coupling term can be quite generic, and the only constraint is that operators $\hat{O}^{(i)}_{\rm L,R}$ are local. The strength of the coupling $g_{LR}$
controls the temperature $(1/\beta)$ of the TFD state. 
Analogous to the discussion in the previous section, 
we can construct oracles for Hamiltonian with left and right coupling and without it.\footnote{
The additional coupling term between left and right can be implemented as a coupling of Pauli operators on individual qubits.
} 
And again we can make use of Algorithm 1 and Theorem 1 of Ref.~\cite{wan2020fast} with initial Hamiltonian $\hat{H}_{\rm L} \otimes \hat{I}_{\rm R} + \hat{I}_{\rm L} \otimes \hat{H}_{\rm R}$, 
which has a simple ground state that is the tensor product of two ground states that we already know how to construct operationally. 
The TFD state, on the other hand, is the ground state of the target Hamiltonian Eq.~\eqref{eq:target} 
assuming the results of Refs.~\cite{Cottrell:2018ash, Wu:2018nrn, Maldacena:2018lmt, Alet:2020ehp} will carry through to the matrix model Hamiltonian. 
The state preparation complexity of the thermofield double will be $O(\frac{K L \beta^2}{\Delta_{\text{gap}}^2} \text{polylog} \frac{1}{\Delta_{\text{gap}}} \frac{1}{\delta})$, recall that $L = \Lambda^8 N^4$.

\subsection{Measuring quantum black holes on the quantum devices}
{}
As we have mentioned before, we could construct the `trivial' vacuum and fuzzy sphere vacua in the BMN matrix model. 
The shape of the wave functions describing these vacua is already a highly nontrivial and interesting target of the quantum simulation.  
However, those are just the tip of the iceberg: quantum simulation could do much more than constructing those states. 

In fact, one of our dreams is to probe the black hole dynamics using the quantum simulation of matrix models. For instance, let us consider physics near the trivial vacuum of the BMN matrix model, with sufficiently small $\mu$, i.e.,~close to the BFSS matrix model. 
Suppose the ground state is prepared by using the adiabatic state preparation method. 
By performing a unitary transformation close to the identity, we can add a small amount of energy to the system. 
Then we can follow the unitary time evolution and see how the system thermalizes.  
We expect that the system thermalizes toward the Schwarzschild black hole in M-theory or black zero-brane in type IIA superstring theory, depending on the energy added to the system \cite{Itzhaki:1998dd}. 
Because the quantum simulation allows us to access the quantum state, 
it might be possible to see how the black hole geometry is realized by the matrix degrees of freedom.
One natural possibility is that the Schwarzschild black hole is realized as the partially-deconfined phase~\cite{Hanada:2016pwv,Berenstein:2018lrm,Hanada:2018zxn,Hanada:2019czd,Hanada:2020uvt}.
This possibility may be testable to some extent with the classical simulation as well, as demonstrated in Ref.~\cite{Watanabe:2020ufk} in a simpler matrix model, and hence, may serve as a benchmark for the power of the quantum simulation. 
Depending on the choice of $\mu$ and energy, the black hole can be unstable. It can be a resonance which eventually evaporates by emitting the Hawking radiation. 
It is extremely important to see such formation and evaporation of quantum black hole based on the first principle. 

It is also interesting to consider QFT embedded in the matrix model and study the problems involving holographic scattering and bulk locality. Namely, when the semiclassical gravity dual (`bulk geometry') is expected, it is possible to perform the scattering experiments by shooting some excitations from the boundary towards the bulk. 
Revealing details of such experiments might tell us the existence of bulk locality, especially at the sub-AdS scale~\cite{Maldacena:2015iua} (see also Ref.~\cite{Heemskerk:2009pn,ElShowk:2011ag}).
It might also be useful for the construction of the tensor network toy models of holography in the precision of sub-AdS \cite{Yang:2015uoa,Hayden:2016cfa}.

Finally, we wish to mention more concrete proposals. 

The first example is a motion of D0-brane in the black zero-brane geometry~\cite{Hanada:2021ipb}. 
Low-energy states above the trivial vacuum of the BFSS matrix model can describe the black zero-brane in type-IIA superstring theory\cite{Itzhaki:1998dd}. 
(See Ref.~\cite{Costa:2014wya} for a generalization to the BMN matrix model.) By exciting the $(N,N)$-components of matrices, we can obtain a state which is peaked around $X_{I,ij}=y_I\delta_{ij}$ in the coordinate basis and $P_{I,ij}=q_I\delta_{ij}$ in the momentum basis. Such a state admits an interpretation as a bound state of black zero-brane sitting at the origin of the bulk and a D0-brane located at $\vec{y}=(y_1,\cdots,y_9)$ moving with the momentum $\vec{q}=(q_1,\cdots,q_9)$. 
By monitoring how the peak moves in the coordinate basis and momentum basis via the Hamiltonian time evolution, and by comparing it with the dual gravity picture, we might be able to see how the bulk geometry is encoded in the matrix model. 
A natural possibility at strong coupling and large $N$ is that the peak moves following the Dirac-Born-Infeld action for a probe D-brane~\cite{Maldacena:1997re}.

Another interesting problem, which involves quantum chaos and the Lyapunov exponent. Given that we have an example of the vacua in the BMN matrix model or some other matrix models, it might be interesting to compute the Lyapunov exponents by constructing the following correlation functions on the quantum computer
\begin{align}
\left\langle {\hat{O}_1(t)\hat{O}_2(t')\hat{O}_3(t)\hat{O}_4(t')} \right\rangle 
\end{align} 
in the Heisenberg picture, where the operator $\hat{O}$'s could be chosen as simple gauge-invariant operators.
This is the out-of-time-ordered correlator that could reveal the Lyapunov exponent.
A quantum simulation protocol proposed in Ref.~\cite{Swingle:2016var} can be useful for this purpose. Note that the Hamiltonian time evolution has to be controlled very precisely, that would be a challenging task without quantum error correction.
In the strong coupling limit where Einstein gravity gives a precise dual description, the Lyapunov exponent should satisfy the Maldacena-Shenker-Stanford (MSS) bound \cite{Maldacena:2015waa}.
It is interesting to consider the finite-oupling corrections away from the MSS bound which corresponds to the stringy corrections to the gravity dual\footnote{For finite temperature state, one could use the algorithms discussed in \cite{Cottrell:2018ash,Wu:2018nrn,Zhu:2019bri} to 
prepare thermofield double states.} \cite{Shenker:2013pqa,Shenker:2014cwa,Maldacena:2015waa,Kobrin:2020xms}. 

\section{Conclusion and discussion}\label{sec:conclusion}
{}
In this paper, we discussed how matrix models (especially the BMN matrix model) and some classes of quantum field theories could be realized on a digital quantum computer. 
In Section~\ref{sec:BMN} and Section~\ref{sec:QFT-from-BMN}, 
we illustrated how the matrix model could be realized in the Hamiltonian formulation. 
We introduced an explicit regularization scheme, such that it can be realized on a digital quantum computer with a large but finite number of qubits. 
Then in Section~\ref{sec:Quantum-algorithms}, we discussed the actual implementations.
A very standard encoding prescription led to a rather simple form of the Hamiltonian, 
which allows us to use efficient simulation algorithms based on the block-encoding, qubitization, and quantum signal processing. The minimal number of qubits required to encode the BMN matrix model with U($N$) gauge group and cutoff $\Lambda$ is $9N^2\log_2\Lambda+8N^2$ and our protocol uses additional $\log_2 L$ qubits for $L\sim\Lambda^8 N^4$ ancilla states for block-encoding and qubitization step.  
The circuit complexity for adiabatic state preparation via the fast digital Wan-Kim algorithm of Ref.~\cite{wan2020fast} is 
\begin{equation}
    O\left(\frac{\Lambda^4   (\log \Lambda)  N^4 \beta^2}{\Delta_{\text{gap}}^2} \text{polylog} \frac{1}{\delta \Delta_{\text{gap}}}\right)
\end{equation}
and the circuit complexity for approximation time-independent Hamiltonian evolution unitary via qubitization and quantum signal processing Ref.~\cite{low2016hamiltonian, low2017optimal} is 
\begin{equation}
    O\left(\Lambda^4   (\log \Lambda)  N^4 \Big[ \beta t + \log \frac{1}{\varepsilon}\Big] \right).
\end{equation}
Note that $\delta$ is the error in state preparation, $\varepsilon$ is the error in unitary, and $\Delta_{gap}$ is set by the mass term $\Delta_{gap}\sim \mu$ at small coupling limit.\footnote{In the strong coupling limit, $\Delta_{\text{gap}}$ is some function of $\mu$ and $g$ that is not known analytically. $\beta$ is dominated by the quartic interaction term and is $\beta \sim g^2N^4\Lambda^6$.} The cutoff $\Lambda$ should be chosen appropriately depending on the physics under consideration, and in Sec.~\ref{sec:cutoff-QFT-from-BMN}, we provided a rough discussion on several possible choices. 

Here we comment on possible implications from this work, combined with quantum Church-Turing Thesis. The quantum Church-Turing Thesis states that any physical process that happens in the real world could be simulated in a quantum computer. We could write it in a more formal way:
{\it Any calculation that cannot be done efficiently by a quantum circuit cannot be done efficiently by any physical system consistent with the laws of physics}.\footnote{There are some differences between different technical definitions of this thesis. Here, we use the version such that the word \emph{efficiently} means that the quantum Turing machine could compute the task in certain complexity classes. }
Our work shows that some supersymmetric QFTs which have natural realizations in superstring/M-theory are simulatable, 
and hence, they are not excluded by the quantum Church-Turing Thesis.  
We do not yet know generic supersymmetric QFT, such as the minimal supersymmetric standard model, can be simulated efficiently. 
In principle, most theories can be simulated if the parameter-fine-tuning is allowed, 
and it would be important to understand the computational complexity of the parameter tuning.

In addition to the ones discussed in this paper, what kind of supersymmetric theories can be simulated 
without involving the parameter fine tuning on a digital quantum computer? 
Kaplan, Katz, and Unsal gave a regularization scheme of the spatial lattice with the continuous time for supersymmetric theories with four, eight, and sixteen supercharges in one, two, and three spatial dimensions~\cite{Kaplan:2002wv},  
and showed that theories in one or two spatial dimensions do not require parameter fine tuning.
It is likely that their lattice Hamiltonian can be simulated efficiently on a universal quantum computer.\footnote{
Strictly speaking, their formulation has a subtlety associated with the moduli stabilization problem. 
This is analogous to the problem associated with the flat direction in BFSS matrix model.
There are several ways to resolve this issue, for example we can introduce the flux deformation to tame the instability.
See Ref.~\cite{orbifold-paper} for more details. 
}
By adding the flux deformation to the (1+1)-d theory and taking the fuzzy-sphere vacuum, the (3+1)-d theory might be obtained, as previously done for the theories on the two-dimensional Euclidean lattice~\cite{Hanada:2010kt,Hanada:2010gs,Hanada:2011qx}. 
In those theories, we can discuss the complexities of the calculations in a quantitative manner based on actual regularizations. 
It might be interesting to understand deeper about potential implications of supersymmetry in the context of the quantum Church-Turing Thesis, or discuss their relations between some recent discussions about quantum simulation capabilities and quantum black holes \cite{Bouland:2019pvu,Susskind:2020kti,Kim:2020cds,Yoshida:2020wpd}.

In this paper, we pointed out that the use of the equivalence between the matrix model and QFTs can
simplify the implementation of the latter on a quantum computer. 
Let us take this opportunity to briefly discuss other `non-lattice' approaches in the context of quantum simulation and quantum computation \cite{Preskill:2018fag,cyber}. 
There also exists a similar approach to the method proposed in this paper about simulating quantum field theories using Hamiltonian simulation, which is called conformal truncation. The method of conformal truncation is proposed as an alternative of lattices for studying generic strongly-coupled quantum field theories: unlike simulating field theories by turning on couplings from free theories in the lattice, conformal truncation solves field theories from another side of the RG flow: turning on operators away from conformal field theories. The Hamiltonian we arrive at, in this case, might be non-local. One could discuss quantum simulation based on conformal truncation in Ref.~\cite{Liu:2020eoa}. 
One could also simulate quantum field theories using consistency relations existing already in quantum field theories. This is called the bootstrap approach. One example of the quantum setting for bootstrap problems is developed in Ref.~\cite{bao2019quantum}, which depends on a theoretical speedup of semi-definite programming algorithms in a quantum computer. 

\section*{Acknowledgement}
{}
We thank Dominic Berry, Alex Buser, Andrew Childs, Raghav Jha, 
Isaac Kim, Alexei Kitaev, Andras Gilyen, Alexey Milekhin, John Preskill, Douglas Stanford, Fumihiko Sugino and Kianna Wan for related discussions. 
We specifically thank Yuan Su for his comments on the draft. H.G. is supported by the Simons Foundation through the It from Qubit collaboration. 
M.~Hanada was supported by the STFC Ernest Rutherford Grant ST/R003599/1.
He also thanks Yukawa Institute for Theoretical Physics for hospitality during his stay in the summer of 2020.
M.~Honda is supported by MEXT Q-LEAP.
JL is supported in part by the Institute for Quantum Information and Matter (IQIM), an NSF Physics Frontiers Center (NSF Grant PHY-1125565) with support from the Gordon and Betty Moore Foundation (GBMF-2644), by the Walter Burke Institute for Theoretical Physics, and by Sandia Quantum Optimization \& Learning \& Simulation, DOE Award \#DE-NA0003525.

\appendix
\section{Relation between the Fock basis and the coordinate basis}
\label{sec:coordinate-basis}
{}
So far, we have used the Fock space basis of harmonic oscillators
while there are also other standard bases
such as coordinate basis and momentum basis. 
Here we discuss a relation between the Fock space basis and coordinate basis after the truncation.

\subsubsection*{Single harmonic oscillator}
{}
Let us start with the simple harmonic oscillator without the truncation
\begin{\eq}
\hat{H}_{\rm osc}= \frac{1}{2} \hat{p}^2 +\frac{\omega^2}{2} \hat{x}^2 \quad
{\rm with}\ [\hat{x},\hat{p} ] = i .
\end{\eq}
As well known, the wave function of 
the $n$-th excited state $|n\rangle$ is given by the Hermite polynomial:
\begin{eqnarray}
\phi_n(x)
\equiv \langle x|n\rangle
=
\frac{1}{\pi^{1/4}\sqrt{2^n n!} } H_n( \sqrt{\omega} x ) e^{-\frac{\omega }{2} x^2 } .
\label{x-n-inner-product}
\end{eqnarray}
Therefore the position state $|x\rangle$ is expanded as 
\begin{eqnarray}
|x\rangle
=\sum_{n=0}^\infty \phi_n^\ast(x)|n\rangle .
\label{eq:stateX}
\end{eqnarray}
The wave function satisfies
\begin{\eq}
x\phi_n (x) 
= \frac{1}{\sqrt{2\omega}} \left( \sqrt{n+1}\phi_{n+1}(x) +\sqrt{n}\phi_{n-1} (x) \right) ,
\end{\eq}
which is equivalent to
\begin{\eq}
\hat{x} |n\rangle
= \frac{1}{\sqrt{2\omega}} \left( \sqrt{n+1}|n+1\rangle +\sqrt{n}|n-1 \rangle \right) .
\label{eq:x_n}
\end{\eq}

Now let us make the truncation on the Fock space
corresponding to take only the states $|n \rangle$ with $n=0,1,\cdots ,\Lambda -1$ as basis.
With the truncation, the relation \eqref{eq:x_n} 
for the position operator $\hat{x}=\frac{1}{\sqrt{2\omega}}(a^\dag +a)$
is no longer true for the highest energy state:
\begin{\eq}
\hat{x} |n\rangle
= 
\begin{cases}
\frac{1}{\sqrt{2\omega}} \left( \sqrt{n+1}|n+1\rangle +\sqrt{n}|n-1 \rangle \right) 
& {\rm for}\ n<\Lambda -1 \cr
\frac{\Lambda -1}{\sqrt{2\omega}}  |\Lambda -2 \rangle  
& {\rm for}\ n=\Lambda -1 
\end{cases} .
\end{\eq}
Inspired by the relation \eqref{eq:stateX} before the truncation,
it would be natural to define a regularized version of the ``position state'' as
\begin{eqnarray}
|x \rangle_{\rm reg}
\equiv \frac{\sum_{n=0}^{\Lambda -1} \phi_n^\ast(x)|n\rangle}{\sqrt{\sum_{n=0}^{\Lambda-1} |\phi_n(x)|^2}}~.
\label{eq:x_approx}
\end{eqnarray}
However, this state is not an eigenstate of the operator $\hat{x}$ in a precise sense:
\begin{eqnarray}
\hat{x} |x \rangle_{\rm reg}
= x |x\rangle_{\rm reg} 
-\sqrt{ \frac{\Lambda }{ 2\omega\sum_{n=0}^{\Lambda-1} |\phi_n(x)|^2} }
   \phi_\Lambda^\ast (x) |\Lambda -1\rangle .
\end{eqnarray}
The second term is the deviation from the correct relation
and irrelevant if we are interested in problems 
where the highest energy state $|\Lambda -1\rangle$ is important.
\subsubsection*{Matrix model}
{}
The matrix model has $9N^2$ harmonic oscillators
labelled by $I=1,\cdots,9$ and $\alpha=1,\cdots,N^2$.
Before the truncation, we have the ``coordinate basis'' which satisfies
\begin{eqnarray}
\hat{X}_I^\alpha
|X^\alpha_I\rangle_{I\alpha}
=
X_I^\alpha
|X^\alpha_I\rangle_{I\alpha}. 
\end{eqnarray}
By taking the tensor product, we can define
\begin{eqnarray}
|X\rangle
=
\otimes_{I,\alpha}|X^\alpha_I\rangle_{I\alpha}~,
\end{eqnarray}
which satisfies
\begin{eqnarray}
\hat{X}_I^\alpha
|X\rangle
=
X_I^\alpha
|X\rangle. 
\end{eqnarray}
After the truncation,
we can approximate the ``coordinate basis'' by replacing it 
with the state \eqref{eq:x_approx} for each $I,\alpha$.

\subsubsection*{Regularization in the coordinate basis}
{}
A natural regularization scheme in the coordinate basis is to introduce a `lattice', i.e., to restrict the values of $X_I^\alpha$  to be $n\delta_X$, where $n$ is integer between $\pm n_{\rm b}$. By sending $\delta_X$ and $n_b$ to zero and infinity, respectively, in such a way that $n_{\rm b}\delta_X$ becomes infinite, the original coordinate basis is reproduced. The momentum operator $\hat{P}_I^\alpha$ is approximated by the difference operator. Such a basis is used in Ref.~\cite{Jordan:2011ne} for scalar field theories. We do not expect a big difference from the Fock basis; the Hamiltonian can be expressed as a sum of the Pauli strings anyways, with more or less the same number of the terms. 
\section{Construction of $\ket{J_i;{\rm VAC}}_{\rm g.f.}$}\label{sec:how-to-construct-|J>}
Formally, the `gauge-fixed' Hamiltonian takes the form
\begin{eqnarray}
\hat{H}^{(J_i)}_{\rm free}+\hat{H}_{\rm g.f.}
=
\hat{H}_{(Y)}
+
\hat{H}_{(X)}, 
\end{eqnarray}
where $\hat{H}_{(Y)}$ (resp., $\hat{H}_{(X)}$) contains only $\hat{P}_{i=1,2,3}$ and $\hat{Y}_{i=1,2,3}$ (resp., $\hat{P}_{a=4,\cdots,9}$ and $\hat{X}_{a=4,\cdots,9}$). 
Both of them takes the form 
\begin{eqnarray}
\frac{1}{2}\sum_{A}\hat{P}_A^2
+
\frac{1}{2}\sum_{A,B}M_{AB}\hat{Z}_A\hat{Z}_B, 
\end{eqnarray}where $\hat{Z}$ denotes $\hat{Y}$ or $\hat{X}$, and we used $A$ and $B$ to denote all the indices together ($A,B=1,2,3$ for $\hat{Y}_i$ and $4,5,\cdots,9$ for $\hat{X}_a$). 

The mass matrix $M_{AB}$ is real, symmetric and positive definite, and hence, it can be diagonalized by using the orthogonal matrix $O$ as $M=O^{\rm T}DO$, where $D_{AB}=d_A^2\delta_{AB}$ and  $d_A>0$. Hence, by using $\hat{Z}'_A=\sum_B O_{AB}\hat{Z}_B$ and $\hat{P}'_A=\sum_B O_{AB}\hat{P}_B$, 
$\hat{H}_{(Y)}$ and $\hat{H}_{(X)}$  are written as $\frac{1}{2}\sum_A\left(\hat{P}_A^{\prime 2}+d_A^2\hat{Z}_A^{\prime 2}\right)$. By using the annihilation operator 
$\hat{b}_A\equiv\sqrt{\frac{d_A}{2}}\hat{Z}'_A+\frac{i\hat{P}'_A}{\sqrt{2d_A}}$  and the creation operator $\hat{b}^\dagger_A\equiv\sqrt{\frac{d_A}{2}}\hat{Z}'_A-\frac{i\hat{P}'_A}{\sqrt{2d_A}}$  we can define the Fock states. 

We can relate this Fock basis and another Fock basis based on 
$\hat{a}'_A\equiv\sqrt{\frac{\omega_A}{2}}\hat{Z}'_A+\frac{i\hat{P}'_A}{\sqrt{2\omega_A}}$ and $\hat{a}^{\prime\dagger}_A\equiv\sqrt{\frac{\omega_A}{2}}\hat{Z}'_A-\frac{i\hat{P}'_A}{\sqrt{2\omega_A}}$. 
We take $\omega_A$ to be $\frac{\mu}{3}$ for the $A=1,2,3$ and $\frac{\mu}{6}$ for $A=4,5,\cdots,9$.  
The Fock bacuum $\ket{0}_{b_A}$ which satisfies $\hat{b}_A\ket{0}_{b_A}=0$ can be written as $(\hat{a}'_A+c_A\hat{a}^{\prime\dagger}_A)\ket{0}_{b_A}=0$, where $c_A$ is chosen appropriately such that  $\hat{a}'_A+c_A\hat{a}^{\prime\dagger}_A$ agrees with $\hat{b}_A$ up to a multiplicative factor. Up to an multiplicative factor,  
\begin{equation}
\ket{0}_{b_A}\propto e^{-\frac{1}{2}c_A\hat{a}_A^{\prime\dagger 2}}\ket{0}_{a'_A}, 
\end{equation}
where $\ket{0}_{a'_A}$ is the Fock vacuum of of $\hat{a}'_A$, i.e., $\hat{a}'_A\ket{0}_{a'_A}=0$. 

We introduce $\hat{a}_A$ by $\hat{a}_A\equiv\sqrt{\frac{\omega_A}{2}}\hat{Z}_A+\frac{i\hat{P}_A}{\sqrt{2\omega_A}}$. 
Then, by construction, $\hat{a}'_A=\sum_BO_{AB}\hat{a}_B$, $\hat{a}^{\prime\dagger}_A=\sum_BO_{AB}\hat{a}^\dagger_B$. 
Hence the Fock vacuum of $\hat{a}'_A$$\hat{a}'_A$ and that of $\hat{a}_A$ are the same, in the sense
\begin{equation}
\prod_A \ket{0}_{a'_A} = \prod_A \ket{0}_{a_A}.
\end{equation}
Therefore,  
\begin{equation}
\ket{J_i;{\rm VAC}}_{\rm g.f.}\propto 
e^{-\frac{1}{2}\sum_Ac_A\left(\sum_BO_{AB}\hat{a}^\dagger_B\right)^2}
\left(\prod_A\ket{0}_{a_A}\right).
\end{equation}

Furthermore, because of 
$e^{-\frac{i\mu}{3g}\sum_{i,\alpha}\hat{P}_i^\alpha J_i^\alpha}
\hat{X}_i
e^{\frac{i\mu}{3g}\sum_{i,\alpha}\hat{P}_i^\alpha J_i^\alpha}
=\hat{Y}_i$, the creation and annihilation operators in the original basis, $\hat{A}$ and $\hat{A}^\dagger$, satisfy $\hat{a}_A=e^{-\frac{i\mu}{3g}\sum_{i,\alpha}\hat{P}_i^\alpha J_i^\alpha}
\hat{A}_A
e^{\frac{i\mu}{3g}\sum_{i,\alpha}\hat{P}_i^\alpha J_i^\alpha}$
and
$\hat{a}^\dagger_A=e^{-\frac{i\mu}{3g}\sum_{i,\alpha}\hat{P}_i^\alpha J_i^\alpha}
\hat{A}^\dagger_A
e^{\frac{i\mu}{3g}\sum_{i,\alpha}\hat{P}_i^\alpha J_i^\alpha}$. 
The state $\prod_A\ket{0}_{a_A}$ is written as $\prod_A\ket{0}_{a_A}=e^{-\frac{i\mu}{3g}\sum_{i,\alpha}\hat{P}_i^\alpha J_i^\alpha}\prod_A\ket{0}_{A_A}$. 
This completes the construction of $\ket{J_i;{\rm VAC}}_{\rm g.f.}$:
\begin{eqnarray}
\ket{J_i;{\rm VAC}}_{\rm g.f.}
&\propto& 
e^{-\frac{1}{2}\sum_Ac_A\left(\sum_BO_{AB}\hat{a}^\dagger_B\right)^2}
\left(e^{-\frac{i\mu}{3g}\sum_{i,\alpha}\hat{P}_i^\alpha J_i^\alpha}\prod_A\ket{0}_{A_A}\right)
\nonumber\\
&=&
e^{-\frac{i\mu}{3g}\sum_{i,\alpha}\hat{P}_i^\alpha J_i^\alpha}
e^{-\frac{1}{2}\sum_Ac_A\left(\sum_BO_{AB}\hat{A}^\dagger_B\right)^2}
\left(\prod_A\ket{0}_{A_A}\right). 
\end{eqnarray}
\section{The Wan-Kim Algorithm}\label{sec:summary-wan-kim}
{}

In Sec.~\ref{sec:Quantum-algorithms}, we used Algorithm~1 and Theorem~1 from Ref.~\cite{wan2020fast} as a black box to construct the ground state of the matrix model. 
In this section, we give a brief summary of the Algorithm~1 and key ideas behind the proof of Theorem~1. Algorithm 1 prepares the ground state of $\hat{H}_1$ by simulating the adiabatic evolution via $\hat{H}(s) = (1-s) \hat{H}_0 + s \hat{H}_1$. The algorithm takes the low-level oracles that block-encode Hamiltonians $\hat{H}_0$, $\hat{H}_1$ and $\hat{H}' = \hat{H}_1-\hat{H}_0$ as inputs. Wan and Kim use the machinery of quasi-adiabatic continuation to provide a protocol that converges in polylogarithmic time in target state precision error, which is better than previously known adiabatic protocols. The goal of the construction is to approximate the entire adiabatic evolution unitary $\hat{U}(s)$ generated by quasi-adiabatic continuation operator 
\begin{equation}
\hat{D}(s) = -\int_{-\infty}^{+\infty} dt W(t) e^{i \hat{H}(s) t}  \hat{H}' e^{-i \hat{H}(s) t}  
\end{equation}
where $W(t)$ is an odd function satisfying $W(t) \geq 0$ at $t\ge 0$. 
The main insight of the paper is finding a function $W(t)$ that gives a good bound of the adiabatic error $(|| \ket{\Omega(s)}  - U(s) \ket{\Omega(0)}||)$ and is easy to simulate digitally. Digital simulation complexity is closely related to the complexity of integrating function $W(t)$ on a quantum computer. The function that satisfies both of these requirements is 
\begin{align}
W(t) = 
    \begin{cases}
      \int_{t}^{\infty} dt' w(t'), & t \geq 0 \\
      -\int_{-\infty}^{t} dt' w(t'), & t<0 
    \end{cases}
\end{align}
where 
\begin{align}
w(t) = \frac{\Delta}{\sqrt{2\pi}} \text{exp} \Big( - \frac{\Delta^2 t^2}{2} \Big).    
\end{align}
In Theorem 3, they prove that the adiabatic error is bounded as follows;
\begin{align}
|| \ket{\Omega(\tau)}  - U_{\Delta}(\tau) \ket{\Omega(0)}|| \leq  \int_{0}^{\tau} ds \frac{1}{\gamma(s)} \text{exp} \Big(-\frac{\gamma(s)^2}{2 \Delta^2}  \Big) ||H'(s)||
\end{align}
where $\gamma(s)$ denotes the gap of the Hamiltonian $H(s)$, 
$\Delta$ is additional control parameter
and $U_{\Delta}(\tau)$ is the ordered exponential of $\hat{D}(s)$.
In Appendix~C, the authors provide a low complexity circuit for performing the discrete integral of the function $W(t)$. They define the $\log M$-qubit state  
\begin{align}
\ket{W_{\Delta, T, M}} = \sum_{n=1}^M \sqrt{W_n} \ket{n}    
\end{align}
where
\begin{align}
W_n = \frac{1}{\mathcal N_{\Delta, T}} \int_{(n-1)T/M}^{nT/M} dt W(t) \quad \text{and} \quad \mathcal N_{\Delta, T} = \int_0^T dt W(t).
\end{align}
They use a set of controlled rotation operators 
to build a circuit ($\widehat W_{T, M}$) that prepares $\ket{W_{\Delta, T, M}}$ such that for given $\epsilon_2$,  
\begin{align}
|| \ket{W_{\Delta, T, M}} - \widehat W_{T, M} \ket{0} ||    \leq \epsilon_2 
\end{align}
(Lemma 7, \cite{wan2020fast}). They prove that $\widehat W_{T, M}$ can be implemented with $O([\Delta^2 T^2 + \log (N/\epsilon_2)]^2)$ ancilla qubits and total number of $O([\Delta^2 T^2 + \log (N/\epsilon_2)]^2 \log M)$ gates. With those properties and low-level oracles for $H_1$, $H_0$ and $H'$ authors successfully construct the block-encoding unitary for quasi-adiabatic continuation operator $D(s)$, which is the final ingredient needed to prove Theorem~1.

\section{Adiabatic state preperation, LCU decomposition, and oracles}
\label{sec:alternative_adiabatic_state_preparation}
{}

\subsection{Naive adiabatic state preparation}
\label{sec:naive_adiabatic_state_preparation}
{}
We start by reviewing how the simplest adiabatic state preparation works based on the adiabatic theorem.

The Hamiltonian consists of the free part and the interaction part,  
\begin{align}
\hat{H} = \hat{H}_{\rm{free}} + \hat{H}_{\rm{int}}(g)~,
\end{align}
where $\hat{H}_{\rm{int}}$ disappears when the coupling constant $g$ is set to zero. 
The free part $\hat{H}_{\rm{free}}$ does not contain $g$. 
For the adiabatic state preparation, the coupling $g$ to depend on time $t$, and define $\hat{H}(t)$ by 
\begin{align}
\hat{H}(t) = \hat{H}_{\rm{free}} + \hat{H}_{\rm{int}}(g(t)). 
\label{time-dependent-Hamiltonian-BMN}
\end{align}

In the weak-coupling limit $g\to 0$, we know the ground state precisely.\footnote{
Note that we cannot take $g$ to be exactly zero if we consider nontrivial a fuzzy sphere vacuum 
$\hat{X}^\alpha_{1,2,3}|{\rm fuzzy\ sphere}\rangle
\simeq
\frac{\mu}{3g}J^\alpha_{1,2,3}|{\rm fuzzy\ sphere}\rangle$.  
}
We take $g(t=t_{\rm init})$ to be parametrically small, 
and take the initial state of the simulation to be the analytically-known ground state 
which we denote by $\ket{\Omega(t_{\rm init})}$. 

From $t=t_{\rm init}$ to $t=t_{\rm fin}$, we gradually increase the coupling $g(t)$, 
such that the final value $g(t=t_{\rm fin})$ becomes the value we are interested. 
Then, if the change is sufficiently slow, the state at time $t$ is the ground state of $\hat{H}(t)$, due to the adiabatic theorem. 

More precisely, the difference between the ground state of $\hat{H}(t_{\rm fin})$ denoted by $\ket{\Omega(t_{\rm fin})}$
and the actual state $\ket{\psi (t_{\rm fin})}$ is 
\begin{eqnarray}
\left\| {\left| {\psi \left( {{t_f}} \right)} \right\rangle \left\langle {\psi \left( {{t_f}} \right)} \right| - \left| {\Omega \left( {{t_f}} \right)} \right\rangle \left\langle {\Omega \left( {{t_f}} \right)} \right|} \right\|\sim \left| {\frac{1}{{\Delta _{{\rm{gap }}}^2}}\frac{{d\hat{H}(t)}}{{dt}}} \right|\sim \left| {\frac{1}{{\Delta t\Delta _{{\rm{gap }}}^2}}\frac{{d\hat{H}(s)}}{{ds}}} \right|~,
\end{eqnarray}
where $\Delta_\text{gap}$ is the scale of the mass gap of the Hamiltonian during the whole time-dependent process, 
$\Delta t=t_{\rm fin}-t_{\rm init}$, 
and $s=(t-t_i)/\Delta t$. (We recommend a nice review \cite{albash2018adiabatic} about this theorem.)

Then the remaining problem is if we can perform the time-evolution with the time-dependent Hamiltonian $\hat{H}(t)$ on the quantum computer. 
For local Hamiltonians, a typical method is to use the product formula to decompose local terms 
and write a factorized product of the exponential (see one of the earliest papers \cite{uni} and a recent paper \cite{tro}). 
One of the most celebrated applications of this algorithm is the Jordan-Lee-Preskill algorithm, 
where they use it to study adiabatic state preparation in the $\lambda \phi^4$ theory in general dimensions; 
see also Refs.~\cite{jordan2016black,Jordan:2011ne,Jordan:2011ci,jordan2014quantum,jordan2017fast,moosavian2018faster,jordan2018bqp,moosavian2019site,chakraborty2020digital}. 

The naive Trotterization is not suitable for the current setup, due to the lack of a manifestly local Hamiltonian in the computational basis. 
Thus, we consider alternative algorithms (a similar situation was discussed in the Hamiltonian truncation formalism of quantum field theories, see \cite{kreshchuk2020quantum,Liu:2020eoa}). In quantum information science, there are alternative algorithms for non-local Hamiltonian evolution, mostly designed for quantum chemistry in the near-term or long-term simulation (for a review, see Ref.~\cite{mcardle2020quantum}).

\subsubsection{Truncated Taylor-series method and linear combination of unitaries (LCU)}\label{sec:truncated-Taylor-Series}
{}
Here, we discuss a relatively simple implementation of the Hamiltonian simulation based on the truncated Taylor-series based on input models of the linear combination of unitaries (LCU), namely, the paper \cite{berry2015simulating}. The discussion is very friendly to people who are not familiar with quantum simulation algorithms. 

We first explain the time-independent-Hamiltonian version and then modify it to the time-dependent-Hamiltonian version. 
That the Hamiltonian is easily expressed as a sum of Pauli strings (as we saw in Sec.~\ref{sec:compact-mapping}) makes the simulation straightforward. 
\subsubsection*{Time-independent-Hamiltonian version}
{}
In this method, the time evolution is written as
\begin{eqnarray}
e^{-i\hat{H}t}
=
\left(
e^{-i\hat{H}t/r}
\right)^r, 
\label{Taylor-series-truncation-1}
\end{eqnarray}
and $e^{-i\hat{H}t/r}$ is approximated by truncating the Taylor expansion at some order, 
\begin{eqnarray}
e^{-i\hat{H}t/r}
\simeq
\sum_{k=1}^K
\frac{1}{k!}\left(\frac{-i\hat{H}t}{r}\right)^k. 
\label{Taylor-series-truncation-2}
\end{eqnarray}
The value of positive integer $r$ is chosen later, in such a way that a technical assumption \eqref{eq:s=2} needed for an efficient method is satisfied. 

In the BMN matrix model, the Hamiltonian is written as a sum of Pauli strings $\hat{\Pi}_i$ as eq.\eqref{Hamiltonian-as-sum-of-Pauli-chains}. 
By substituting \eqref{Hamiltonian-as-sum-of-Pauli-chains} to \eqref{Taylor-series-truncation-2} and then plugging it into \eqref{Taylor-series-truncation-1}, 
we can rewrite the write hand side of \eqref{Taylor-series-truncation-1} as a sum of the products of the Pauli strings. 
The products of Pauli strings are again Pauli strings. 
Hence $e^{-i\hat{H}t/r}$ can be expressed as 
\begin{eqnarray}
e^{-i\hat{H}t/r}
\simeq
\sum_{i=1}^{m}\beta_i\hat{\tilde{\Pi}}_i, 
\qquad
\beta_i>0, 
\label{time-evolution-as-some-of-Pauli-strings}
\end{eqnarray}
where $\hat{\tilde{\Pi}}$'s are again Pauli strings. Note that this decomposition depends on $t$, $r$ and $K$. 
The values of $\beta_i$'s can be evaluated without using a quantum computer, and they are used as a part of the inputs for the quantum simulation.  
We choose $r$ so that the value of $s$ defined by 
\begin{eqnarray}
s
\equiv
\sum_{i=1}^m\beta_i
=
\sum_{k=0}^K\frac{1}{k!}\left(\frac{t}{r}\sum_{i=1}^L\alpha_i\right)^k ~,
\end{eqnarray}
becomes $2$:
\begin{eqnarray}
s=2.
\label{eq:s=2}
\end{eqnarray} 
When $K$ is sufficiently large, this is equivalent to $\frac{t}{r}\sum_{i=1}^L\alpha_i= \log 2$ with a good precision. 
This condition can always be satisfied, by allowing the identity $\hat{I}$ as one of Pauli string $\hat{Pi}_i$ in \eqref{Hamiltonian-as-sum-of-Pauli-chains}
and tuning its coefficient $\alpha_i$.\footnote{
This only shifts the zero-point of the energy. Because the theory under consideration does not have gravity (though in the dual description, gravity does exist!), such shift does not affect physics. 
}

In order to utilize this decomposition, we introduce ancilla states $\ket{i}$ ($i=1,\cdots,m$), and define an operator $\hat{V}$ acting on $\mathbb{C}^m$ times the Hilbert space as 
\begin{eqnarray}
\hat{V}\left(
\ket{i}\otimes\ket{\psi}
\right)
=
\ket{i}\otimes\left(\hat{\tilde{\Pi}}_i\ket{\psi}\right). 
\label{truncated-Taylor-def-of-V}
\end{eqnarray}
This operator $\hat{V}$ is unitary because the Pauli strings are unitary. 
We also prepare a state $\ket{B}=\sum_{i=1}^m\sqrt{\frac{\beta_i}{s}}\ket{i}$, where $s=2$ by assumption, 
and unitary-and-Hermitian operator $\hat{R}=2\ket{B}\bra{B}-\hat{I}$. Then, we can check that  
\begin{eqnarray}
-\bra{B}\hat{V}\hat{R}\hat{V}^\dagger\hat{R}\hat{V}\ket{B}
\simeq
\left(
\frac{3}{s}-\frac{4}{s^3}
\right)e^{-i\hat{H}t/r}
=
e^{-i\hat{H}t/r}.  
\end{eqnarray} 
Therefore, for any state $\ket{\psi}$, 
\begin{eqnarray}
-\hat{V}\hat{R}\hat{V}^\dagger\hat{R}\hat{V}\left(\ket{B}\otimes\ket{\psi}\right)
\simeq
\ket{B}
\otimes
\left(
e^{-i\hat{H}t/r}
\ket{\psi}
\right),   
\end{eqnarray}
and
 \begin{eqnarray}
\left(
-\hat{V}\hat{R}\hat{V}^\dagger\hat{R}\hat{V}
\right)^r
\left(\ket{B}\otimes\ket{\psi}\right)
\simeq
\ket{B}
\otimes
\left(
e^{-i\hat{H}t}
\ket{\psi}
\right).    
\end{eqnarray}
\subsubsection*{Time-dependent-Hamiltonian version}
{}
The generalization to the time-dependent Hamiltonian is tedious but straightforward. 
$e^{-i\hat{H}t/r}$ should be replaced with the Dyson series
\begin{eqnarray}
{\cal T}e^{-i\int_{(n-1)t/r}^{nt/r} dt\hat{H}(t)}
\qquad
n=1,2,\cdots,r, 
\end{eqnarray}
and \eqref{Taylor-series-truncation-2} is replaced by 
\begin{eqnarray}
{\cal T}e^{-i\int_{(n-1)t/r}^{nt/r} dt\hat{H}(t)}
\simeq
\sum_{k=1}^K
\frac{(-i)^k}{k!}
\int dt_1\cdots dt_k
{\cal T}\left(
\hat{H}(t_1)\cdots \hat{H}(t_k)
\right). 
\end{eqnarray}
Here ${\cal T}$ stands for the time ordering, i.e.,~operators at a later time come left.  
Because the time-dependence is only in the coupling constant $g(t)$
in \eqref{time-dependent-Hamiltonian-BMN}, 
this can be solved before using a quantum computer, and the same form as 
\eqref{time-evolution-as-some-of-Pauli-strings}, with different values of the coefficients $\beta_i$, can be obtained. 

A full discussion about the time-dependent Hamiltonian simulation algorithms is included in \cite{berry2019time}, where the main theorem is given in Theorem 10'. 
\subsubsection{Utilizing the oracles and sparseness}\label{sec:Oracle-based-algorithm}
{}
The truncated Taylor series method explained in Sec.~\ref{sec:truncated-Taylor-Series} does not fully take advantage of the sparseness of the Hamiltonian; 
as we can see from Sec.~\ref{sec:compact-mapping}, there are many cancellations among Pauli strings such that only a small number of nonzero entries remain, 
but if we treat each Pauli string separately as in \eqref{truncated-Taylor-def-of-V}, this cancellation is not utilized at all. 
By using another decomposition of the Hamiltonian, a more efficient simulation might be achieved. 
Then, in principle,  how efficient can the simulation be? 
In order to answer this question, one has to make some assumptions regarding the available oracles.  
An example of the oracles is $\hat{V}$ in \eqref{truncated-Taylor-def-of-V}; one assumes the existence of some oracles
with which the sparseness can be fully utilized, and count the number of the gates and oracles necessary for the simulations.   

Here, we discuss another input model that could manifest the sparseness. There has been substantial effort toward efficient quantum simulation algorithms utilizing the sparseness of the Hamiltonian. 
As one of such algorithms, we introduce the rescaled Dyson series method with sparse oracles designed in Refs.~\cite{berry2019time}. We will only provide the statement, and we provide some additional information in the appendix~\ref{sec:Berry-et-al-algorithm} for self-completeness. The following algorithm takes advantage of sparsity, which is discussed before, pretty generic in matrix models.

We start with the definition of sparsity. 
For a given Hamiltonian $H$, the sparsity of the Hamiltonian $d$ is the maximal number of non-zero entries it could have in any row or column.
As we have seen in Sec.~\ref{sec:sparseness}
the sparsity is $d\sim N^4$ in the BMN matrix model. 

Then the statement proven in Ref.~\cite{berry2019time}, 
with some standard assumptions regarding the oracles (see the appendix \ref{sec:Berry-et-al-algorithm} for details), is 
\begin{theorem}[Efficient simulation for time-dependent sparse Hamiltonians]
Suppose that the time-dependent Hamiltonian $\hat{H}(t)$ is $d$-sparse during the whole time $[t_{\rm init},t_{\rm fin}]$.
Then, there exists an algorithm \cite{berry2019time} such that  
\begin{align}
O\left( {d{{\left\| H \right\|}_{\max ,1}}\frac{{\log \left( {d{{\left\| H \right\|}_{\max ,1}}/\epsilon } \right)}}{{\log \log \left( {d{{\left\| H \right\|}_{\max ,1}}/\epsilon } \right)}}} \right)
\end{align}
queries towards the oracles, and
\begin{align}
\tilde{O}\left( {d{{\left\| H \right\|}_{\max ,1}}n_q} \right)
\end{align}
2-qubit gates are used with error $\epsilon$. 
\end{theorem}
This theorem is the Theorem 10 of Ref.~\cite{berry2019time}. Here $n_q$ is the number of qubits in the Hilbert space, and 
${\left\| H \right\|_{\max ,1}}$ is the one-norm (defined in the $L^1$ space of the Hilbert space) for the maximal matrix element of the Hamiltonian, 
which is defined by\footnote{The notation ${\left\| A \right\|_{p,q}}$ means that for a time-dependent matrix ${A(\tau )}$, we have ${\left\| A \right\|_{p,q}} \equiv {\left( {\int {d\tau } {{\left( {{\mathop{\rm Tr}\nolimits} \left( {{{\left| {A(\tau )} \right|}^p}} \right)} \right)}^{q/p}}} \right)^{1/q}}$.}
\begin{align}
{\left\| H \right\|_{\max ,1}} \equiv \int_{t_{\rm init}}^{t_{\rm fin}} dt\ {\rm max}_{j,k}\left(\left| H_{jk}(t) \right|\right). 
\end{align}
The notation $\tilde{O}$ means that we are ignoring logarithmic factors. This algorithm is sufficient for us when doing time evolution for our adiabatic state preparation purpose. 

Finally, it is worth notice that although we only have logarithmic dependence about precision during time evolution, there is polynomial dependence of adiabatic errors when applying the algorithm towards adiabatic state preparation. The latter might be significantly improved according to the work by Wan and Kim \cite{wan2020fast}, which we have discussed in the main text.

\subsection{Hamiltonian simulation based on the rescaled Dyson series}
\label{sec:Berry-et-al-algorithm}
{}
Here we discuss some details about the algorithm appearing in Appendix.~\ref{sec:Oracle-based-algorithm}. Let us start by repeating the definition of the sparsity:
\begin{definition}
For a given Hamiltonian $H$, the sparsity of the Hamiltonian $d$ is the maximal number of non-zero entries it could have in any row or column.
\end{definition}
\noindent
For the BMN matrix model in the Fock basis, the sparsity is $d\sim N^4$. 

Now, we start introducing our oracles. We assume that $H(\tau)$ is at most $d$-sparse at any $\tau \in [0,T]$.
(This is actually the case in Sec.~\ref{sec:adiabatic_state_preparation}.) 
We consider a set of basis states with four labels $\ket{\tau,j,k,z}=\ket{\tau}\ket{j}\ket{k}\ket{z}$, 
where $\tau$ is the (discretized) time, $j$ and $k$ are integers which run from 1 to ${\rm dim}{\cal H}$ where ${\cal H}$ is the Hilbert space of the system under consideration
(and hence the Hamiltonian is a ${\rm dim}{\cal H}\times{\rm dim}{\cal H}$ matrix), and $z\in\mathbb{C}$ is a complex number which is expressed by using binaries with some accuracy.
We define two oracles ${\cal O}_{\rm loc}$ and ${\cal O}_{\rm val}$ to help us access the Hamiltonian. 
They are defined by\footnote{
If we keep all labels explicitly, ${\cal O}_{\rm loc}\left(\ket{\tau}\ket{j}\ket{s}\ket{z}\right)  = \ket{\tau}\ket{j}\ket{{\rm col}(j,s)}\ket{z}$. 
} 
\begin{eqnarray}
{\cal O}_{\rm loc}\left(\ket{j}\ket{s}\right)  = 
\ket{j}\ket{{\rm col}(j,s)}~,
\end{eqnarray}
and 
\begin{eqnarray}
{\cal O}_{\rm val}\left| {\tau ,j,k,z} \right\rangle  = \left| {\tau ,j,k,z \oplus {H_{jk}}(\tau )} \right\rangle ~.
\end{eqnarray}
Here, the notation $\text{col}(j,s)$ is used to denote the location at the $j$-th row and the $s$-th non-zero element in it. 
(It is used only for $1\le s\le d$.)
Hence ${\cal O}_{\rm loc}$ tells us the location of nonzero elements in the Hamiltonian, 
and ${\cal O}_{\rm val}$ tells us the actual values of the nonzero elements. 

For further optimization, extra oracles ${\cal O}_{\rm norm}$ and ${\cal O}_{\rm var}$ that are closely related to the rescaled Dyson series are used. 
The former is defined as 
\begin{eqnarray}
{\cal O}_{\rm norm}
\left(
\ket{\tau}\ket{z}
\right)
= 
\ket{\tau}\ket{z\oplus\left\| H(\tau) \right\|_{\max}},  
\end{eqnarray}
where $||H(\tau)||_{\rm max}\equiv{\rm max}_{jk}|H_{jk}(\tau)|$. 
This is used to compute the max-norm.
The latter requires another state $\ket{w}$, with $w\in\mathbb{C}$. 
By using 
\begin{align}
w = f(t) \equiv \int_0^t d \tau \left\| H(\tau) \right\|_{\rm max},  
\end{align}
it is defined by 
\begin{eqnarray}
&{\cal O}_{\rm var}
\left(
\ket{w}\ket{z}
\right)
 = \ket{w}\ket{z \oplus {f^{-1}}(w)}.  
\end{eqnarray}

The cost and feasibility of the actual implementation of such oracles depend on the Hamiltonian and the architecture. 
Given that the matrix models have rather simple Hamiltonians, it would not be unrealistic to assume the existence of such oracles
as a starting point for the discussion of efficient quantum simulations. For constructions of some oracles, see \cite{childs2017toward}, G4 and \cite{babbush2018encoding}, III A and III B. 

The analysis of query complexity includes the determination of how many queries we address for given oracles, and moreover, how many additional gates we need in this process. 
By using the oracles defined above, the following theorem can be shown: 
\begin{theorem}
(Theorem 10 in \cite{berry2019time}) 
Suppose that the time-dependent Hamiltonian $H(\tau)$ acting on the $2^{n_q}$-dimensional Hilbert space consisting of $n_q$ qubits is $d$-sparse. 
Suppose also that there is an upper bound on the max-norm, ${\left\| {H\left( {{f^{ - 1}}(\varsigma )} \right)} \right\|_{\max }}$, that is positive, and continuously differentiable. Then, there exists an algorithm such that the Hamiltonian evolution could be simulated using 
\begin{align}
O\left( {d{{\left\| H \right\|}_{\max ,1}}\frac{{\log \left( {d{{\left\| H \right\|}_{\max ,1}}/\epsilon } \right)}}{{\log \log \left( {d{{\left\| H \right\|}_{\max ,1}}/\epsilon } \right)}}} \right)
\end{align}
queries towards $\mathcal{O}_{\operatorname{loc}}$, $\mathcal{O}_{\operatorname{val}}$, $\mathcal{O}_{\operatorname{var}}$ and $\mathcal{O}_{\operatorname{norm}}$, and we also need to use 
\begin{align}
\tilde{O}\left( {d{{\left\| H \right\|}_{\max ,1}}n_q} \right)
\end{align}
additional gates. By $\tilde{O}$ we mean an order estimate up to logarithmic corrections, and by $\epsilon$ we mean the given error. 
\end{theorem}
\noindent
Here we used the notation $||H||_{{\rm max},1}\equiv\int_0^td\tau||H(\tau)||_{\rm max}$. 
This theorem (Theorem 10 in \cite{berry2019time}) is proved using a time-dependent version of the truncated Taylor series. Some related ideas include \cite{spar,berry2015simulating,kieferova2019simulating,low2018hamiltonian}. 
\section{Quantum signal processing}\label{sec:QSP}
Following Refs.~\cite{Babbush:2018mlj,low2017optimal},  
we review the quantum signal processing method,  
which was mentioned in Sec.~\ref{sec:QSP-real-time}.  
\subsection{Proof of \eqref{eq:chebyshev}}\label{sec:chebyshev-proof}
{}
The $n$-th order Chebyshev polynomial of the first kind is defined by 
\begin{eqnarray}
T_n(x)=\cos(nt), 
\qquad
x=\cos t. 
\end{eqnarray}
It is straightforward to check a recurrence formula
\begin{eqnarray}
T_{n+1}(x)
=
2xT_n(x)
-
T_{n-1}(x), 
\end{eqnarray}
and
\begin{eqnarray}
T_0(x)=1,
\qquad
T_1(x)=x, 
\qquad
T_2(x)=2x^2-1. 
\end{eqnarray}

Let us prove \eqref{eq:chebyshev} by using the mathematical induction. 
We can directly check \eqref{eq:chebyshev} for $n=0$ and $n=1$. 
Suppose \eqref{eq:chebyshev} holds for $n-1$ and $n$. Then, for $n+1$, we can check \eqref{eq:chebyshev} as follows. 
Firstly, note that
\begin{eqnarray}
\langle G|\hat{R}=\langle G|. 
\end{eqnarray}
Furthermore, by using $\hat{U}^2=\sum_i\ket{i}\bra{i}$, 
we obtain $\hat{U}^2 \ket{G}=\ket{G}$ and $ \bra{G}\hat{U}^2=\bra{G}$. 
By using them, we can show \eqref{eq:chebyshev} for $n+1$ as follows:
\begin{eqnarray}
\bra{G}\hat{W}^{n+1}\ket{G}
&=&
\bra{G}\hat{R}\hat{U}\hat{R}\hat{U}\hat{W}^{n-1}\ket{G}
\nonumber\\
&=&
\bra{G}\hat{R}\hat{U}
\left(
2\ket{G}\bra{G}-\hat{I}
\right)
\hat{U}\hat{W}^{n-1}\ket{G}
\nonumber\\
&=&
2\bra{G}\hat{R}\hat{U}\ket{G}\bra{G}\hat{U}\hat{W}^{n-1}\ket{G}
-
\bra{G}\hat{R}\hat{U}^2\hat{W}^{n-1}\ket{G}
\nonumber\\
&=&
2\bra{G}\hat{R}\hat{U}\ket{G}\bra{G}\hat{R}\hat{U}\hat{W}^{n-1}\ket{G}
-
\bra{G}\hat{U}^2\hat{W}^{n-1}\ket{G}
\nonumber\\
&=&
2\bra{G}\hat{W}\ket{G}\bra{G}\hat{W}\hat{W}^{n-1}\ket{G}
-
\bra{G}\hat{W}^{n-1}\ket{G}
\nonumber\\
&=&
2\frac{\hat{H}}{\lambda}
T_n\Big(\frac{\hat{H}}{\lambda}\Big)
-
T_{n-1}\Big(\frac{\hat{H}}{\lambda}\Big)
\nonumber\\
&=&
T_{n+1}\Big(\frac{\hat{H}}{\lambda}\Big). 
\end{eqnarray}
Therefore, \eqref{eq:chebyshev} holds for any $n$.

\subsection{Quantum signal processing}
{}
By combining \eqref{eq:chebyshev} and the Jacobi-Anger expansion \eqref{Jacobi-Anger}, we obtain 
\begin{equation}
e^{-i\hat{H}t} 
=
\bra{G}
\left(
J_0(-\lambda t) + 2 \sum_{n=1}^\infty i^n  J_{n}(-\lambda t) \hat{W}^{n}
\right)
\ket{G}
\equiv
 \bra{G}
f(\hat{W})
\ket{G}. 
\label{Jacobi-Anger-W}
\end{equation}
The ancilla state $\ket{G}$ can easily be realized. 
Hence, if $f(\hat{W})$ can be realized efficiently, 
the Hamiltonian time evolution can be simulated. 
The quantum signal processing \cite{low2017optimal} provides us with a black box to construct $f(\hat{W})$ 
when $\hat{W}$ is provided as an input.

\subsubsection{1-qubit signal processing}
{}
Let us consider a $2\times 2$ special unitary matrix acting on a qubit, 
which can be written as  
\begin{equation}
\hat{V}(\theta) = A(\theta)\textbf{1} + i B(\theta)\sigma_z + i C(\theta)\sigma_x+  i D(\theta)\sigma_y~,
\end{equation} 
with real coefficients $A(\theta), B(\theta), C(\theta)$ and $D(\theta)$. 
We assume those functions are periodic and the period is $2\pi$, namely $\hat{V}(\theta)=\hat{V}(\theta+2\pi)$. 
We want to find a systematic way to construct such operator $\hat{V}(\theta)$, when the angle $\theta$ is given as the input. 

Any function of this form can be approximated by considering a product of sufficiently many operators of the following form:
\begin{equation}
\hat{R}_{\phi}(\theta) = e^{-i\frac{\phi}{2} \sigma_z}  e^{-i\theta \sigma_x} e^{+ i\frac{\phi}{2} \sigma_z}.  
\end{equation} 
Namely, by choosing sufficiently many parameters $\phi_1,\cdots,\phi_n\in{\mathbb R}$, 
any $2\times 2$ special unitary matrix $\hat{U}$ can be approximated: 
\begin{eqnarray}
\hat{V}(\theta)
\simeq
\hat{R}_{\phi_n}(\theta)
\hat{R}_{\phi_{n-1}}(\theta)
\cdots
\hat{R}_{\phi_1}(\theta). 
\end{eqnarray}
Once the functions $A,B,C$, and $D$ are given, and the number of $\hat{R}_\phi$'s used for the approximation is fixed, 
we can determine the parameters $\phi_1,\cdots,\phi_n$ by using a classical computer. 
\subsubsection{Hamiltonian time evolution}
{}
Let $\ket{w}$ be an eigenstate of $\hat{W}$ with eigenvalue $w=e^{i\theta}$, i.e.,~$\hat{W}\ket{w}=w\ket{w}=e^{i\theta}\ket{w}$. 
Then we want to construct an operator $\hat{V}$ which sends $\ket{w}$ to $f(w)\ket{w}$,
where $f(w)=J_0(-\lambda t) + 2 \sum_{n=1}^\infty i^n  J_{n}(-\lambda t) w^n$. 
Then, because of \eqref{Jacobi-Anger-W}, such $\hat{V}$ can be used to obtain $e^{-i\hat{H}t}$. 
For that purpose, we introduce a control qubit $\ket{b}$ ($b=0,1$), and define a special controlled-$W$ gate $\widehat{\text{CW}}$ as
\begin{eqnarray}
\widehat{\text{CW}}: 
\ket{0}\otimes\ket{w}\mapsto w^{-1}\ket{0}\ket{w}, 
\qquad 
\ket{1}\otimes\ket{w}\mapsto w\ket{1}\otimes\ket{w}, 
\end{eqnarray}
or equivalently,\footnote{
Note that 
$\hat{W}=\hat{R}\hat{U}$,  
$\hat{W}^{-1}=\hat{U}\hat{R}$, 
$\hat{R}^{-1}=\hat{R}$ and $\hat{U}^{-1}=\hat{U}$. 
} 
\begin{eqnarray}
\widehat{\text{CW}}: 
\ket{0}\otimes\ket{\psi}\mapsto \ket{0}\otimes(\hat{W}^{-1}\ket{\psi}), 
\qquad 
\ket{1}\otimes\ket{\psi}\mapsto \ket{1}\otimes(\hat{W}\ket{\psi}). 
\end{eqnarray}
We also use the Hadamard gate 
\begin{eqnarray}
\widehat{\text{Had}}
=
\frac{1}{\sqrt{2}}
\left(
\begin{array}{cc}
1 & 1\\
1 & -1
\end{array}
\right)~,
\end{eqnarray}
acting on the control qubit. 
Then, we define an operator similar to $\hat{R}_\phi(\theta)$ used in the 1-qubit quantum signal processing: 
\begin{equation}
\hat{R}_{\phi}
\equiv
e^{-i\frac{\phi}{2} \sigma_z}  
\cdot
\widehat{\text{Had}}
\cdot
\widehat{\text{CW}}
\cdot
\widehat{\text{Had}}
\cdot
 e^{+ i\frac{\phi}{2} \sigma_z}.  
\end{equation} 
Note that we do not have $\theta$ in the definition; 
$\theta$ is picked up by $\widehat{\text{CW}}$. 
When this operator acts on $\ket{b}\otimes\ket{w=e^{i\theta}}$, 
we obtain 
\begin{equation}
\hat{R}_{\phi}
(\ket{b}\otimes\ket{w=e^{i\theta}})
=
(\hat{R}_{\phi}(\theta)\ket{b})\otimes\ket{w=e^{i\theta}}. 
\end{equation} 

Let us take the parameters $\phi_1,\cdots,\phi_n\in{\mathbb R}$ in such a way that $A(\theta)+iB(\theta)\simeq f(w=e^{i\theta})$. 
We define an operator $\hat{V}$ as 
\begin{eqnarray}
\hat{V}
\equiv
\bra{b=0}
\hat{R}_{\phi_n}
\hat{R}_{\phi_n-1}
\cdots
\hat{R}_{\phi_1}
\ket{b=0}. 
\end{eqnarray}
Then
\begin{eqnarray}
\hat{V}:
\ket{w}
\mapsto
f(w)\ket{w}. 
\end{eqnarray}
It holds for any $\ket{w}$, hence 
\begin{eqnarray}
\hat{V}
=
f(\hat{W}). 
\end{eqnarray}
By combining it with \eqref{Jacobi-Anger-W}, we obtain $e^{-i\hat{H}t}$.


\section{A short introduction on Hamiltonian simulation algorithms}\label{overview}
{}
In this appendix, we give a short introduction and overview of the existing algorithms about the Hamiltonian simulation. One could read Ref.~\cite{mcardle2020quantum} for some more detailed discussions.

As we mentioned before, perhaps the simplest way for simulating Hamiltonian evolution is through Trotterization, namely, to use the Lie-Trotter-Suzuki formula, or the product formula. The idea about Trotterization is that one could decompose the whole Hamiltonian towards local terms and simulate them separately for a relatively short time. This requires us to divide the total time towards short time steps. The more time steps we use, the more accurate result we obtain, but the more gates we need. There are many historical discussions along the line of product formulas, see, for instance, Refs.~\cite{uni,suzuki1976trotter,PhysRevA.78.052325,tro,csahinouglu2020hamiltonian}. An important type of improvement from the original product formula is to make use of randomization. See for instance, Refs.~\cite{childs2018faster,campbell2018random,ouyang2019stochastichamiltonian,chen2020quantum}. Moreover, one can also formulate a time-dependent version of the product formula algorithm. See, for instance, Refs.~\cite{wiebe2011simulating,PoulinTimeDependent}. 

Beyond the naive application of product formulas, we might consider using some more advanced algorithms. Usually, those kinds of algorithms include some black box Hamiltonian input models, which we call oracles, and ask how many queries we need to access the oracles. This type of algorithms includes algorithms based on quantum walks~\cite{childs2011walks,berry2015hamiltonian}, multiproduct formula~\cite{low2019multiproduct,childs2012multiproduct}, Taylor expansion~\cite{berry2012black,berry2015simulating}, fractional-query models~\cite{spar}, Chebyshev polynomial approximations~\cite{subramanian2018implementing}, qubitization~\cite{low2016hamiltonian,low2018hamiltonian}, and quantum signal processing~\cite{LowQSPprx,low2017optimal}. Many elements in the web of such algorithms are conceptually or technically related. 
In terms of input models, there are two common choices. One is through a {\it linear combination of unitaries} (LCU) by a decomposition of the Hamiltonian into a series of unitaries. 
While one could also construct oracles based on accessing matrix entries, which usually involves some assumptions about sparseness. 
For time-dependent situation, progresses are made in Refs.~\cite{berry2015simulating,kieferova2019simulating}. An alternative strategy is to construct oracles that encode the Hamiltonian as a block-diagonal element of an oracle unitary matrix, known as {\it block encoding}. 
See  Refs.~\cite{low2016hamiltonian,low2018hamiltonian} and Ref.~\cite{wan2020fast}, where block encoding is used for time-independent simulation as well as digital adiabatic state preparation. 

Those oracle-based methods are widely used in quantum computational chemistry~\cite{mcardle2020quantum}. For a general energy-level system, a helpful generic introduction is given in Ref.~\cite{sawaya2020resource}. 

In this paper, we will mainly consider two different algorithms. For real-time dynamics with the static Hamiltonian, we construct a protocol based on block-encoding, qubitization, and quantum signal processing~\cite{low2017optimal}. This example is particularly important since (similar to Ref.~\cite{Babbush:2018mlj}) we have an explicit oracle construction for the matrix models.  
For the adiabatic state preparation, we will review and use the Wan-Kim algorithm~\cite{wan2020fast}. The Hamiltonian simulation industry is under rapid construction, and some better algorithms may appear in the future. In Appendices~\ref{sec:alternative_adiabatic_state_preparation} and~\ref{sec:Berry-et-al-algorithm}, we provide alternative algorithms based on LCU decomposition, rescaled Dyson series, and the naive adiabatic state preparation.

\bibliographystyle{utphys}
\bibliography{BFSS}

\providecommand{\href}[2]{#2}\begingroup\raggedright\begin{thebibliography}{100}

\bibitem{Wilson:1974sk}
K.~G. Wilson, ``{Confinement of Quarks},''
  \href{http://dx.doi.org/10.1103/PhysRevD.10.2445}{{\em Phys. Rev.} {\bfseries
  D10} (1974) 2445--2459}.
[,45(1974); ,319(1974)].

\bibitem{Creutz:1980zw}
M.~Creutz, ``{Monte Carlo Study of Quantized SU(2) Gauge Theory},''
  \href{http://dx.doi.org/10.1103/PhysRevD.21.2308}{{\em Phys. Rev. D}
  {\bfseries 21} (1980) 2308--2315}.

\bibitem{Maldacena:1997re}
J.~M. Maldacena, ``{The Large N limit of superconformal field theories and
  supergravity},'' \href{http://dx.doi.org/10.1023/A:1026654312961}{{\em Int.
  J. Theor. Phys.} {\bfseries 38} (1999) 1113--1133},
  \href{http://arxiv.org/abs/hep-th/9711200}{{\ttfamily arXiv:hep-th/9711200}}.

\bibitem{Hanada:2018fnp}
M.~Hanada, ``{Markov Chain Monte Carlo for Dummies},''
  \href{http://arxiv.org/abs/1808.08490}{{\ttfamily arXiv:1808.08490
  [hep-th]}}.

\bibitem{Joseph:2019zer}
A.~Joseph, \href{http://dx.doi.org/10.1007/978-3-030-46044-0}{``{Markov Chain
  Monte Carlo Methods in Quantum Field Theories: A Modern Primer},''}
\newblock 12, 2019.
\newblock \href{http://arxiv.org/abs/1912.10997}{{\ttfamily arXiv:1912.10997
  [hep-th]}}.

\bibitem{Aoki:2008sm}
{\bfseries PACS-CS} Collaboration, S.~Aoki {\em et~al.}, ``{2+1 Flavor Lattice
  QCD toward the Physical Point},''
  \href{http://dx.doi.org/10.1103/PhysRevD.79.034503}{{\em Phys. Rev. D}
  {\bfseries 79} (2009) 034503},
  \href{http://arxiv.org/abs/0807.1661}{{\ttfamily arXiv:0807.1661 [hep-lat]}}.

\bibitem{Durr:2008zz}
S.~Durr {\em et~al.}, ``{Ab-Initio Determination of Light Hadron Masses},''
  \href{http://dx.doi.org/10.1126/science.1163233}{{\em Science} {\bfseries
  322} (2008) 1224--1227}, \href{http://arxiv.org/abs/0906.3599}{{\ttfamily
  arXiv:0906.3599 [hep-lat]}}.

\bibitem{Ishii:2006ec}
N.~Ishii, S.~Aoki, and T.~Hatsuda, ``{The Nuclear Force from Lattice QCD},''
  \href{http://dx.doi.org/10.1103/PhysRevLett.99.022001}{{\em Phys. Rev. Lett.}
  {\bfseries 99} (2007) 022001},
  \href{http://arxiv.org/abs/nucl-th/0611096}{{\ttfamily
  arXiv:nucl-th/0611096}}.

\bibitem{Anagnostopoulos:2007fw}
K.~N. Anagnostopoulos, M.~Hanada, J.~Nishimura, and S.~Takeuchi, ``{Monte Carlo
  studies of supersymmetric matrix quantum mechanics with sixteen supercharges
  at finite temperature},''
  \href{http://dx.doi.org/10.1103/PhysRevLett.100.021601}{{\em Phys. Rev.
  Lett.} {\bfseries 100} (2008) 021601},
  \href{http://arxiv.org/abs/0707.4454}{{\ttfamily arXiv:0707.4454 [hep-th]}}.

\bibitem{Catterall:2008yz}
S.~Catterall and T.~Wiseman, ``{Black hole thermodynamics from simulations of
  lattice Yang-Mills theory},''
  \href{http://dx.doi.org/10.1103/PhysRevD.78.041502}{{\em Phys. Rev. D}
  {\bfseries 78} (2008) 041502},
  \href{http://arxiv.org/abs/0803.4273}{{\ttfamily arXiv:0803.4273 [hep-th]}}.

\bibitem{wan2020fast}
K.~Wan and I.~Kim, ``Fast digital methods for adiabatic state preparation,''
  {\em arXiv preprint arXiv:2004.04164} (2020) .

\bibitem{low2016hamiltonian}
G.~H. Low and I.~L. Chuang, ``Hamiltonian simulation by qubitization,'' {\em
  arXiv preprint arXiv:1610.06546} (2016) .

\bibitem{low2017optimal}
G.~H. Low and I.~L. Chuang, ``Optimal hamiltonian simulation by quantum signal
  processing,'' {\em Physical review letters} {\bfseries 118} no.~1, (2017)
  010501.

\bibitem{Nambu:1961tp}
Y.~Nambu and G.~Jona-Lasinio, ``{Dynamical Model of Elementary Particles Based
  on an Analogy with Superconductivity. 1.},''
  \href{http://dx.doi.org/10.1103/PhysRev.122.345}{{\em Phys. Rev.} {\bfseries
  122} (1961) 345--358}.

\bibitem{Nielsen:1980rz}
H.~B. Nielsen and M.~Ninomiya, ``{Absence of Neutrinos on a Lattice. 1. Proof
  by Homotopy Theory},'' \href{http://dx.doi.org/10.1016/0550-3213(81)90361-8,
  10.1016/0550-3213(82)90011-6}{{\em Nucl. Phys.} {\bfseries B185} (1981) 20}.
[Erratum: Nucl. Phys.B195,541(1982); ,533(1980)].

\bibitem{Neuberger:1997fp}
H.~Neuberger, ``{Exactly massless quarks on the lattice},''
  \href{http://dx.doi.org/10.1016/S0370-2693(97)01368-3}{{\em Phys. Lett.}
  {\bfseries B417} (1998) 141--144},
\href{http://arxiv.org/abs/hep-lat/9707022}{{\ttfamily arXiv:hep-lat/9707022
  [hep-lat]}}.

\bibitem{Kaplan:1992bt}
D.~B. Kaplan, ``{A Method for simulating chiral fermions on the lattice},''
  \href{http://dx.doi.org/10.1016/0370-2693(92)91112-M}{{\em Phys. Lett.}
  {\bfseries B288} (1992) 342--347},
\href{http://arxiv.org/abs/hep-lat/9206013}{{\ttfamily arXiv:hep-lat/9206013
  [hep-lat]}}.

\bibitem{Kaplan:2002wv}
D.~B. Kaplan, E.~Katz, and M.~Unsal, ``{Supersymmetry on a spatial lattice},''
  \href{http://dx.doi.org/10.1088/1126-6708/2003/05/037}{{\em JHEP} {\bfseries
  05} (2003) 037}, \href{http://arxiv.org/abs/hep-lat/0206019}{{\ttfamily
  arXiv:hep-lat/0206019}}.

\bibitem{Cohen:2003xe}
A.~G. Cohen, D.~B. Kaplan, E.~Katz, and M.~Unsal, ``{Supersymmetry on a
  Euclidean space-time lattice. 1. A Target theory with four supercharges},''
  \href{http://dx.doi.org/10.1088/1126-6708/2003/08/024}{{\em JHEP} {\bfseries
  08} (2003) 024}, \href{http://arxiv.org/abs/hep-lat/0302017}{{\ttfamily
  arXiv:hep-lat/0302017}}.

\bibitem{Cohen:2003qw}
A.~G. Cohen, D.~B. Kaplan, E.~Katz, and M.~Unsal, ``{Supersymmetry on a
  Euclidean space-time lattice. 2. Target theories with eight supercharges},''
  \href{http://dx.doi.org/10.1088/1126-6708/2003/12/031}{{\em JHEP} {\bfseries
  12} (2003) 031}, \href{http://arxiv.org/abs/hep-lat/0307012}{{\ttfamily
  arXiv:hep-lat/0307012}}.

\bibitem{Kaplan:2005ta}
D.~B. Kaplan and M.~Unsal, ``{A Euclidean lattice construction of
  supersymmetric Yang-Mills theories with sixteen supercharges},''
  \href{http://dx.doi.org/10.1088/1126-6708/2005/09/042}{{\em JHEP} {\bfseries
  09} (2005) 042}, \href{http://arxiv.org/abs/hep-lat/0503039}{{\ttfamily
  arXiv:hep-lat/0503039}}.

\bibitem{Sugino:2003yb}
F.~Sugino, ``{A Lattice formulation of superYang-Mills theories with exact
  supersymmetry},'' \href{http://dx.doi.org/10.1088/1126-6708/2004/01/015}{{\em
  JHEP} {\bfseries 01} (2004) 015},
  \href{http://arxiv.org/abs/hep-lat/0311021}{{\ttfamily
  arXiv:hep-lat/0311021}}.

\bibitem{Sugino:2004qd}
F.~Sugino, ``{SuperYang-Mills theories on the two-dimensional lattice with
  exact supersymmetry},''
  \href{http://dx.doi.org/10.1088/1126-6708/2004/03/067}{{\em JHEP} {\bfseries
  03} (2004) 067}, \href{http://arxiv.org/abs/hep-lat/0401017}{{\ttfamily
  arXiv:hep-lat/0401017}}.

\bibitem{Sugino:2004uv}
F.~Sugino, ``{Various super Yang-Mills theories with exact supersymmetry on the
  lattice},'' \href{http://dx.doi.org/10.1088/1126-6708/2005/01/016}{{\em JHEP}
  {\bfseries 01} (2005) 016},
  \href{http://arxiv.org/abs/hep-lat/0410035}{{\ttfamily
  arXiv:hep-lat/0410035}}.

\bibitem{Catterall:2003wd}
S.~Catterall, ``{Lattice supersymmetry and topological field theory},''
  \href{http://dx.doi.org/10.1088/1126-6708/2003/05/038}{{\em JHEP} {\bfseries
  05} (2003) 038}, \href{http://arxiv.org/abs/hep-lat/0301028}{{\ttfamily
  arXiv:hep-lat/0301028}}.

\bibitem{Catterall:2004np}
S.~Catterall, ``{A Geometrical approach to N=2 super Yang-Mills theory on the
  two dimensional lattice},''
  \href{http://dx.doi.org/10.1088/1126-6708/2004/11/006}{{\em JHEP} {\bfseries
  11} (2004) 006}, \href{http://arxiv.org/abs/hep-lat/0410052}{{\ttfamily
  arXiv:hep-lat/0410052}}.

\bibitem{Kaplan:1983sk}
D.~B. Kaplan, ``{Dynamical Generation of Supersymmetry},''
  \href{http://dx.doi.org/10.1016/0370-2693(84)91172-9}{{\em Phys. Lett. B}
  {\bfseries 136} (1984) 162--164}.

\bibitem{Curci:1986sm}
G.~Curci and G.~Veneziano, ``{Supersymmetry and the Lattice: A
  Reconciliation?},''
  \href{http://dx.doi.org/10.1016/0550-3213(87)90660-2}{{\em Nucl. Phys. B}
  {\bfseries 292} (1987) 555--572}.

\bibitem{Hanada:2007ti}
M.~Hanada, J.~Nishimura, and S.~Takeuchi, ``{Non-lattice simulation for
  supersymmetric gauge theories in one dimension},''
  \href{http://dx.doi.org/10.1103/PhysRevLett.99.161602}{{\em Phys. Rev. Lett.}
  {\bfseries 99} (2007) 161602},
\href{http://arxiv.org/abs/0706.1647}{{\ttfamily arXiv:0706.1647 [hep-lat]}}.

\bibitem{Catterall:2007fp}
S.~Catterall and T.~Wiseman, ``{Towards lattice simulation of the gauge theory
  duals to black holes and hot strings},''
  \href{http://dx.doi.org/10.1088/1126-6708/2007/12/104}{{\em JHEP} {\bfseries
  12} (2007) 104},
\href{http://arxiv.org/abs/0706.3518}{{\ttfamily arXiv:0706.3518 [hep-lat]}}.

\bibitem{Hanada:2013rga}
M.~Hanada, Y.~Hyakutake, G.~Ishiki, and J.~Nishimura, ``{Holographic
  description of quantum black hole on a computer},''
  \href{http://dx.doi.org/10.1126/science.1250122}{{\em Science} {\bfseries
  344} (2014) 882--885}, \href{http://arxiv.org/abs/1311.5607}{{\ttfamily
  arXiv:1311.5607 [hep-th]}}.

\bibitem{Berkowitz:2016jlq}
E.~Berkowitz, E.~Rinaldi, M.~Hanada, G.~Ishiki, S.~Shimasaki, and P.~Vranas,
  ``{Precision lattice test of the gauge/gravity duality at large-$N$},''
  \href{http://dx.doi.org/10.1103/PhysRevD.94.094501}{{\em Phys. Rev. D}
  {\bfseries 94} no.~9, (2016) 094501},
  \href{http://arxiv.org/abs/1606.04951}{{\ttfamily arXiv:1606.04951
  [hep-lat]}}.

\bibitem{Asano:2016kxo}
Y.~Asano, V.~G. Filev, S.~Kovavcik, and D.~O'Connor, ``{A computer test of
  holographic avour dynamics. Part II},''
  \href{http://dx.doi.org/10.1007/JHEP03(2018)055}{{\em JHEP} {\bfseries 03}
  (2018) 055}, \href{http://arxiv.org/abs/1612.09281}{{\ttfamily
  arXiv:1612.09281 [hep-th]}}.

\bibitem{Hanada:2016jok}
M.~Hanada, ``{What lattice theorists can do for superstring/M-theory},''
  \href{http://dx.doi.org/10.1142/S0217751X16430065}{{\em Int. J. Mod. Phys. A}
  {\bfseries 31} no.~22, (2016) 1643006},
  \href{http://arxiv.org/abs/1604.05421}{{\ttfamily arXiv:1604.05421
  [hep-lat]}}.

\bibitem{Kogut:1974ag}
J.~B. Kogut and L.~Susskind, ``{Hamiltonian Formulation of Wilson's Lattice
  Gauge Theories},''
\href{http://dx.doi.org/10.1103/PhysRevD.11.395}{{\em Phys. Rev.} {\bfseries
  D11} (1975) 395--408}.

\bibitem{Zohar:2014qma}
E.~Zohar and M.~Burrello, ``{Formulation of lattice gauge theories for quantum
  simulations},'' \href{http://dx.doi.org/10.1103/PhysRevD.91.054506}{{\em
  Phys. Rev. D} {\bfseries 91} no.~5, (2015) 054506},
  \href{http://arxiv.org/abs/1409.3085}{{\ttfamily arXiv:1409.3085
  [quant-ph]}}.

\bibitem{Eguchi:1982nm}
T.~Eguchi and H.~Kawai, ``{Reduction of Dynamical Degrees of Freedom in the
  Large N Gauge Theory},''
\href{http://dx.doi.org/10.1103/PhysRevLett.48.1063}{{\em Phys. Rev. Lett.}
  {\bfseries 48} (1982) 1063}.

\bibitem{GonzalezArroyo:1982ub}
A.~Gonzalez-Arroyo and M.~Okawa, ``{A Twisted Model for Large $N$ Lattice Gauge
  Theory},''
\href{http://dx.doi.org/10.1016/0370-2693(83)90647-0}{{\em Phys. Lett.}
  {\bfseries 120B} (1983) 174--178}.

\bibitem{GonzalezArroyo:1982hz}
A.~Gonzalez-Arroyo and M.~Okawa, ``{The Twisted Eguchi-Kawai Model: A Reduced
  Model for Large N Lattice Gauge Theory},''
\href{http://dx.doi.org/10.1103/PhysRevD.27.2397}{{\em Phys. Rev.} {\bfseries
  D27} (1983) 2397}.

\bibitem{Myers:1999ps}
R.~C. Myers, ``{Dielectric branes},''
  \href{http://dx.doi.org/10.1088/1126-6708/1999/12/022}{{\em JHEP} {\bfseries
  12} (1999) 022},
\href{http://arxiv.org/abs/hep-th/9910053}{{\ttfamily arXiv:hep-th/9910053
  [hep-th]}}.

\bibitem{Hanada:2010kt}
M.~Hanada, S.~Matsuura, and F.~Sugino, ``{Two-dimensional lattice for
  four-dimensional N=4 supersymmetric Yang-Mills},''
  \href{http://dx.doi.org/10.1143/PTP.126.597}{{\em Prog. Theor. Phys.}
  {\bfseries 126} (2011) 597--611},
  \href{http://arxiv.org/abs/1004.5513}{{\ttfamily arXiv:1004.5513 [hep-lat]}}.

\bibitem{Hanada:2010gs}
M.~Hanada, ``{A proposal of a fine tuning free formulation of 4d N = 4 super
  Yang-Mills},'' \href{http://dx.doi.org/10.1007/JHEP11(2010)112}{{\em JHEP}
  {\bfseries 11} (2010) 112}, \href{http://arxiv.org/abs/1009.0901}{{\ttfamily
  arXiv:1009.0901 [hep-lat]}}.

\bibitem{Danshita:2016xbo}
I.~Danshita, M.~Hanada, and M.~Tezuka, ``{Creating and probing the
  Sachdev-Ye-Kitaev model with ultracold gases: Towards experimental studies of
  quantum gravity},'' \href{http://dx.doi.org/10.1093/ptep/ptx108}{{\em PTEP}
  {\bfseries 2017} no.~8, (2017) 083I01},
  \href{http://arxiv.org/abs/1606.02454}{{\ttfamily arXiv:1606.02454
  [cond-mat.quant-gas]}}.

\bibitem{Brown:2019hmk}
A.~R. Brown, H.~Gharibyan, S.~Leichenauer, H.~W. Lin, S.~Nezami, G.~Salton,
  L.~Susskind, B.~Swingle, and M.~Walter, ``{Quantum Gravity in the Lab:
  Teleportation by Size and Traversable Wormholes},''
  \href{http://arxiv.org/abs/1911.06314}{{\ttfamily arXiv:1911.06314
  [quant-ph]}}.

\bibitem{Landsman2019}
K.~A. Landsman, C.~Figgatt, T.~Schuster, N.~M. Linke, B.~Yoshida, N.~Y. Yao,
  and C.~Monroe, ``Verified quantum information scrambling,''
  \href{http://dx.doi.org/10.1038/s41586-019-0952-6}{{\em Nature} {\bfseries
  567} no.~7746, (2019) 61--65}.
  \url{https://doi.org/10.1038/s41586-019-0952-6}.

\bibitem{PhysRevLett.124.240505}
M.~K. Joshi, A.~Elben, B.~Vermersch, T.~Brydges, C.~Maier, P.~Zoller, R.~Blatt,
  and C.~F. Roos, ``Quantum information scrambling in a trapped-ion quantum
  simulator with tunable range interactions,''
  \href{http://dx.doi.org/10.1103/PhysRevLett.124.240505}{{\em Phys. Rev.
  Lett.} {\bfseries 124} (Jun, 2020) 240505}.
  \url{https://link.aps.org/doi/10.1103/PhysRevLett.124.240505}.

\bibitem{Babbush:2018mlj}
R.~Babbush, D.~W. Berry, and H.~Neven, ``{Quantum Simulation of the
  Sachdev-Ye-Kitaev Model by Asymmetric Qubitization},''
  \href{http://dx.doi.org/10.1103/PhysRevA.99.040301}{{\em Phys. Rev. A}
  {\bfseries 99} no.~4, (2019) 040301},
  \href{http://arxiv.org/abs/1806.02793}{{\ttfamily arXiv:1806.02793
  [quant-ph]}}.

\bibitem{Garcia-Alvarez:2016wem}
L.~Garcia-Alvarez, I.~Egusquiza, L.~Lamata, A.~del Campo, J.~Sonner, and
  E.~Solano, ``{Digital Quantum Simulation of Minimal AdS/CFT},''
  \href{http://dx.doi.org/10.1103/PhysRevLett.119.040501}{{\em Phys. Rev.
  Lett.} {\bfseries 119} no.~4, (2017) 040501},
  \href{http://arxiv.org/abs/1607.08560}{{\ttfamily arXiv:1607.08560
  [quant-ph]}}.

\bibitem{Xu:2020shn}
S.~Xu, L.~Susskind, Y.~Su, and B.~Swingle, ``{A Sparse Model of Quantum
  Holography},'' \href{http://arxiv.org/abs/2008.02303}{{\ttfamily
  arXiv:2008.02303 [cond-mat.str-el]}}.

\bibitem{Maldacena:2016upp}
J.~Maldacena, D.~Stanford, and Z.~Yang, ``{Conformal symmetry and its breaking
  in two dimensional Nearly Anti-de-Sitter space},''
  \href{http://dx.doi.org/10.1093/ptep/ptw124}{{\em PTEP} {\bfseries 2016}
  no.~12, (2016) 12C104}, \href{http://arxiv.org/abs/1606.01857}{{\ttfamily
  arXiv:1606.01857 [hep-th]}}.

\bibitem{Landsman:2018jpm}
K.~Landsman, C.~Figgatt, T.~Schuster, N.~Linke, B.~Yoshida, N.~Yao, and
  C.~Monroe, ``{Verified Quantum Information Scrambling},''
  \href{http://dx.doi.org/10.1038/s41586-019-0952-6}{{\em Nature} {\bfseries
  567} no.~7746, (2019) 61--65},
  \href{http://arxiv.org/abs/1806.02807}{{\ttfamily arXiv:1806.02807
  [quant-ph]}}.

\bibitem{Kruchkov:2019idx}
A.~Kruchkov, A.~Patel, P.~Kim, and S.~Sachdev, ``{Thermoelectric power of
  Sachdev-Ye-Kitaev islands: Probing Bekenstein-Hawking entropy in quantum
  matter experiments},'' \href{http://arxiv.org/abs/1912.02835}{{\ttfamily
  arXiv:1912.02835 [cond-mat.str-el]}}.

\bibitem{Liu:2020sqb}
J.~Liu, ``{Scrambling and decoding the charged quantum information},''
  \href{http://arxiv.org/abs/2003.11425}{{\ttfamily arXiv:2003.11425
  [quant-ph]}}.

\bibitem{Berenstein:2002jq}
D.~E. Berenstein, J.~M. Maldacena, and H.~S. Nastase, ``{Strings in flat space
  and pp waves from N=4 superYang-Mills},''
  \href{http://dx.doi.org/10.1088/1126-6708/2002/04/013}{{\em JHEP} {\bfseries
  04} (2002) 013},
\href{http://arxiv.org/abs/hep-th/0202021}{{\ttfamily arXiv:hep-th/0202021
  [hep-th]}}.

\bibitem{Banks:1996vh}
T.~Banks, W.~Fischler, S.~H. Shenker, and L.~Susskind, ``{M theory as a matrix
  model: A Conjecture},''
  \href{http://dx.doi.org/10.1103/PhysRevD.55.5112}{{\em Phys. Rev.} {\bfseries
  D55} (1997) 5112--5128},
\href{http://arxiv.org/abs/hep-th/9610043}{{\ttfamily arXiv:hep-th/9610043
  [hep-th]}}.

\bibitem{deWit:1988wri}
B.~de~Wit, J.~Hoppe, and H.~Nicolai, ``{On the Quantum Mechanics of
  Supermembranes},'' \href{http://dx.doi.org/10.1016/0550-3213(88)90116-2}{{\em
  Nucl.\ Phys.\ B} {\bfseries 305} (1988) 545}.

\bibitem{Dasgupta:2002hx}
K.~Dasgupta, M.~M. Sheikh-Jabbari, and M.~Van~Raamsdonk, ``{Matrix perturbation
  theory for M theory on a PP wave},''
  \href{http://dx.doi.org/10.1088/1126-6708/2002/05/056}{{\em JHEP} {\bfseries
  05} (2002) 056}, \href{http://arxiv.org/abs/hep-th/0205185}{{\ttfamily
  arXiv:hep-th/0205185}}.

\bibitem{Kim:2003rza}
N.~Kim, T.~Klose, and J.~Plefka, ``{Plane wave matrix theory from N=4
  superYang-Mills on R x S**3},''
  \href{http://dx.doi.org/10.1016/j.nuclphysb.2003.08.019}{{\em Nucl. Phys. B}
  {\bfseries 671} (2003) 359--382},
  \href{http://arxiv.org/abs/hep-th/0306054}{{\ttfamily arXiv:hep-th/0306054}}.

\bibitem{Maldacena:2018vsr}
J.~Maldacena and A.~Milekhin, ``{To gauge or not to gauge?},''
  \href{http://dx.doi.org/10.1007/JHEP04(2018)084}{{\em JHEP} {\bfseries 04}
  (2018) 084},
\href{http://arxiv.org/abs/1802.00428}{{\ttfamily arXiv:1802.00428 [hep-th]}}.

\bibitem{Jordan:2011ne}
S.~P. Jordan, K.~S. Lee, and J.~Preskill, ``Quantum algorithms for quantum
  field theories,'' {\em Science} {\bfseries 336} no.~6085, (2012) 1130--1133.

\bibitem{Klco:2018zqz}
N.~Klco and M.~J. Savage, ``{Digitization of scalar fields for quantum
  computing},'' \href{http://dx.doi.org/10.1103/PhysRevA.99.052335}{{\em Phys.
  Rev. A} {\bfseries 99} no.~5, (2019) 052335},
  \href{http://arxiv.org/abs/1808.10378}{{\ttfamily arXiv:1808.10378
  [quant-ph]}}.

\bibitem{Milekhin:2020zpg}
A.~Milekhin, ``{Quantum error correction and large $N$},''
  \href{http://arxiv.org/abs/2008.12869}{{\ttfamily arXiv:2008.12869
  [hep-th]}}.

\bibitem{Berkowitz:2018qhn}
E.~Berkowitz, M.~Hanada, E.~Rinaldi, and P.~Vranas, ``{Gauged And Ungauged: A
  Nonperturbative Test},''
  \href{http://dx.doi.org/10.1007/JHEP06(2018)124}{{\em JHEP} {\bfseries 06}
  (2018) 124},
\href{http://arxiv.org/abs/1802.02985}{{\ttfamily arXiv:1802.02985 [hep-th]}}.

\bibitem{Dasgupta:2002ru}
K.~Dasgupta, M.~M. Sheikh-Jabbari, and M.~Van~Raamsdonk, ``{Protected
  multiplets of M theory on a plane wave},''
  \href{http://dx.doi.org/10.1088/1126-6708/2002/09/021}{{\em JHEP} {\bfseries
  09} (2002) 021}, \href{http://arxiv.org/abs/hep-th/0207050}{{\ttfamily
  arXiv:hep-th/0207050}}.

\bibitem{Ishii:2008ib}
T.~Ishii, G.~Ishiki, S.~Shimasaki, and A.~Tsuchiya, ``{N=4 Super Yang-Mills
  from the Plane Wave Matrix Model},''
  \href{http://dx.doi.org/10.1103/PhysRevD.78.106001}{{\em Phys. Rev.}
  {\bfseries D78} (2008) 106001},
\href{http://arxiv.org/abs/0807.2352}{{\ttfamily arXiv:0807.2352 [hep-th]}}.

\bibitem{Witten:1995im}
E.~Witten, ``{Bound states of strings and p-branes},''
  \href{http://dx.doi.org/10.1016/0550-3213(95)00610-9}{{\em Nucl. Phys. B}
  {\bfseries 460} (1996) 335--350},
  \href{http://arxiv.org/abs/hep-th/9510135}{{\ttfamily arXiv:hep-th/9510135}}.

\bibitem{Itzhaki:1998dd}
N.~Itzhaki, J.~M. Maldacena, J.~Sonnenschein, and S.~Yankielowicz,
  ``{Supergravity and the large N limit of theories with sixteen
  supercharges},'' \href{http://dx.doi.org/10.1103/PhysRevD.58.046004}{{\em
  Phys. Rev. D} {\bfseries 58} (1998) 046004},
  \href{http://arxiv.org/abs/hep-th/9802042}{{\ttfamily arXiv:hep-th/9802042}}.

\bibitem{Maldacena:2002rb}
J.~M. Maldacena, M.~M. Sheikh-Jabbari, and M.~Van~Raamsdonk, ``{Transverse
  five-branes in matrix theory},''
  \href{http://dx.doi.org/10.1088/1126-6708/2003/01/038}{{\em JHEP} {\bfseries
  01} (2003) 038},
\href{http://arxiv.org/abs/hep-th/0211139}{{\ttfamily arXiv:hep-th/0211139
  [hep-th]}}.

\bibitem{Minwalla:1999px}
S.~Minwalla, M.~Van~Raamsdonk, and N.~Seiberg, ``{Noncommutative perturbative
  dynamics},'' \href{http://dx.doi.org/10.1088/1126-6708/2000/02/020}{{\em
  JHEP} {\bfseries 02} (2000) 020},
  \href{http://arxiv.org/abs/hep-th/9912072}{{\ttfamily arXiv:hep-th/9912072}}.

\bibitem{Matusis:2000jf}
A.~Matusis, L.~Susskind, and N.~Toumbas, ``{The IR / UV connection in the
  noncommutative gauge theories},''
  \href{http://dx.doi.org/10.1088/1126-6708/2000/12/002}{{\em JHEP} {\bfseries
  12} (2000) 002}, \href{http://arxiv.org/abs/hep-th/0002075}{{\ttfamily
  arXiv:hep-th/0002075}}.

\bibitem{Hanada:2014ima}
M.~Hanada and H.~Shimada, ``{On the continuity of the commutative limit of the
  4d N=4 non-commutative super Yang--Mills theory},''
  \href{http://dx.doi.org/10.1016/j.nuclphysb.2015.01.016}{{\em Nucl. Phys. B}
  {\bfseries 892} (2015) 449--474},
  \href{http://arxiv.org/abs/1410.4503}{{\ttfamily arXiv:1410.4503 [hep-th]}}.

\bibitem{Asano:2017nxw}
Y.~Asano, G.~Ishiki, S.~Shimasaki, and S.~Terashima, ``{Spherical transverse
  M5-branes from the plane wave matrix model},''
  \href{http://dx.doi.org/10.1007/JHEP02(2018)076}{{\em JHEP} {\bfseries 02}
  (2018) 076},
\href{http://arxiv.org/abs/1711.07681}{{\ttfamily arXiv:1711.07681 [hep-th]}}.

\bibitem{Asano:2017xiy}
Y.~Asano, G.~Ishiki, S.~Shimasaki, and S.~Terashima, ``{Spherical transverse
  M5-branes in matrix theory},''
  \href{http://dx.doi.org/10.1103/PhysRevD.96.126003}{{\em Phys. Rev.}
  {\bfseries D96} no.~12, (2017) 126003},
\href{http://arxiv.org/abs/1701.07140}{{\ttfamily arXiv:1701.07140 [hep-th]}}.

\bibitem{Kawai:2009vb}
H.~Kawai, S.~Shimasaki, and A.~Tsuchiya, ``{Large N reduction on group
  manifolds},'' \href{http://dx.doi.org/10.1142/S0217751X10049396}{{\em Int. J.
  Mod. Phys. A} {\bfseries 25} (2010) 3389--3406},
  \href{http://arxiv.org/abs/0912.1456}{{\ttfamily arXiv:0912.1456 [hep-th]}}.

\bibitem{Ishiki:2008te}
G.~Ishiki, S.-W. Kim, J.~Nishimura, and A.~Tsuchiya, ``{Deconfinement phase
  transition in N=4 super Yang-Mills theory on R x S**3 from supersymmetric
  matrix quantum mechanics},''
  \href{http://dx.doi.org/10.1103/PhysRevLett.102.111601}{{\em Phys. Rev.
  Lett.} {\bfseries 102} (2009) 111601},
  \href{http://arxiv.org/abs/0810.2884}{{\ttfamily arXiv:0810.2884 [hep-th]}}.

\bibitem{Ishiki:2009sg}
G.~Ishiki, S.-W. Kim, J.~Nishimura, and A.~Tsuchiya, ``{Testing a novel large-N
  reduction for N=4 super Yang-Mills theory on R x S**3},''
  \href{http://dx.doi.org/10.1088/1126-6708/2009/09/029}{{\em JHEP} {\bfseries
  09} (2009) 029}, \href{http://arxiv.org/abs/0907.1488}{{\ttfamily
  arXiv:0907.1488 [hep-th]}}.

\bibitem{Honda:2010nx}
M.~Honda, G.~Ishiki, S.-W. Kim, J.~Nishimura, and A.~Tsuchiya, ``{Supersymmetry
  non-renormalization theorem from a computer and the AdS/CFT
  correspondence},'' \href{http://dx.doi.org/10.22323/1.105.0253}{{\em PoS}
  {\bfseries LATTICE2010} (2010) 253},
  \href{http://arxiv.org/abs/1011.3904}{{\ttfamily arXiv:1011.3904 [hep-lat]}}.

\bibitem{Honda:2011qk}
M.~Honda, G.~Ishiki, J.~Nishimura, and A.~Tsuchiya, ``{Testing the AdS/CFT
  correspondence by Monte Carlo calculation of BPS and non-BPS Wilson loops in
  4d N=4 super-Yang-Mills theory},''
  \href{http://dx.doi.org/10.22323/1.139.0244}{{\em PoS} {\bfseries
  LATTICE2011} (2011) 244}, \href{http://arxiv.org/abs/1112.4274}{{\ttfamily
  arXiv:1112.4274 [hep-lat]}}.

\bibitem{Honda:2013nfa}
M.~Honda, G.~Ishiki, S.-W. Kim, J.~Nishimura, and A.~Tsuchiya, ``{Direct test
  of the AdS/CFT correspondence by Monte Carlo studies of N=4 super Yang-Mills
  theory},'' \href{http://dx.doi.org/10.1007/JHEP11(2013)200}{{\em JHEP}
  {\bfseries 11} (2013) 200}, \href{http://arxiv.org/abs/1308.3525}{{\ttfamily
  arXiv:1308.3525 [hep-th]}}.

\bibitem{Honda:2012ni}
M.~Honda and Y.~Yoshida, ``{Localization and Large N reduction on $S^3$ for the
  Planar and M-theory limit},''
  \href{http://dx.doi.org/10.1016/j.nuclphysb.2012.07.022}{{\em Nucl. Phys. B}
  {\bfseries 865} (2012) 21--53},
  \href{http://arxiv.org/abs/1203.1016}{{\ttfamily arXiv:1203.1016 [hep-th]}}.

\bibitem{Asano:2012gt}
Y.~Asano, G.~Ishiki, T.~Okada, and S.~Shimasaki, ``{Large-N reduction for
  $\mathcal{N}$ =2 quiver Chern-Simons theories on $S^3$ and localization in
  matrix models},'' \href{http://dx.doi.org/10.1103/PhysRevD.85.106003}{{\em
  Phys. Rev. D} {\bfseries 85} (2012) 106003},
  \href{http://arxiv.org/abs/1203.0559}{{\ttfamily arXiv:1203.0559 [hep-th]}}.

\bibitem{Bhanot:1982sh}
G.~Bhanot, U.~M. Heller, and H.~Neuberger, ``{The Quenched Eguchi-Kawai
  Model},'' \href{http://dx.doi.org/10.1016/0370-2693(82)90106-X}{{\em Phys.
  Lett. B} {\bfseries 113} (1982) 47--50}.

\bibitem{Parisi:1982gp}
G.~Parisi, ``{A Simple Expression for Planar Field Theories},''
  \href{http://dx.doi.org/10.1016/0370-2693(82)90849-8}{{\em Phys. Lett. B}
  {\bfseries 112} (1982) 463--464}.

\bibitem{Gross:1982at}
D.~J. Gross and Y.~Kitazawa, ``{A Quenched Momentum Prescription for Large N
  Theories},'' \href{http://dx.doi.org/10.1016/0550-3213(82)90278-4}{{\em Nucl.
  Phys. B} {\bfseries 206} (1982) 440--472}.

\bibitem{Kim:2006wg}
N.~Kim and J.-H. Park, ``{Massive super Yang-Mills quantum mechanics:
  Classification and the relation to supermembrane},''
  \href{http://dx.doi.org/10.1016/j.nuclphysb.2006.10.005}{{\em Nucl. Phys. B}
  {\bfseries 759} (2006) 249--282},
  \href{http://arxiv.org/abs/hep-th/0607005}{{\ttfamily arXiv:hep-th/0607005}}.

\bibitem{VanRaamsdonk:2001jd}
M.~Van~Raamsdonk, ``{The Meaning of infrared singularities in noncommutative
  gauge theories},''
  \href{http://dx.doi.org/10.1088/1126-6708/2001/11/006}{{\em JHEP} {\bfseries
  11} (2001) 006}, \href{http://arxiv.org/abs/hep-th/0110093}{{\ttfamily
  arXiv:hep-th/0110093}}.

\bibitem{Azeyanagi:2007su}
T.~Azeyanagi, M.~Hanada, T.~Hirata, and T.~Ishikawa, ``{Phase structure of
  twisted Eguchi-Kawai model},''
  \href{http://dx.doi.org/10.1088/1126-6708/2008/01/025}{{\em JHEP} {\bfseries
  01} (2008) 025}, \href{http://arxiv.org/abs/0711.1925}{{\ttfamily
  arXiv:0711.1925 [hep-lat]}}.

\bibitem{Azeyanagi:2008bk}
T.~Azeyanagi, M.~Hanada, and T.~Hirata, ``{On Matrix Model Formulations of
  Noncommutative Yang-Mills Theories},''
  \href{http://dx.doi.org/10.1103/PhysRevD.78.105017}{{\em Phys. Rev. D}
  {\bfseries 78} (2008) 105017},
  \href{http://arxiv.org/abs/0806.3252}{{\ttfamily arXiv:0806.3252 [hep-th]}}.

\bibitem{Hanada:2016bnb}
M.~Hanada, ``{Yang-Mills theory on noncommutative space: does it exist?},''
  \href{http://dx.doi.org/10.22323/1.263.0105}{{\em PoS} {\bfseries CORFU2015}
  (2016) 105}, \href{http://arxiv.org/abs/1604.04662}{{\ttfamily
  arXiv:1604.04662 [hep-th]}}.

\bibitem{Hanada:2009hd}
M.~Hanada, L.~Mannelli, and Y.~Matsuo, ``{Large-N reduced models of
  supersymmetric quiver, Chern-Simons gauge theories and ABJM},''
  \href{http://dx.doi.org/10.1088/1126-6708/2009/11/087}{{\em JHEP} {\bfseries
  11} (2009) 087},
\href{http://arxiv.org/abs/0907.4937}{{\ttfamily arXiv:0907.4937 [hep-th]}}.

\bibitem{somma2005quantum}
R.~D. Somma, ``Quantum computation, complexity, and many-body physics,'' {\em
  arXiv preprint quant-ph/0512209} (2005) .

\bibitem{mcardle2019digital}
S.~McArdle, A.~Mayorov, X.~Shan, S.~Benjamin, and X.~Yuan, ``Digital quantum
  simulation of molecular vibrations,'' {\em Chemical science} {\bfseries 10}
  no.~22, (2019) 5725--5735.

\bibitem{bravyi2002fermionic}
S.~B. Bravyi and A.~Y. Kitaev, ``Fermionic quantum computation,'' {\em Annals
  of Physics} {\bfseries 298} no.~1, (2002) 210--226.

\bibitem{seeley2012bravyi}
J.~T. Seeley, M.~J. Richard, and P.~J. Love, ``The bravyi-kitaev transformation
  for quantum computation of electronic structure,'' {\em The Journal of
  chemical physics} {\bfseries 137} no.~22, (2012) 224109.

\bibitem{berry2019time}
D.~W. Berry, A.~M. Childs, Y.~Su, X.~Wang, and N.~Wiebe, ``Time-dependent
  hamiltonian simulation with l1-norm scaling,'' {\em arXiv preprint
  arXiv:1906.07115} (2019) .

\bibitem{Cottrell:2018ash}
W.~Cottrell, B.~Freivogel, D.~M. Hofman, and S.~F. Lokhande, ``{How to Build
  the Thermofield Double State},''
  \href{http://dx.doi.org/10.1007/JHEP02(2019)058}{{\em JHEP} {\bfseries 02}
  (2019) 058}, \href{http://arxiv.org/abs/1811.11528}{{\ttfamily
  arXiv:1811.11528 [hep-th]}}.

\bibitem{Wu:2018nrn}
J.~Wu and T.~H. Hsieh, ``{Variational Thermal Quantum Simulation via
  Thermofield Double States},''
  \href{http://dx.doi.org/10.1103/PhysRevLett.123.220502}{{\em Phys. Rev.
  Lett.} {\bfseries 123} no.~22, (2019) 220502},
  \href{http://arxiv.org/abs/1811.11756}{{\ttfamily arXiv:1811.11756
  [cond-mat.str-el]}}.

\bibitem{Maldacena:2018lmt}
J.~Maldacena and X.-L. Qi, ``{Eternal traversable wormhole},''
  \href{http://arxiv.org/abs/1804.00491}{{\ttfamily arXiv:1804.00491
  [hep-th]}}.

\bibitem{Alet:2020ehp}
F.~Alet, M.~Hanada, A.~Jevicki, and C.~Peng, ``{Entanglement and Confinement in
  Coupled Quantum Systems},'' \href{http://arxiv.org/abs/2001.03158}{{\ttfamily
  arXiv:2001.03158 [hep-th]}}.

\bibitem{Hanada:2016pwv}
M.~Hanada and J.~Maltz, ``{A proposal of the gauge theory description of the
  small Schwarzschild black hole in AdS$_5\times$S$^5$},''
  \href{http://dx.doi.org/10.1007/JHEP02(2017)012}{{\em JHEP} {\bfseries 02}
  (2017) 012}, \href{http://arxiv.org/abs/1608.03276}{{\ttfamily
  arXiv:1608.03276 [hep-th]}}.

\bibitem{Berenstein:2018lrm}
D.~Berenstein, ``{Submatrix deconfinement and small black holes in AdS},''
  \href{http://dx.doi.org/10.1007/JHEP09(2018)054}{{\em JHEP} {\bfseries 09}
  (2018) 054}, \href{http://arxiv.org/abs/1806.05729}{{\ttfamily
  arXiv:1806.05729 [hep-th]}}.

\bibitem{Hanada:2018zxn}
M.~Hanada, G.~Ishiki, and H.~Watanabe, ``{Partial Deconfinement},''
  \href{http://dx.doi.org/10.1007/JHEP03(2019)145}{{\em JHEP} {\bfseries 03}
  (2019) 145}, \href{http://arxiv.org/abs/1812.05494}{{\ttfamily
  arXiv:1812.05494 [hep-th]}}. [Erratum: JHEP 10, 029 (2019)].

\bibitem{Hanada:2019czd}
M.~Hanada, A.~Jevicki, C.~Peng, and N.~Wintergerst, ``{Anatomy of
  Deconfinement},'' \href{http://dx.doi.org/10.1007/JHEP12(2019)167}{{\em JHEP}
  {\bfseries 12} (2019) 167}, \href{http://arxiv.org/abs/1909.09118}{{\ttfamily
  arXiv:1909.09118 [hep-th]}}.

\bibitem{Hanada:2020uvt}
M.~Hanada, H.~Shimada, and N.~Wintergerst, ``{Color Confinement and
  Bose-Einstein Condensation},''
  \href{http://arxiv.org/abs/2001.10459}{{\ttfamily arXiv:2001.10459
  [hep-th]}}.

\bibitem{Watanabe:2020ufk}
H.~Watanabe, G.~Bergner, N.~Bodendorfer, S.~Shiba~Funai, M.~Hanada, E.~Rinaldi,
  A.~Schafer, and P.~Vranas, ``{Partial Deconfinement at Strong Coupling on the
  Lattice},'' \href{http://arxiv.org/abs/2005.04103}{{\ttfamily
  arXiv:2005.04103 [hep-th]}}.

\bibitem{Maldacena:2015iua}
J.~Maldacena, D.~Simmons-Duffin, and A.~Zhiboedov, ``{Looking for a bulk
  point},'' \href{http://dx.doi.org/10.1007/JHEP01(2017)013}{{\em JHEP}
  {\bfseries 01} (2017) 013}, \href{http://arxiv.org/abs/1509.03612}{{\ttfamily
  arXiv:1509.03612 [hep-th]}}.

\bibitem{Heemskerk:2009pn}
I.~Heemskerk, J.~Penedones, J.~Polchinski, and J.~Sully, ``{Holography from
  Conformal Field Theory},''
  \href{http://dx.doi.org/10.1088/1126-6708/2009/10/079}{{\em JHEP} {\bfseries
  10} (2009) 079}, \href{http://arxiv.org/abs/0907.0151}{{\ttfamily
  arXiv:0907.0151 [hep-th]}}.

\bibitem{ElShowk:2011ag}
S.~El-Showk and K.~Papadodimas, ``{Emergent Spacetime and Holographic CFTs},''
  \href{http://dx.doi.org/10.1007/JHEP10(2012)106}{{\em JHEP} {\bfseries 10}
  (2012) 106}, \href{http://arxiv.org/abs/1101.4163}{{\ttfamily arXiv:1101.4163
  [hep-th]}}.

\bibitem{Yang:2015uoa}
Z.~Yang, P.~Hayden, and X.-L. Qi, ``{Bidirectional holographic codes and
  sub-AdS locality},'' \href{http://dx.doi.org/10.1007/JHEP01(2016)175}{{\em
  JHEP} {\bfseries 01} (2016) 175},
  \href{http://arxiv.org/abs/1510.03784}{{\ttfamily arXiv:1510.03784
  [hep-th]}}.

\bibitem{Hayden:2016cfa}
P.~Hayden, S.~Nezami, X.-L. Qi, N.~Thomas, M.~Walter, and Z.~Yang,
  ``{Holographic duality from random tensor networks},''
  \href{http://dx.doi.org/10.1007/JHEP11(2016)009}{{\em JHEP} {\bfseries 11}
  (2016) 009}, \href{http://arxiv.org/abs/1601.01694}{{\ttfamily
  arXiv:1601.01694 [hep-th]}}.

\bibitem{Hanada:2021ipb}
M.~Hanada, ``{Bulk geometry in gauge/gravity duality and color degrees of
  freedom},'' \href{http://dx.doi.org/10.1103/PhysRevD.103.106007}{{\em Phys.
  Rev. D} {\bfseries 103} no.~10, (2021) 106007},
  \href{http://arxiv.org/abs/2102.08982}{{\ttfamily arXiv:2102.08982
  [hep-th]}}.

\bibitem{Costa:2014wya}
M.~S. Costa, L.~Greenspan, J.~Penedones, and J.~Santos, ``{Thermodynamics of
  the BMN matrix model at strong coupling},''
  \href{http://dx.doi.org/10.1007/JHEP03(2015)069}{{\em JHEP} {\bfseries 03}
  (2015) 069},
\href{http://arxiv.org/abs/1411.5541}{{\ttfamily arXiv:1411.5541 [hep-th]}}.

\bibitem{Swingle:2016var}
B.~Swingle, G.~Bentsen, M.~Schleier-Smith, and P.~Hayden, ``{Measuring the
  scrambling of quantum information},''
  \href{http://dx.doi.org/10.1103/PhysRevA.94.040302}{{\em Phys. Rev. A}
  {\bfseries 94} no.~4, (2016) 040302},
  \href{http://arxiv.org/abs/1602.06271}{{\ttfamily arXiv:1602.06271
  [quant-ph]}}.

\bibitem{Maldacena:2015waa}
J.~Maldacena, S.~H. Shenker, and D.~Stanford, ``{A bound on chaos},''
  \href{http://dx.doi.org/10.1007/JHEP08(2016)106}{{\em JHEP} {\bfseries 08}
  (2016) 106}, \href{http://arxiv.org/abs/1503.01409}{{\ttfamily
  arXiv:1503.01409 [hep-th]}}.

\bibitem{Zhu:2019bri}
D.~Zhu, S.~Johri, N.~Linke, K.~Landsman, N.~Nguyen, C.~Alderete, A.~Matsuura,
  T.~Hsieh, and C.~Monroe, ``{Generation of Thermofield Double States and
  Critical Ground States with a Quantum Computer},''
  \href{http://arxiv.org/abs/1906.02699}{{\ttfamily arXiv:1906.02699
  [quant-ph]}}.

\bibitem{Shenker:2013pqa}
S.~H. Shenker and D.~Stanford, ``{Black holes and the butterfly effect},''
  \href{http://dx.doi.org/10.1007/JHEP03(2014)067}{{\em JHEP} {\bfseries 03}
  (2014) 067}, \href{http://arxiv.org/abs/1306.0622}{{\ttfamily arXiv:1306.0622
  [hep-th]}}.

\bibitem{Shenker:2014cwa}
S.~H. Shenker and D.~Stanford, ``{Stringy effects in scrambling},''
  \href{http://dx.doi.org/10.1007/JHEP05(2015)132}{{\em JHEP} {\bfseries 05}
  (2015) 132}, \href{http://arxiv.org/abs/1412.6087}{{\ttfamily arXiv:1412.6087
  [hep-th]}}.

\bibitem{Kobrin:2020xms}
B.~Kobrin, Z.~Yang, G.~D. Kahanamoku-Meyer, C.~T. Olund, J.~E. Moore,
  D.~Stanford, and N.~Y. Yao, ``{Many-Body Chaos in the Sachdev-Ye-Kitaev
  Model},'' \href{http://arxiv.org/abs/2002.05725}{{\ttfamily arXiv:2002.05725
  [hep-th]}}.

\bibitem{orbifold-paper}
A.~J.~Buser, H.~Gharibyan, M.~Hanada, M.~Honda, and J.~Liu, ``{Quantum
  simulation of gauge theory via orbifold lattice},''
  \href{http://arxiv.org/abs/2011.06576}{{\ttfamily arXiv:2011.06576
  [hep-th]}}.

\bibitem{Hanada:2011qx}
M.~Hanada, S.~Matsuura, and F.~Sugino, ``{Non-perturbative construction of 2D
  and 4D supersymmetric Yang-Mills theories with 8 supercharges},''
  \href{http://dx.doi.org/10.1016/j.nuclphysb.2011.12.014}{{\em Nucl. Phys. B}
  {\bfseries 857} (2012) 335--361},
  \href{http://arxiv.org/abs/1109.6807}{{\ttfamily arXiv:1109.6807 [hep-lat]}}.

\bibitem{Bouland:2019pvu}
A.~Bouland, B.~Fefferman, and U.~Vazirani, ``{Computational pseudorandomness,
  the wormhole growth paradox, and constraints on the AdS/CFT duality},''
  \href{http://arxiv.org/abs/1910.14646}{{\ttfamily arXiv:1910.14646
  [quant-ph]}}.

\bibitem{Susskind:2020kti}
L.~Susskind, ``{Horizons Protect Church-Turing},''
  \href{http://arxiv.org/abs/2003.01807}{{\ttfamily arXiv:2003.01807
  [hep-th]}}.

\bibitem{Kim:2020cds}
I.~Kim, E.~Tang, and J.~Preskill, ``{The ghost in the radiation: Robust
  encodings of the black hole interior},''
  \href{http://arxiv.org/abs/2003.05451}{{\ttfamily arXiv:2003.05451
  [hep-th]}}.

\bibitem{Yoshida:2020wpd}
B.~Yoshida, ``{Remarks on Black Hole Complexity Puzzle},''
  \href{http://arxiv.org/abs/2005.12491}{{\ttfamily arXiv:2005.12491
  [hep-th]}}.

\bibitem{Preskill:2018fag}
J.~Preskill, ``{Simulating quantum field theory with a quantum computer},''
  \href{http://dx.doi.org/10.22323/1.334.0024}{{\em PoS} {\bfseries
  LATTICE2018} (2018) 024}, \href{http://arxiv.org/abs/1811.10085}{{\ttfamily
  arXiv:1811.10085 [hep-lat]}}.

\bibitem{cyber}
J.~Liu, ``Aspects of cyberpunkian quantum field theory,
  \url{https://github.com/junyuphybies/cyber/blob/master/Cyber_QF.pdf}.,''.

\bibitem{Liu:2020eoa}
J.~Liu and Y.~Xin, ``{Quantum simulation of quantum field theories as quantum
  chemistry},'' \href{http://arxiv.org/abs/2004.13234}{{\ttfamily
  arXiv:2004.13234 [hep-th]}}.

\bibitem{bao2019quantum}
N.~Bao and J.~Liu, ``Quantum algorithms for conformal bootstrap,'' {\em Nuclear
  Physics B} {\bfseries 946} (2019) 114702.

\bibitem{albash2018adiabatic}
T.~Albash and D.~A. Lidar, ``Adiabatic quantum computation,'' {\em Reviews of
  Modern Physics} {\bfseries 90} no.~1, (2018) 015002.

\bibitem{uni}
S.~Lloyd, ``Universal quantum simulators,'' {\em Science} (1996) 1073--1078.

\bibitem{tro}
A.~M. Childs, Y.~Su, M.~C. Tran, N.~Wiebe, and S.~Zhu, ``A theory of trotter
  error,'' {\em arXiv preprint arXiv:1912.08854} (2019) .

\bibitem{jordan2016black}
S.~P. Jordan, ``Black holes, quantum mechanics, and the limits of
  polynomial-time computability,'' {\em XRDS: Crossroads, The ACM Magazine for
  Students} {\bfseries 23} no.~1, (2016) 30--33.

\bibitem{Jordan:2011ci}
S.~P. Jordan, K.~S. Lee, and J.~Preskill, ``Quantum computation of scattering
  in scalar quantum field theories,'' {\em arXiv preprint arXiv:1112.4833}
  (2011) .

\bibitem{jordan2014quantum}
S.~P. Jordan, K.~S. Lee, and J.~Preskill, ``Quantum algorithms for fermionic
  quantum field theories,'' {\em arXiv preprint arXiv:1404.7115} (2014) .

\bibitem{jordan2017fast}
S.~P. Jordan, ``Fast quantum computation at arbitrarily low energy,'' {\em
  Physical Review A} {\bfseries 95} no.~3, (2017) 032305.

\bibitem{moosavian2018faster}
A.~H. Moosavian and S.~Jordan, ``Faster quantum algorithm to simulate fermionic
  quantum field theory,'' {\em Physical Review A} {\bfseries 98} no.~1, (2018)
  012332.

\bibitem{jordan2018bqp}
S.~P. Jordan, H.~Krovi, K.~S. Lee, and J.~Preskill, ``Bqp-completeness of
  scattering in scalar quantum field theory,'' {\em Quantum} {\bfseries 2}
  (2018) 44.

\bibitem{moosavian2019site}
A.~H. Moosavian, J.~R. Garrison, and S.~P. Jordan, ``Site-by-site quantum state
  preparation algorithm for preparing vacua of fermionic lattice field
  theories,'' {\em arXiv preprint arXiv:1911.03505} (2019) .

\bibitem{chakraborty2020digital}
B.~Chakraborty, M.~Honda, T.~Izubuchi, Y.~Kikuchi, and A.~Tomiya, ``Digital
  quantum simulation of the schwinger model with topological term via adiabatic
  state preparation,'' {\em arXiv preprint arXiv:2001.00485} (2020) .

\bibitem{kreshchuk2020quantum}
M.~Kreshchuk, W.~M. Kirby, G.~Goldstein, H.~Beauchemin, and P.~J. Love,
  ``Quantum simulation of quantum field theory in the light-front
  formulation,'' {\em arXiv preprint arXiv:2002.04016} (2020) .

\bibitem{mcardle2020quantum}
S.~McArdle, S.~Endo, A.~Aspuru-Guzik, S.~C. Benjamin, and X.~Yuan, ``Quantum
  computational chemistry,'' {\em Reviews of Modern Physics} {\bfseries 92}
  no.~1, (2020) 015003.

\bibitem{berry2015simulating}
D.~W. Berry, A.~M. Childs, R.~Cleve, R.~Kothari, and R.~D. Somma, ``Simulating
  hamiltonian dynamics with a truncated taylor series,'' {\em Physical review
  letters} {\bfseries 114} no.~9, (2015) 090502.

\bibitem{childs2017toward}
A.~M. Childs, D.~Maslov, Y.~Nam, N.~J. Ross, and Y.~Su, ``Toward the first
  quantum simulation with quantum speedup,'' {\em arXiv preprint
  arXiv:1711.10980} (2017) .

\bibitem{babbush2018encoding}
R.~Babbush, C.~Gidney, D.~W. Berry, N.~Wiebe, J.~McClean, A.~Paler, A.~Fowler,
  and H.~Neven, ``Encoding electronic spectra in quantum circuits with linear t
  complexity,'' {\em Physical Review X} {\bfseries 8} no.~4, (2018) 041015.

\bibitem{spar}
D.~W. Berry, A.~M. Childs, R.~Cleve, R.~Kothari, and R.~D. Somma, ``Exponential
  improvement in precision for simulating sparse hamiltonians,'' in {\em Forum
  of Mathematics, Sigma}, vol.~5, Cambridge University Press.
\newblock 2017.

\bibitem{kieferova2019simulating}
M.~Kieferova, A.~Scherer, and D.~W. Berry, ``Simulating the dynamics of
  time-dependent hamiltonians with a truncated dyson series,'' {\em Physical
  Review A} {\bfseries 99} no.~4, (2019) 042314.

\bibitem{low2018hamiltonian}
G.~H. Low, ``Hamiltonian simulation with nearly optimal dependence on spectral
  norm,'' {\em arXiv preprint arXiv:1807.03967} (2018) .

\bibitem{suzuki1976trotter}
M.~Suzuki, ``Generalized trotter's formula and systematic approximants of
  exponential operators and inner derivations with applications to many-body
  problems,'' \href{http://dx.doi.org/10.1007/BF01609348}{{\em Communications
  in Mathematical Physics} {\bfseries 51} no.~2, (Jun, 1976) 183--190}.
  \url{https://doi.org/10.1007/BF01609348}.

\bibitem{PhysRevA.78.052325}
W.~D\"ur, M.~J. Bremner, and H.~J. Briegel, ``Quantum simulation of interacting
  high-dimensional systems: The influence of noise,''
  \href{http://dx.doi.org/10.1103/PhysRevA.78.052325}{{\em Phys. Rev. A}
  {\bfseries 78} (Nov, 2008) 052325}.
  \url{https://link.aps.org/doi/10.1103/PhysRevA.78.052325}.

\bibitem{csahinouglu2020hamiltonian}
B.~{\c{S}}ahino{\u{g}}lu and R.~D. Somma, ``Hamiltonian simulation in the low
  energy subspace,'' {\em arXiv preprint arXiv:2006.02660} (2020) .

\bibitem{childs2018faster}
A.~M. Childs, A.~Ostrander, and Y.~Su, ``Faster quantum simulation by
  randomization,'' {\em arXiv preprint arXiv:1805.08385} (2018) .

\bibitem{campbell2018random}
E.~Campbell, ``A random compiler for fast hamiltonian simulation,'' {\em arXiv
  preprint arXiv:1811.08017} (2018) .

\bibitem{ouyang2019stochastichamiltonian}
Y.~Ouyang, D.~R. White, and E.~Campbell, ``Compilation by stochastic
  hamiltonian sparsification,''
  \href{http://arxiv.org/abs/arXiv:1910.06255}{{\ttfamily arXiv:1910.06255}}.

\bibitem{chen2020quantum}
C.-F. Chen, R.~Kueng, J.~A. Tropp, {\em et~al.}, ``Quantum simulation via
  randomized product formulas: Low gate complexity with accuracy guarantees,''
  {\em arXiv preprint arXiv:2008.11751} (2020) .

\bibitem{wiebe2011simulating}
N.~Wiebe, D.~W. Berry, P.~H{\o}yer, and B.~C. Sanders, ``Simulating quantum
  dynamics on a quantum computer,''
  \href{http://dx.doi.org/https://doi.org/10.1088/1751-8113/44/44/445308}{{\em
  Journal of Physics A: Mathematical and Theoretical} {\bfseries 44} no.~44,
  (2011) 445308}.

\bibitem{PoulinTimeDependent}
D.~Poulin, A.~Qarry, R.~Somma, and F.~Verstraete, ``Quantum simulation of
  time-dependent hamiltonians and the convenient illusion of hilbert space,''
  \href{http://dx.doi.org/10.1103/PhysRevLett.106.170501}{{\em Phys. Rev.
  Lett.} {\bfseries 106} (Apr, 2011) 170501}.
  \url{https://link.aps.org/doi/10.1103/PhysRevLett.106.170501}.

\bibitem{childs2011walks}
A.~M. Childs and R.~Kothari, ``Simulating sparse hamiltonians with star
  decompositions,'' in {\em Theory of Quantum Computation, Communication, and
  Cryptography}, W.~van Dam, V.~M. Kendon, and S.~Severini, eds., pp.~94--103.
\newblock Springer Berlin Heidelberg, Berlin, Heidelberg, 2011.

\bibitem{berry2015hamiltonian}
D.~W. Berry, A.~M. Childs, and R.~Kothari, ``Hamiltonian simulation with nearly
  optimal dependence on all parameters,'' in {\em Foundations of Computer
  Science (FOCS), 2015 IEEE 56th Annual Symposium on}, pp.~792--809, IEEE.
\newblock 2015.

\bibitem{low2019multiproduct}
G.~H. Low, V.~Kliuchnikov, and N.~Wiebe, ``Well-conditioned multiproduct
  hamiltonian simulation,''
  \href{http://arxiv.org/abs/arXiv:1907.11679}{{\ttfamily arXiv:1907.11679}}.

\bibitem{childs2012multiproduct}
A.~M. Childs and N.~Wiebe, ``Hamiltonian simulation using linear combinations
  of unitary operations,'' {\em Quantum Info. Comput.} {\bfseries 12}
  no.~11-12, (Nov., 2012) 901--924.
  \url{http://dl.acm.org/citation.cfm?id=2481569.2481570}.

\bibitem{berry2012black}
D.~W. Berry and A.~M. Childs, ``Black-box hamiltonian simulation and unitary
  implementation,'' {\em Quantum Info. Comput.} {\bfseries 12} no.~1-2, (Jan.,
  2012) 29--62. \url{http://dl.acm.org/citation.cfm?id=2231036.2231040}.

\bibitem{subramanian2018implementing}
S.~Subramanian, S.~Brierley, and R.~Jozsa, ``Implementing smooth functions of a
  hermitian matrix on a quantum computer,'' {\em arXiv preprint
  arXiv:1806.06885} (2018) .

\bibitem{LowQSPprx}
G.~H. Low, T.~J. Yoder, and I.~L. Chuang, ``Methodology of resonant equiangular
  composite quantum gates,''
  \href{http://dx.doi.org/10.1103/PhysRevX.6.041067}{{\em Phys. Rev. X}
  {\bfseries 6} (Dec, 2016) 041067}.
  \url{https://link.aps.org/doi/10.1103/PhysRevX.6.041067}.

\bibitem{sawaya2020resource}
N.~P. Sawaya, T.~Menke, T.~H. Kyaw, S.~Johri, A.~Aspuru-Guzik, and G.~G.
  Guerreschi, ``Resource-efficient digital quantum simulation of d-level
  systems for photonic, vibrational, and spin-s hamiltonians,'' {\em npj
  Quantum Information} {\bfseries 6} no.~1, (2020) 1--13.

\end{thebibliography}\endgroup
\end{document}